\documentclass[aps,prd,superscriptaddress]{revtex4}

\usepackage{graphicx,color}
\usepackage{amsmath,amssymb,amsfonts}
\usepackage{multirow}
\usepackage{epsf}
\usepackage{color}

\def\CO{{\cal O}}
\newcommand{\ba}{\begin{eqnarray}}
\newcommand{\ea}{\end{eqnarray}}
\newcommand{\ban}{\begin{eqnarray*}}
\newcommand{\ean}{\end{eqnarray*}}
\newcommand{\be}{\begin{equation}}
\newcommand{\ee}{\end{equation}}

\newcommand {\tr} {{\rm Tr}\,}

\newcommand{\bea}{\ba}
\newcommand{\eea}{\ea}

\def\({\left (}
\def\){\right )}
\def\[{\left [}
\def\[{\right ]}

\def\tr{\mathrm{tr}}

\input amssym.def
\input amssym.tex
\usepackage{fancyhdr}
\RequirePackage{hyperref}

\begin{document}

\title{Holographic Thermalization}

\author{V.~Balasubramanian}
\affiliation{David Rittenhouse Laboratory, University of Pennsylvania, Philadelphia, PA 19104, USA}
\author{A.~Bernamonti}
\affiliation{Theoretische Natuurkunde, Vrije Universiteit Brussel, and
International Solvay Institutes,
Pleinlaan 2, B-1050 Brussels, Belgium}
\author{J.~de~Boer}
\affiliation{Institute for Theoretical Physics, University of Amsterdam,
Science Park 904, Postbus 94485, 1090 GL
Amsterdam, The Netherlands }
\author{N.~Copland}
\affiliation{Theoretische Natuurkunde, Vrije Universiteit Brussel, and
International Solvay Institutes,
Pleinlaan 2, B-1050 Brussels, Belgium}
\author{B.~Craps}
\affiliation{Theoretische Natuurkunde, Vrije Universiteit Brussel, and
International Solvay Institutes,
Pleinlaan 2, B-1050 Brussels, Belgium}
\author{E.~Keski-Vakkuri}
\affiliation{Helsinki Institute of Physics \& Department of Physics,
P.O.Box 64, FIN-00014 University of Helsinki, Finland, and Department of Physics and Astronomy,
Uppsala University, SE-751 08 Uppsala, Sweden}
\author{B.~M\"uller}
\affiliation{Department of Physics \& CTMS, Duke University, Durham, NC 27708, USA}
\author{A.~Sch\"afer}
\affiliation{Institut f\"{u}r Theoretische Physik, Universit\"{a}t Regensburg, D-93040 Regensburg, Germany}
\author{M.~Shigemori}
\affiliation{Kobayashi-Maskawa Institute for the Origin of Particles and the Universe,
Nagoya University, Nagoya 464-8602, Japan}
\author{W.~Staessens}
\affiliation{Theoretische Natuurkunde, Vrije Universiteit Brussel, and
International Solvay Institutes,
Pleinlaan 2, B-1050 Brussels, Belgium}

%\date{Updated 10 May 2010}

\begin{abstract}
Using the AdS/CFT correspondence, we probe the scale-dependence of thermalization in strongly coupled field theories following a quench, via calculations of two-point functions, Wilson loops and entanglement entropy in $d=2,3,4$.   In the saddlepoint approximation these probes are computed in AdS space in terms of invariant geometric objects -- geodesics, minimal surfaces and minimal volumes.    Our calculations for two-dimensional field theories are analytical.   In our strongly coupled setting, all probes in all dimensions share certain universal features in their thermalization: (1) a slight delay in the onset of thermalization, (2) an apparent non-analyticity at the endpoint of thermalization, (3) top-down thermalization where the UV thermalizes first.   For homogeneous initial conditions the entanglement entropy thermalizes slowest, and sets a timescale for equilibration that saturates a causality bound over the range of scales studied.  The growth rate of entanglement entropy density is nearly volume-independent for small volumes, but slows for larger volumes.
\end{abstract}

\maketitle

\tableofcontents

%%%%%%%%%%%%%%%%%%%%%%%%%%%%%%%%%%%%%%%%%%%%%%%%%%%%%%%%%%%%%%%%%%%%%%%

\section{Introduction}

%%%%%%%%%%%%%%%%%%%%%%%%%%%%%%%%%%%%%%%%%%%%%%%%%%%%%%%%%%%%%%%%%%%%%%%

\subsection{Motivation}

The observed nearly inviscid hydrodynamic expansion of the hot QCD
matter produced in nuclear collisions at the Relativistic Heavy Ion
Collider (RHIC) indicates that matter produced in these nuclear
reactions is strongly coupled
\cite{Whitepapers:2005,Gyulassy:2004zy,Harris:1996zx}. The inability of
perturbation theory to account for this phenomenon has motivated studies
of nonequilibrium dynamics in analytically tractable, strongly coupled
gauge theories. The prototype of such theories is the maximally
supersymmetric Yang-Mills theory in the large-$N_c$, large 't~Hooft
coupling limit, which is holographically dual to the classical limit of
a superstring theory on the Anti-de Sitter space AdS$_5\times$S$^5$
background \cite{Maldacena:1997re}.  Hydrodynamical evolution with a
minimal shear viscosity emerges naturally as late-time behavior in the
longitudinal expansion of a thermal gauge plasma in this model
\cite{Policastro:2001yc,Janik:2005zt}.  In the dual supergravity theory,
the thermal state of the gauge theory is represented by a black brane in
the asymptotic AdS$_5$ space, and the near-equilibrium dynamics of the
gauge theory giving rise to hydrodynamic behavior is mapped onto the
dynamics of perturbations of the AdS$_5$--black brane metric.
A nice recent review on the holographic study of hot QCD matter, focused
on equilibrium and near-equilibrium aspects, is
\cite{CasalderreySolana:2011us}.

While the near-equilibrium dynamics of the strongly coupled super-Yang-Mills (SYM) theory is well studied, the process of thermalization itself is still poorly understood. The RHIC data demand that the time scale for equilibration of matter is considerably shorter than expected in the framework of perturbative approaches to thermalization \cite{Baier:2000sb,Mueller:2005un}. The dearth of other nonperturbative tools for the description of the short pre-hydrodynamic stage of the nuclear collision motivates the use of the AdS/CFT correspondence to study thermalization of strongly coupled plasmas. Phenomenologically, the central rapidity region of a high energy collision turns out to be boost invariant to a good approximation, so it is natural to consider boost invariant configurations \cite{Janik:2005zt,Kovchegov:2007pq,Chesler:2008hg,Beuf:2009cx}. One interesting question is how thermal equilibrium is reached in such systems. This is the question we will address in the present paper, albeit in the case of translationally invariant plasmas.

We investigate the thermalization process not only in the strongly coupled $(3+1)=4-$dimensional super-Yang-Mills theory, but also in the analogous lower (2 and 3) dimensional field theories. This has several motivations. The first reason is that the $2-$dimensional version, which is dual to classical (super-)gravity on AdS$_3$ space, admits analytical solutions and thus allows us to explore thermalization over a wide range of parameters. A second motivation is that some version of the $2-$dimensional dual theory, $(1+1)$ dimensional super-Yang-Mills theory, can be formulated on a lattice and solved nonperturbatively \cite{Catterall:2010fx,Hanada:2010gs}. Finally, the comparison of theories in different dimensions makes it easier to distinguish generic aspects of thermalization of strongly coupled quantum field theories from aspects that are special features of $4-$dimensional field theories.

%%%%%%%%%%%%%%%%%%%%%%%%%%%%%%%%%%%%%%%%%%%%%%%%%%%%%%%%%%%%%%%%%%%%%%%

\subsection{Holographic models of thermalization}

%%%%%%%%%%%%%%%%%%%%%%%%%%%%%%%%%%%%%%%%%%%%%%%%%%%%%%%%%%%%%%%%%%%%%%%

\subsubsection{Thermalization scenarios}

What is appropriately called the thermalization time may not only depend on the probe of the state of the field but also on the initial field configuration that evolves toward thermal equilibrium. Here one can distinguish between two broad classes of scenarios. The first class considers small perturbations around thermal equilibrium and asks how equilibrium is reached. For a gauge theory with a holographic dual, this approach amounts to studying the decay of small perturbations of an AdS black brane geometry. This is conveniently studied in terms of the quasi-normal modes of fields propagating in the black brane background \cite{Chandrasekhar:1975zza}. The imaginary parts of the eigenvalues of these modes describe the thermal relaxation rates of various excitations, such as anisotropic perturbations of the stress-energy tensor. In the context of relativistic heavy-ion collisions, this approach is relevant for the study of viscous corrections to hydrodynamics and other transport processes in the presence of a thermal gauge theory plasma (see {\em e.g.}\ \cite{Hubeny:2010ry} for a review).

The second class of scenarios considers the thermalization of an initial field configuration, which is generally far from equilibrium.
Through the AdS/CFT correspondence, the approach to thermal equilibrium in the boundary gauge theory is related to the process of black hole formation in the bulk. In early work, before the motivation from RHIC experiments, the goal was
to understand how black hole formation from a gravitational collapse of a shell of matter would be encoded in the field theory in the boundary \cite{Danielsson:1999zt,Giddings:1999zu,Danielsson:1999fa,Giddings:2001ii}. Also in more recent work, the initial conditions have usually been defined directly in the dual gravity theory -- in most, if not all, of these cases the precise form of the initial condition is not known in terms of gauge theory excitations. Examples of such configurations are colliding gravitational shock waves \cite{Kang:2004jd}, sheets of fundamental strings that may provide an AdS model of densely packed flux tubes \cite{Lin:2006rf}, and sudden perturbations of the metric near the boundary that propagate into the bulk \cite{Chesler:2008hg,Bhattacharyya:2009uu}. Motivated by the properties of the initial state of a relativistic heavy-ion collision, which contains two highly energetic nuclei, these scenarios have in common that they inject energy into the AdS geometry at high momentum or short distance scales.

Some of these recent studies of gravitational collapse in AdS$_5$ started from translation invariant, but locally anisotropic perturbations of the metric near the boundary and followed their propagation into the bulk, ultimately resulting in the formation of an event horizon \cite{Chesler:2008hg}. The initial anisotropy of the metric is dissipated by the black brane, asymptotically resulting in viscous, boost-invariant scaling hydrodynamics on the boundary \cite{Beuf:2009cx,Janik:2006ft}. Gravitational collapse in AdS$_{d+1}$ induced by a scalar field perturbation was studied in \cite{Bhattacharyya:2009uu}, most explicitly for the case of AdS$_4$, where the bulk equations of motion for gravity coupled to a massless scalar field were solved perturbatively for a small amplitude scalar perturbation that propagates in from the AdS boundary. The perturbation was taken to be translationally invariant along the spatial directions of the boundary and to vanish outside of a short time interval. An interesting technical result is that for this translationally invariant collapse in AdS$_4$, the metric outside the infalling shell of matter coincides with a black brane metric to first non-trivial order in the perturbation \cite{Bhattacharyya:2009uu}.  This has the consequence that expectation values of local observables thermalize essentially instantaneously in the thermally quenched field theory dual to the infalling shell background.

Similarly to the approach of \cite{Bhattacharyya:2009uu}, in \cite{Lin:2008rw} the equilibrating field configuration was modeled by a homogeneous, infalling thin mass shell. The motivation for choosing such a configuration was that it naturally arises in the AdS dual description of the asymptotic limit of a transversely extended ensemble of flux tubes \footnote{Hence, in distinction to the matter shells in early work, the shell has tension which in part contributes to the collapse dynamics.}.  The latter represent the energy density deposited by the colliding leading gauge charges (i.e.\ the valence quarks of the colliding nuclei), and the AdS description involves a sufficiently simple geometry to admit analytic or semi-analytic solutions.  
Our model does not allow us to study the approach to local isotropy and hydrodynamical behavior, but it enables us to investigate the approach of the field to a thermal configuration in momentum space and to measure the growth of its entanglement entropy, as discussed in Sections~\ref{s:3surf} and \ref{s:dynahyper}. In this sense, our present study is complementary to those mentioned above which focus on the approach to the hydrodynamical limit.  Note that the work \cite{Lin:2008rw} considered a quasi-static approximation with metrics that correspond to stationary ``snapshots'' of the dynamic geometry containing the infalling mass shell.  Here we study both the dynamic and the quasi-stationary geometries and show that there are some significant differences in the pictures that they yield of thermalization. Various aspects of the dynamic case of a thin infalling shell have been investigated in \cite{Hubeny:2007xt,AbajoArrastia:2010yt,Albash:2010mv}. The effect of a dynamical boundary condition has also been studied in the background of a moving mirror \cite{Erdmenger:2011jb}. New techniques for the evaluation of various Green functions in non-equilibrium
backgrounds have been developed in \cite{CaronHuot:2011dr}, in addition to the complex contour approach by
\cite{Skenderis:2008dh}. The counterpart to Chandrasekhar limit for a gravitational collapse of
a degenerate star to a black hole in AdS/CFT has been investigated in \cite{Arsiwalla:2010bt}. Finally, thermalization after a quantum quench in gauge theory has also been modeled by a sudden change in couplings or other background fields than the bulk metric \cite{Das:2010yw, Hashimoto:2010wv}. In probe brane approach such a change then can manifest as time-dependence in the induced metric on the brane.

Concretely, we focus on the ($d+1$)-dimensional infalling shell geometry described in Poincar\'e coordinates by the Vaidya metric
\be
ds^2 = \frac{1}{z^2}\left[-\left(1 - m(v)z^d\right) dv^2 - 2 dz\, dv + d\mathbf{x}^2 \right] \,,
\label{eq:Vaidya}
\ee
where $v$ labels ingoing null trajectories and we have set the AdS radius equal to 1. The boundary space-time is located at $z=0$ and $\mathbf{x}=(x_1,\dots,x_{d-1})$ correspond to the spatial coordinates on the boundary.
For constant $m(v) =M$, the coordinate transformation
\be
dv=dt-\frac{dz}{1- M\, z^d}
\label{eq:dt_and_dv}
\ee
brings the metric \eqref{eq:Vaidya} in the form
\be
ds^2 = \frac{1}{z^2} \left[-\left(1 - M z^d\right) dt^2 + \frac{ dz^2}{1-M z^d} + d\mathbf{x}^2 \right]\,.
\ee
On the boundary, the coordinates $v$ and $t$ coincide.
We take the mass function of the infalling shell to be
\be
m(v)=\frac{M}{2}\,\left( 1+\tanh\frac{v}{v_0} \right) ,
\label{eq:mv}
\ee
where $v_0$ parametrizes the thickness of the shell falling along $v=0$. We will often be interested in the zero thickness limit $v_0 \to 0$. The Vaidya metric describes a shell composed of tensionless null dust, which represents an analytical simplification of the tensionful shell models studied in \cite{Lin:2008rw} and of the model of \cite{Bhattacharyya:2009uu}.

\begin{figure}[htbp]
\begin{center}
\includegraphics[width=0.3 \textwidth]{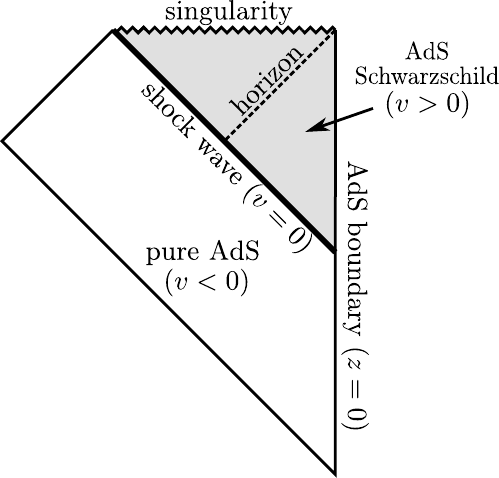}
\end{center}
\caption{The causal structure of the Vaidya spacetime (in the $v_0 \to 0$ limit) shown in the
Poincar\'{e} patch of AdS space.  In this presentation, the asymptotic
boundary (vertical line on the right hand side) is planar, and the null
lines on the left hand side of the diagram represent the Poincar\'{e}
horizon.
} \label{fig:VaidyaSpacetime}
\end{figure}

The geometry outside the infalling shell (which corresponds to $v>0$ in the zero thickness limit) is identical to that of an AdS black brane geometry with Hawking temperature $T=dM^{1/d}/4\pi$, while the geometry inside the shell ($v<0$ when $v_0 \to 0$) is the same as that of pure
AdS geometry.  The causal structure of the Vaidya space-time is shown in Fig.~\ref{fig:VaidyaSpacetime}.

%%%%%%%%%%%%%%%%%%%%%%%%%%%%%%%%%%%%%%%%%%%%%%%%%%%%%%%%%%%%%%%%%%%%%%%

\subsubsection{Probes of thermalization}

In order to probe the dynamics of the thermalization process, one can study a variety of observables of the boundary gauge theory. Expectation values of local gauge-invariant operators, such as the energy-momentum tensor and its derivatives provide valuable information about the applicability of viscous hydrodynamics, but they cannot be used to explore deviations from thermal equilibrium in detail.   Nonlocal observables such as pair correlation functions, Wilson loop expectation values, and entanglement entropy provide much more information about progress towards thermalization.   (They are also relevant to the physics probed in relativistic heavy ion collisions, e.~g.\ through the jet quenching parameter $\hat{q}$ \cite{Kovner:2001vi,Liu:2006he} and the color screening length.)

To illustrate, consider spatially homogeneous states of a weakly interacting massless scalar field. The energy-momentum tensor can be expressed as
\be
T^{\mu\nu} = \int \frac{d{\bf k}}{k^0}\, k^\mu k^\nu n({\bf k}) ,
\label{eq:Tmunu}
\ee
where $n({\bf k}) =  \langle a^{\dagger}_{\bf k} a_{\bf k} \rangle$ denotes the occupation number of a momentum mode of the scalar field.   (There is no spatial dependence of the stress tensor because the field configuration is homogeneous here.) It is obvious from (\ref{eq:Tmunu}) that this observable contains only limited information about the particle distribution. In particular, $T^{\mu\nu}$ cannot inform us whether thermal equilibrium has been reached; it only tells us whether the pressure is locally isotropic. This is not even a sufficient condition for the matching of the field theory to hydrodynamics, because the equation of state relies on the assumption of thermal occupation numbers for the various modes.  It is also insufficient to answer many other questions, {\em e.g.}\ about the expected spectrum of radiation emitted by collisions among the scalar particles. The equal-time two-point function, on the other hand, is given by
\be
G({\bf x}) =  \int \frac{d{\bf k}}{k^0}\, [n({\bf k})+1] \exp(i{\bf k}\cdot{\bf x}) ,
\label{eq:Gx}
\ee
which allows us to extract detailed information about the particle distribution and thus probe its closeness to a thermal distribution in detail.  Note that even the two-point function is insufficient to probe for the phase relationship between the occupation amplitudes of different field modes. This information can, however, be obtained by an analysis of the four-point function of the field.

In an interacting quantum field theory, the two-point function is determined not only by the mode occupation numbers $n({\bf k})$ but also by the spectral density function $\sigma(k^0,{\bf k})$. For the free field, $\sigma(k)=2\pi\delta(k^2-m^2)$, but more generally the spectral function is determined by the exact self-energy $\Pi(k^0,{\bf k})$ as follows:
\be
\sigma(k^0,{\bf k}) = \frac{2\,{\rm Im}\, \Pi(k^0,{\bf k})}
 {\left[(k^0)^2-{\bf k}^2-m^2-{\rm Re}\,\Pi(k^0,{\bf k})\right]^2 + {\rm Im}\, \Pi(k^0,{\bf k})^2} .
\label{eq:sigma(k0,k)}
\ee
Since the self-energy is a function of the temperature, so is the spectral density function.  For a weakly coupled field, the medium dependence is reflected, {\em e.g.}\ , in medium modifications of particle masses and widths. In the strongly coupled gauge theory, the vacuum spectral function has the generic form \cite{Kovtun:2006pf}
\be
\sigma(k^0,{\bf k})\sim ((k^0)^2-{\bf k}^2)^\nu \theta((k^0)^2 - {\bf k}^2) ,
\label{eq:strongsigma}
\ee
where the exponent $\nu$ depends on the dimension of the considered field. This becomes a sigmoidal function at finite temperature. The mode occupation number $n({\bf k})$ and the spectral function $\sigma(k^0,{\bf k})$ combine multiplicatively to yield the true, physical excitation probability of a momentum mode ${\bf k}$. The time evolution of the equal-time two-point function probes the approach of this excitation probability to thermal equilibrium.

In the AdS/CFT correspondence there is a geometric intuition for why expectation values of local gauge-invariant operators are insensitive to details of the progress towards thermalization -- they are only sensitive to phenomena near the AdS boundary. Thus they do not probe the details of phenomena occurring near the thermal scale.
By contrast, nonlocal objects, such as two-point correlators of gauge-invariant operators and expectation values of Wilson loops, are dual to AdS  quantities that probe deeper into the space-time and further away from the boundary.  For example, at strong 't Hooft coupling, it is possible to approximate the path integral for the connected two-point correlator as a sum over all possible AdS geodesics connecting the two endpoints, which are placed on the AdS boundary \cite{Balasubramanian:1999zv}. (For certain subtleties regarding this statement, see \cite{Louko:2000tp,  Fidkowski:2003nf, Festuccia:2005pi}.) The geodesics probe the interior of AdS space, which is dual to the statement that they are sensitive to a wide range of energy scales in the boundary field theory.   As we will see, Wilson loop expectation values \cite{Maldacena:1998im} and entanglement entropy \cite{Ryu:2006bv,Hubeny:2007xt,Nishioka:2009un} are also related to minimal lengths, surfaces and volumes of various kinds in AdS\@.  These also extend into the bulk of AdS and hence probe a range of energy scales.

%%%%%%%%%%%%%%%%%%%%%%%%%%%%%%%%%%%%%%%%%%%%%%%%%%%%%%%%%%%%%%%%%%%%%%%

\subsection{Overview of this work}

We consider two-, three-, and four-dimensional field theories dual to gravity in asymptotically AdS$_3$, AdS$_4$, and AdS$_5$ spacetimes, respectively. The motivation for studying all these cases (rather than only the case of four-dimensional field theories potentially relevant for heavy-ion collisions) is that we are interested in generic conclusions on strongly coupled field theories with gravity duals. Lower dimensional theories are technically much simpler: analytic computations are possible in AdS$_3$.   Furthermore, our basic methods are similar in all dimensions -- we calculate field theory observable that can be related to invariant geometric objects -- minimal lengths, surfaces and volumes -- in AdS spacetimes of different dimension.
Our results show that many conclusions are not sensitive to the dimension.

Sec.~\ref{Probes} discusses three measures of thermalization -- two-point functions, Wilson loop expectation values, and entanglement entropy. We compute these observables at thermal equilibrium, and, in later sections, use deviations from these results as  probes of incomplete thermalization.  Sec.~\ref{thermalization} relates thermal quenches in two-, three- and four-dimensional theories, to collapsing shells of matter in AdS$_3$, AdS$_4$ and AdS$_5$ that form black branes.   The black brane formation represents the process of equilibration and thermalization of energy injected in the dual field theory.  In these dynamical backgrounds we compute nonlocal observables as a function of time, and track the rate at which thermalization progresses for each observable at different spatial scales in the field theory.  Sec.~\ref{Discussion} concludes the paper by summarizing our findings.  An Appendix describes how the results would differ in a quasistatic approximation. A brief account of this work has appeared in
\cite{Balasubramanian:2010ce}.

One broad finding is that these models exhibit a ``top-down'' thermalization mechanism, see also \cite{Lin:2008rw}. We would like to stress that this is very much in contrast with the standard ``bottom-up'' paradigm \cite{Baier:2000sb} based on perturbative gauge theory.  In the ``bottom-up'' scenario, hard quanta of the gauge field do not equilibrate directly by randomizing their momenta in two-body collisions, but they do so by radiating softer quanta, which gradually fill up the thermal phase space and equilibrate by collisions among themselves. In other words, infrared modes are the first to thermalize, and the thermalization proceeds gradually (``from the bottom up'') to more energetic modes of the gauge field. The bottom-up scenario is closely linked to the infrared divergence of the splitting functions of gauge bosons and fermions in the perturbative gauge theory. This contrasts with the ``democratic'' splitting properties of excitations in the strongly coupled SYM theory, which favor an approximately equal sharing of energy and momentum \cite{Hatta:2008tx,Iancu:2008sp}. We should therefore expect that the thermalization process proceeds quite differently in the strongly coupled gauge theory. The phase space for radiation and, at strong coupling, also the spectral weight (\ref{eq:strongsigma}) is largest for the most energetic quanta. 
They then divide their energy and momentum approximately equally among their siblings, leading to a rapid cascading down to a thermal distribution. This consideration suggests that the thermalization process more closely resembles a ``top-down'' scenario, in which energetic gauge field modes equilibrate first and soft modes last.   Further broad conclusions appear in Sec.~\ref{Discussion}.

It is appropriate here to comment on the differences and
overlap between the current work and the recent related work in the
literature.  In \cite{Hubeny:2007xt,AbajoArrastia:2010yt}, geodesics (or
equivalently, entanglement entropy) in AdS$_3$ Vaidya were studied
numerically for a thin but finite shell.  In contrast, we also study
geodesics in AdS$_3$ Vaidya analytically for an infinitesimally thin
shell.  In \cite{Albash:2010mv}, Wilson loops (or equivalently,
entanglement entropy) were studied in AdS$_4$ Vaidya, for the case of
circular and rectangular loops.  We do not only confirm their results
but also study geodesics in AdS$_4$ Vaidya, which have not been closely
investigated in the context of holographic thermalization as far as we
know. Moreover, we study AdS$_5$ Vaidya geometries for the first time by
computing geodesics, circular and rectangular Wilson lines, and
entanglement entropy for spherical volumes on the boundary.

%%%%%%%%%%%%%%%%%%%%%%%%%%%%%%%%%%%%%%%%%%%%%%%%%%%%%%%%%%%%%%%%%%%%%%%

\section{Probes of thermality}\label{Probes}

We are interested in processes where energy is injected into a strongly coupled field theory and equilibrates over time.   Thus we need probes that take {\it instantaneous} snapshots of the state of the system which can be compared to the thermally equilibrated state.  As discussed above, expectation values of {\it local} gauge-invariant operators are insufficient.  Thus, in this section we describe three nonlocal probes -- two point functions, Wilson loop expectation values, and entanglement entropy -- which can be elegantly computed using geometric techniques in the AdS/CFT correspondence.  We will show how these quantities computed in strongly coupled theory at thermal equilibrium differ from the corresponding quantities in the vacuum.   In Sec.~\ref{thermalization} we will use the same quantities as probes of thermalization following a quench, by comparing the instantaneous values of the probes with the values at thermal equilibrium.

%%%%%%%%%%%%%%%%%%%%%%%%%%%%%%%%%%%%%%%%%%%%%%%%%%%%%%%%%%%%%%%%%%%%%%%

\subsection{Equal-time two point functions}

We begin our exploration with the equal-time two-point functions, which can be analytically calculated in two-dimensional conformal field theories at thermal equilibrium. We first remind the reader how the thermal occupation probabilities of momentum modes enter into the two-point function and how they are related to the spatial form of the equal-time Wightman function. We then explain how we generalize this approach to higher dimensional theories, where we can easily calculate the Wightman function in the saddle-point approximation.

%%%%%%%%%%%%%%%%%%%%%%%%%%%%%%%%%%%%%%%%%%%%%%%%%%%%%%%%%%%%%%%%%%%%%%%

\subsubsection{Two-dimensional free scalar field in thermal equilibrium}

We would like to understand how we can extract information about the thermalization process from the time dependence of the Wightman function
\be
G_{\cal O}^>(t,x;t',x') = \langle {\cal O}(t,x) {\cal O}(t',x') \rangle ,
\ee
where $\cal O$ is a local operator of dimension $\Delta$.  As a warmup exercise we consider the free
massless scalar field in two dimensions with mode expansion
\be
\phi(t,x) = \phi (x^-,x^+) = \int_{k^+\geq 0} \frac{dk^+}{2\pi} \left[\frac{1}{k^+} e^{-ik^+x^-}a_{k^+}^{\dagger} + cc\right] +
\int_{k^-\geq 0} \frac{dk^-}{2\pi} \left[\frac{1}{k^-} e^{-ik^-x^+}\tilde{a}_{k^-}^{\dagger} + cc\right].
\ee
We defined the operators $a_k^+$ as dimensionless, which implies the commutation relation $[a_{k^+},a_{k'^+}^\dagger] = k^+\delta(k^+-k'^+)$, etc. Employing the free spectral function $\sigma_0(k) = 2\pi\delta(k^2) = 2\pi\delta(k^+k^-)$, this can be written as
\be
\phi(x^-,x^+) = \int \frac{d^2k}{(2\pi)^2} \sigma_0(k) e^{ik^+x^- + ik^-x^+} b_{k^+,k^-} ,
\ee
where for $k^+,k^- \geq 0$:
$
b_{k^+,0} = a_{k^+}, \quad b_{-k^+,0} = a_{k^+}^{\dagger} , \qquad
b_{0,k^-} = \tilde{a}_{k^-}, \quad b_{0,-k^-} = \tilde{a}_{k^-}^{\dagger} .
$
Consider the dimension one operators  $\partial^+ \phi$ or $\partial^- \phi$. The vacuum two point function of $\partial^\pm \phi$ is time and space translation invariant, so we consider
\be
G_\pm^>(x^\pm) = \langle \partial^\pm \phi(0) \partial^\pm \phi(x^\pm) \rangle = \int_{k^\pm\geq 0} \frac{d^2k}{(2\pi)^2} (k^\pm)^2 \sigma_0(k) e^{-i k^+ x^- -i k^- x^+} =  \int_{k^\pm\geq 0} \frac{dk^\pm}{2\pi} k^+ e^{-i k^\pm x^\mp} \sim \frac{1}{(x^\mp)^2}.
\ee
For nonzero temperature, the spectral function of the free field remains unchanged, and we  find
\be
G_+^>(x^\pm;T) =  \int_{k^+>0} dk^+\, k^+ \left( e^{-ik^+x^-} (n(k^+)+1)+ e^{ik^+x^-} n(k^+) \right) ,
\ee
where $n(k)=(e^{\beta k}-1)^{-1}$ is the standard (Bose) thermal occupation number. This can be rewritten as
\be \label{weakcouplingscalar}
\int_{k^+} dk^+\, e^{-i k^+ x_+} \frac{e^{\beta k^+}}{e^{\beta k^+}-1} k^+
\sim \frac{1}{\beta^2 \sinh^2 \left( \frac{\pi x_+}{\beta}\right)}.
\ee
This is exactly the result expected from conformal invariance. The finite temperature computation is a computation on a Euclidean cylinder, which can be mapped to the complex plane using the exponential map, and we already know the answer on the plane. Undoing the coordinate transformation leads to the above two-point function.

%%%%%%%%%%%%%%%%%%%%%%%%%%%%%%%%%%%%%%%%%%%%%%%%%%%%%%%%%%%%%%%%%%%%%%%

\subsubsection{Strongly interacting scalar field theory in equilibrium from AdS/CFT}
\label{wightmaneq}

For an interacting field theory the spectral function changes with temperature.  As a preparatory example of an interacting two-dimensional thermal field theory defined by a gauge-gravity duality we consider the AdS$_3$ case, i.e.\ the BTZ black hole, and make use of the results presented in \cite{Son:2002sd}.  Writing $\omega = (k^+ + k^-)/\sqrt{2}$ and $k = (k^+ - k^-)/\sqrt{2}$, the thermal Wightman function for a scalar field in such a theory is \cite{Son:2002sd}:
\be
G^>(k) = i \left( 1 - \coth \frac{\omega}{2T} \right) {\rm Im}\ G_R(k) = \frac{-2i}{e^{\omega/T}-1}\ {\rm Im}\ G_R(k) ,
\ee
where $G_R$ is the retarded Green's function and $-2~{\rm Im}\ G_R$ yields the spectral function. The precise form of the thermal retarded Green function depends on the dimension $\Delta$ of the field in the operator. For $\Delta = 1$, we have \cite{Son:2002sd}
\be
G_R(\omega,k) =  \frac{1}{2\pi} \left[ \psi\left( \frac{1}{2} -i\frac{\omega-k}{4\pi T}\right) + \psi\left( \frac{1}{2} -i\frac{\omega+k}{4\pi T}\right) \right] .
\ee
Using
$
{\rm Im}\ \psi( \frac{1}{2}+iy) = \frac{\pi}{2} \tanh \pi y ,
$
gives the spectral density of states as
\be
\sigma(\omega,k) = - 2{\rm Im}\ G_R(\omega,k) = \frac{\sinh\frac{\omega}{2T}}{\cosh\frac{\omega}{2T} + \cosh\frac{k}{2T}} .
\ee

What can we learn from the equal-time Wightman function? One strategy is to use the vacuum subtracted Wightman function which at thermal equilibrium depends only on $\Delta t=t'-t$
\be
G^>_{{\rm sub}}(\Delta t,k;T) = G^>(\Delta t,k;T) - G^>(\Delta t,k;T=0) = \int \frac{d\omega}{2\pi}~ e^{-i\omega \Delta t}~\big[ G^>(\omega,k;T) - G^>(\omega,k;T=0) \big].
\ee
Using the Bose occupation number density $n_B(\omega) = (e^{\omega/T}-1)^{-1}$, the equal-time Wightman function ($\Delta t=0$) is then
\ba
G^>_{{\rm sub}}(0,k;T) &=&  i \int_{-\infty}^{\infty} \frac{d\omega}{2\pi} n_B(\omega) \sigma
(\omega,k;T)  + i \int_{-\infty}^{\infty} \frac{d\omega}{2\pi} \theta(-\omega) \sigma(\omega,k;0)  \nonumber \\
&=&
2 i \int_{0}^{\infty} \frac{d\omega}{2\pi} n_B(\omega) \sigma(\omega,k;T)
  + i \int_{0}^{\infty} \frac{d\omega}{2\pi} [\sigma(\omega,k;T) - \sigma(\omega,k;0) ] .
\ea
Inserting the explicit expressions, we obtain after some manipulations
\ba
G^>_{{\rm sub}}(0,k;T)
&=&
i~\int_0^{\infty} \frac{d\omega}{2\pi} ~\left[ \left( \frac{2}{e^{\omega/T}-1}+1 \right)
~ \frac{\sinh\frac{\omega}{2T}}{\cosh\frac{\omega}{2T} + \cosh\frac{k}{2T}}~-1\right]
+  i~\int_0^{k} \frac{d\omega}{2\pi}
\nonumber\\
&=&
-i~\int_0^{\infty} \frac{d\omega}{2\pi} ~
\frac{\cosh\frac{k}{2T}}{\cosh\frac{\omega}{2T} + \cosh\frac{k}{2T}}
+  i~\frac{k}{2\pi}
\ea
Introducing the notation $\cosh a = \coth(k/2T)$, and using the integral identity
\be
\int_0^{\infty} \frac{d\omega}{2T} ~ \left[ \sinh a \cosh\frac{\omega}{2T} + \cosh a \right]^{-1}
= \frac{1}{2} \ln \left|\frac{1+\cosh a}{1-\cosh a} \right|
= \frac{k}{2T}  ,
\ee
the final result reads
\be
G^>_{{\rm sub}}(k,T) = G^>(0,k;T) - G^>(0,k;T=0) = - \frac{i}{2\pi}~\frac{k}{e^{k/T}-1} .
\label{thermalads3wight}
\ee
Having derived the functional form of the equal-time Wightman function for a thermally equilibrated state in AdS$_3$,
we ask what we could learn from the same function at non-equilibrium background. In that case time translation
invariance is lost. As in the above, we could define a vacuum subtracted Wightman function $G^>_{{\rm sub}}(t,k)$ now depending on the equal-time $t=t'$, and expect that thermalization implies $G^>_{{\rm sub}}(t,k)\rightarrow G^>_{{\rm sub}}(k;T)$. This gives us a way to
estimate equilibration times for different momentum scales $k$, by analyzing how this limiting behavior is approached.
The result (\ref{thermalads3wight}), that the thermal contribution to the equal-time Wightman function of the two-dimensional field theory is proportional to the Bose distribution, justifies a more complete investigation of the equilibration properties of the Wightman function in two and more dimensions.
For this purpose, we now turn to the methods made available by the holographic duality between strongly coupled supersymmetric gauge theories and gravity theories. We will also analyze thermalization times at different scales,
in the spirit of the above motivation, but the details will differ somewhat.

%%%%%%%%%%%%%%%%%%%%%%%%%%%%%%%%%%%%%%%%%%%%%%%%%%%%%%%%%%%%%%%%%%%%%%%

\subsubsection{The geodesic approximation}\label{geodesicapprox}

While the approach described above would give detailed information regarding scale-dependent thermalization, there is a technical challenge for strongly coupled field theories, even if they have a dual description in terms of gravity in an AdS space. The easiest case is two-dimensional field theory with a three-dimensional gravity dual.  In this case, the thermal Wightman function in momentum space is known in closed form for operators of general conformal dimension $\Delta$ ($\Delta = 1$ was analyzed above) \cite{Son:2002sd}.   However, the integral with respect to $\omega$ that is needed to compute the equal-{\it time} Green function is sensitive to the ultraviolet completion of the theory and thus requires careful regulation.   Thus, it would be more convenient to directly compute the equal-time thermal Wightman function from the AdS$_3$ theory.   The next simplest case involves three-dimensional field theories with an AdS$_4$ dual.  In this case there is not even a closed form expression for the thermal Wightman function.  Thus we cannot pursue the approach discussed above analytically.    Furthermore, we are really interested in computing the equal-time Wightman function in a non-equilibrium setting in which matter collapsing in AdS to form a black hole models the thermalization of energy injected into the strongly interacting field theory.  In this situation, there are no known analytic solutions for the Green functions in any dimension, and we have to resort to approximation schemes and/or numerical analysis.

An insightful approach is to probe the non-equilibrium state of the strongly coupled field theory with a very heavy operator whose Wightman function can be approximated in terms of AdS geodesics as described below.   According to the AdS/CFT dictionary, a massive scalar field $\varphi (z, t, {\bf x})$ with mass $m$ in ($d+1$) dimensions is dual to an operator ${\cal O} (t,{ \bf x})$ of conformal dimension
$\Delta(m,d) = \frac{1}{2}\left(d  + \sqrt{d^2 + 4 m^2}\right)$  \cite{Maldacena:1997re}.   In coordinates where the boundary of AdS is at $ z =0$, the scalar field $\varphi$ behaves like
\ba
\varphi (z,t, {\bf x}) = z^{d-\Delta} \phi_{(0)} (t,{\bf x}) + \ldots + z^{\Delta} \phi_{(1)} (t, {\bf x}) + \ldots,
\ea
where the ellipsis corresponds to subleading terms in the $z$-expansion. In CFT language we interpret the non-normalizable mode $\phi_{(0)} (t,{\bf x})$ as the source to which ${\cal O} (t,{\bf x})$ couples and the normalizable mode $\phi_{(1)} (t,{\bf x})$ as the vacuum expectation value of the renormalized operator  ${\cal O}^{\rm ren} (t, {\bf x})$ \cite{Balasubramanian:1998sn},
\ba
\phi_{(1)} (t,{\bf x}) \sim \langle {\cal O}^{\rm ren} (t,{\bf x})\rangle.
\ea
The bare operator ${\cal O} (t,{\bf x})$ and the renormalized operator ${\cal O}^{\rm ren} (t,{\bf x})$ are related as \footnote{Working with renormalized operators, instead of bare operators, ensures finite expressions when the cut-off $z_0$ is removed, since renormalized operators do not depend on the cut-off $z_0$. A useful heuristic treatment can be found in {\em e.g.}\  \cite{McGreevy:2009xe}.}
\ba
{\cal O} (t,{\bf x}) = z_0^{\Delta} {\cal O}^{\rm ren} (t,{\bf x}),
\ea
where $z_0$ represents an IR bulk cut-off. The two-point function $\langle {\cal O} (t',{\bf x}) {\cal O} (t,\bf x') \rangle$ is often computed using the on-shell (supergravity) action for $\varphi$, when the action is explicitly given and solutions to the wave equation can be readily computed.

However, for our purposes it is more convenient to follow  \cite{Balasubramanian:1999zv} and to observe that the equal-time Green function can be computed via a path integral as
\be
\langle {\cal O }(t,{\bf x}) {\cal O }(t,{\bf x'}) \rangle
= \int {\cal D} {\cal P} \,  e^{i \Delta L({\cal P})}
\approx \sum_{{\rm geodesics}} e^{- \Delta \cal L} \, .
\ee
The first expression sums over all paths that begin and end at the boundary points $(t,\bf x)$ and $(t,\bf x')$, with $L({\cal P})$ being the proper length of the path ($L({\cal P})$ is imaginary for space-like trajectories). The second expression is a saddlepoint approximation to the path sum as a sum over geodesics \cite{Balasubramanian:1999zv} (here $ \cal L$ is the real length of the geodesic between the boundary points). The latter approximation is effective when the probe operator is heavy so that $\Delta \gg 1$.
It is easy, to check, for example, that in the zero temperature theory modeled by empty AdS space, this formula gives the correct conformally invariant equal-time two point function of the renormalized operator  \cite{Balasubramanian:1999zv}
\ba \label{pureAdScorrel}
\langle {\cal O}^{\rm ren}(t,0) {\cal O}^{\rm ren} (t,x_0) \rangle = \frac{1}{ x_0^{2 \Delta}}.
\ea
Some care is necessary in evaluating Lorentzian correlators using the geodesic approximation, because careful consideration of steepest descent contours of integration is generally necessary   \cite{Festuccia:2005pi, Fidkowski:2003nf, Louko:2000tp}.

The geodesic length $\cal L$ diverges due to contributions near the AdS boundary. Therefore, we define a renormalized length $\delta {\cal L } \equiv {\cal L} + 2 \ln (z_0/2)$, in terms of the cut-off $z_0$, by removing the divergent part of the geodesic length in pure AdS (see Eq. \eqref{LpureAdS} below). The renormalized equal-time two-point function is
\be \label{ren_two-point}
\langle {\cal O }(t, {\bf x}){\cal O}(t, {\bf x'}) \rangle_{ren} \sim e^{- \Delta  \delta {\cal L}} \,.
\ee

\begin{figure}[h]
\begin{center}
\includegraphics[width=0.4 \textwidth]{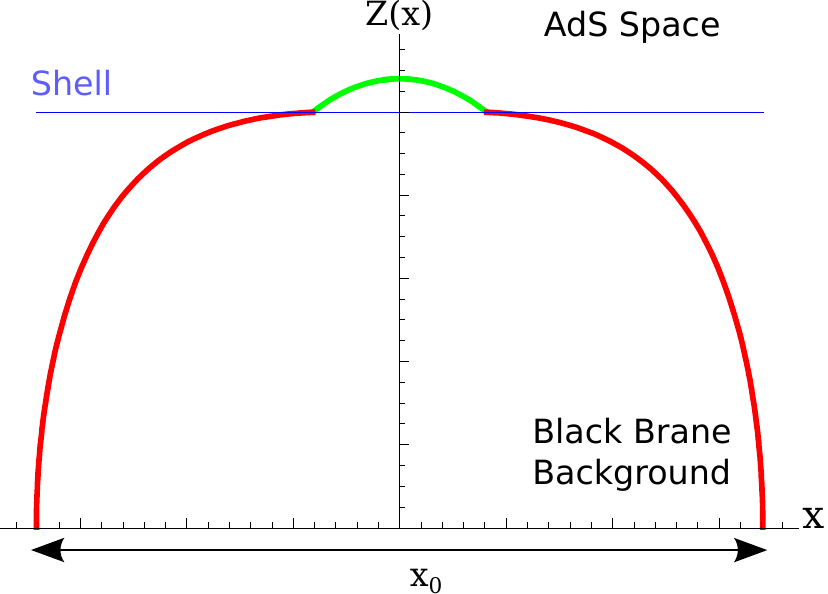} \hfil
\includegraphics[width=0.4 \textwidth]{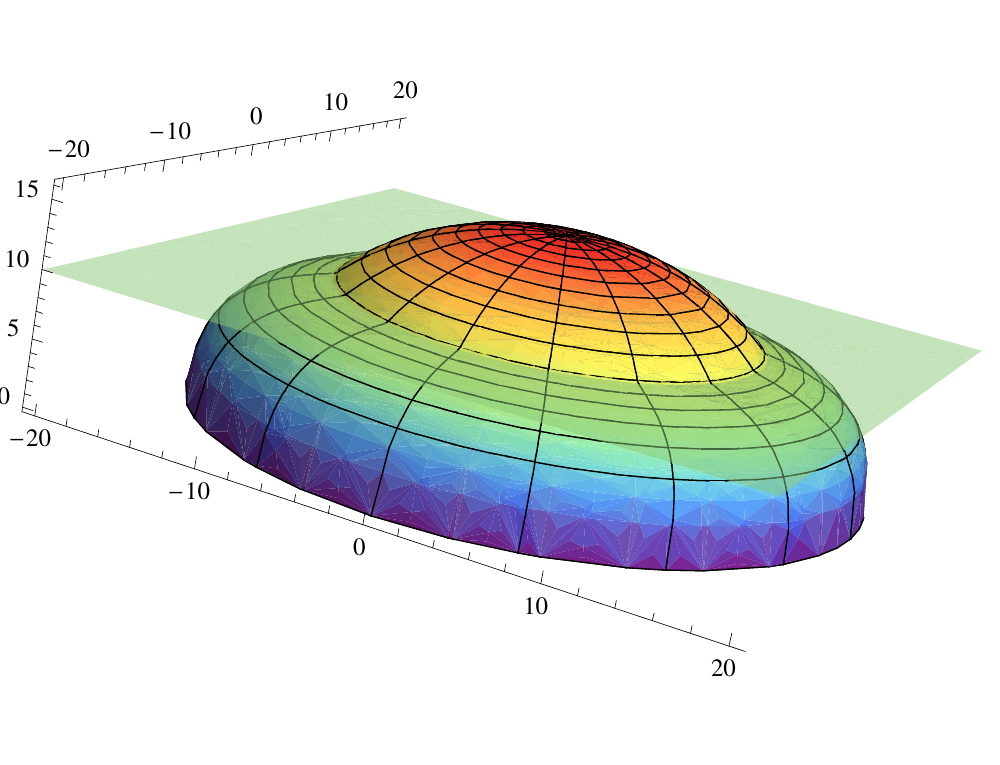} \\
({\bf A}) \hfil ({\bf B})
\caption{({\bf A}) An example space-like geodesic that starts and ends on the boundary of AdS ($z=0$) with a separation $x_0$.  Outside the shell, the geodesic propagates in a black brane geometry, while inside it propagates in an empty AdS geometry.  The shell refracts the geodesic.  The Wightman function at scales associated to geodesics that do not penetrate the shell of matter will be thermalized. ({\bf B}) The minimal surface in AdS space associated to a circular Wilson loop.  The shell of matter (indicated in light green) refracts the surface.  Loops with associated surfaces that never penetrate the shell of matter will be thermalized. Both figures illustrate a quasistatic situation where the geodesic or minimal surface lies entirely at a fixed time. When the shell is dynamically falling into AdS, the geodesic or minimal surface, while remaining space-like, may not lie entirely within an equal-time surface. In both figures we are at late time when the shell is close to where the event horizon would be, so that the `refraction' at the shell is clearly visible.
}
\label{FigCrossingTheShell}
\end{center}
\end{figure}

The geodesic approach to computing equal-time correlators gives a clear intuition for how thermalization proceeds in a strongly coupled theory with an AdS dual (Fig.~\ref{FigCrossingTheShell}).   Consider injecting energy homogeneously into the field theory -- we model this in the AdS theory by dropping in a shell of matter from infinity.   As the shell progresses inward into AdS space, the spacetime outside the shell will be well described by the AdS-black brane metric (except within the shell itself), while the metric inside the shell will be well described by the empty AdS metric.   Now consider the two-point function calculation using geodesics.  If the two points are close together on the boundary, the associated geodesic will not penetrate very much into the bulk space, and thus will not cross the matter shell.  Thus, at these separations the correlation function should look thermal, because a thermal field theory is modeled by a black brane background in AdS\@.     At larger boundary separations, the associated geodesic will penetrate the shell and be ``refracted'' by it (Fig.~\ref{FigCrossingTheShell}) leading to deviations from thermality.   As time passes, the shell will penetrate deeper into the bulk and thus an ever larger range of spatial scales in the field theory will have associated geodesics that do not penetrate the shell and hence have thermal correlators.    Thus we can come to a qualitative conclusion that thermalization proceeds top-down in this setup -- ultraviolet, i.e.\ short distance, degrees of freedom thermalize first.   In subsequent sections we will calculate the rates at which different spatial scales thermalize.

%%%%%%%%%%%%%%%%%%%%%%%%%%%%%%%%%%%%%%%%%%%%%%%%%%%%%%%%%%%%%%%%%%%%%%%

\subsubsection{Two-point function for two-dimensional field theories in equilibrium: analytic computation}\label{BTZanalytic}

In order to gain intuition for the two-point function in the geodesic approximation, we return to the case of two-dimensional boundary field theories. In order to obtain the  thermal boundary-to-boundary Wightman function, we need to study space-like geodesics in the thermal black brane geometry in three dimensions.  The metric is obtained by setting $m(v)= M$ in \eqref{eq:Vaidya}, and is given by
\begin{align}
 ds^2=-(r^2-r_H^2)dt^2+{dr^2\over r^2-r_H^2}+r^2dx^2\,,\qquad
\textrm{with} \quad t=v-{1\over 2r_H}\ln{|r-r_H|\over r+r_H}\,, \qquad r\equiv \frac{1}{ z}\,.
 \label{eq:BTZ}
\end{align}
This is a black brane geometry with a horizon at $r_H\equiv \sqrt{M}$
\footnote{Since we are taking the coordinate $x$ to be noncompact,
in terms of the metric the spacetime \eqref{eq:BTZ} is identical to empty
AdS$_3$. Furthermore, $r=0$, which is a Milne-type singularity for
compact $x$, is not a singularity.  However, we still refer to $r=0$ as
the ``singularity''.  As the geodesics we consider in the current
paper never reach $r=0$, this does not matter to us.}.  The Vaidya geometry
in three dimensions with a constant $m(v) = M$ can be put in this form.

Parametrizing the geodesic by the geodesic length $\lambda$, the
geodesic equations in the geometry \eqref{eq:BTZ} are
\bea
- r_H E&=& -\left( r^2 - r_H^2\right) \dot t\,, \label{GE1}\\
r_H J& = & r^2 \dot x\,, \label{GE2}\\
1 &=&  -( r^2 - r_H^2 ) \dot t^2 +\frac{\dot r^2}{r^2 - r_H^2} + r^2 \dot x^2  \label{GE3} \,.
\eea
where $E,J$ are constants and $\dot{\ }\equiv d/d\lambda$. Plugging (\ref{GE1}) and (\ref{GE2}) into (\ref{GE3}) and solving for $\dot r$,
\be
\dot r
= \pm {1\over r}\sqrt{r^4 + (-1+E^2 - J^2)r_H^2r^2+ J^2 r_H^4
}\,.
\ee
Integrating this, we obtain
\be \label{r_out(lambda)}
r(\lambda)^2 = \frac {r_H^2} {4} \left[ e^{\lambda- \lambda_0} +B_+ e^{-(\lambda - \lambda_0)} \right] \left[e^{\lambda-\lambda_0} + B_- e^{-(\lambda -\lambda_0)} \right]\,
\ee
or, equivalently,
\be \label{r_out(lambda)2}
r(\lambda)^2-r_H^2 = \frac {r_H^2} {4} \left[ e^{\lambda- \lambda_0} + A_+ e^{-(\lambda - \lambda_0)} \right] \left[e^{\lambda-\lambda_0} + A_- e^{-(\lambda -\lambda_0)} \right]\,.
\ee
Here, $\lambda_0$ is an integration constant which we will henceforth
set to zero, $\lambda_0=0$, by absorbing it into the definition of the
parameter $\lambda$.  Also, we defined the combinations
\be
\label{eq:ABdef}
A_{\pm} \equiv J^2 - (1 \pm E)^2\,, \qquad B_{\pm} \equiv (J\pm1)^2 - E^2\,.
\ee
The geodesic reaches the boundary $r=\infty$ (or $z=0$) as
$\lambda\to+\infty$.

\begin{figure}[htbp]
\begin{center}
\includegraphics[width=0.25 \textwidth]{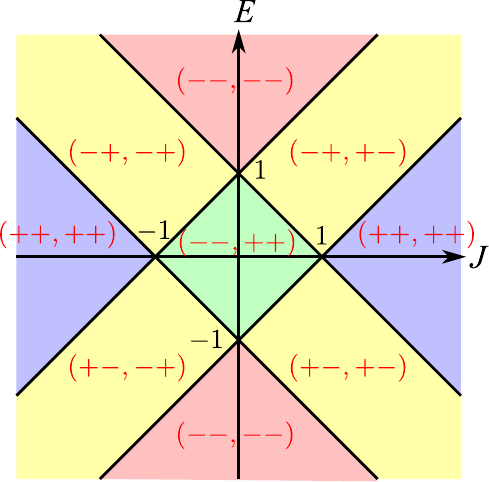}
\caption{The values of $E,J$ and the signs of $A_\pm,B_\pm$.  The signs
 of $A_\pm,B_\pm$ are written in the form $(A_+\,A_-,B_+\,B_-)$.  See text for
 the behavior of geodesics in each regime of parameters.}
\label{fig:A+A-B+B-}
\end{center}
\end{figure}

\begin{figure}[h]
\begin{tabular}{cccc}
\includegraphics[width=4cm, height=3cm,clip]{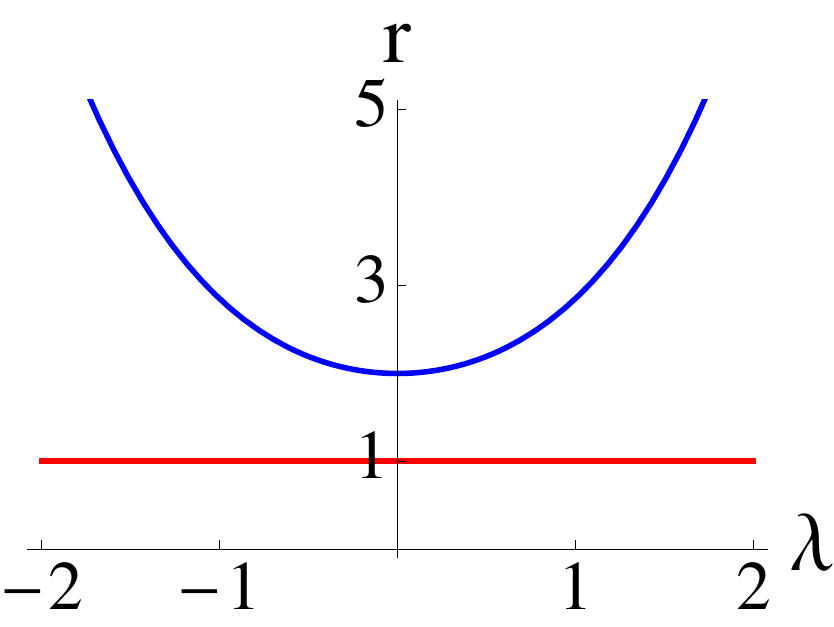}
&
\includegraphics[width=4cm, height=3cm,clip]{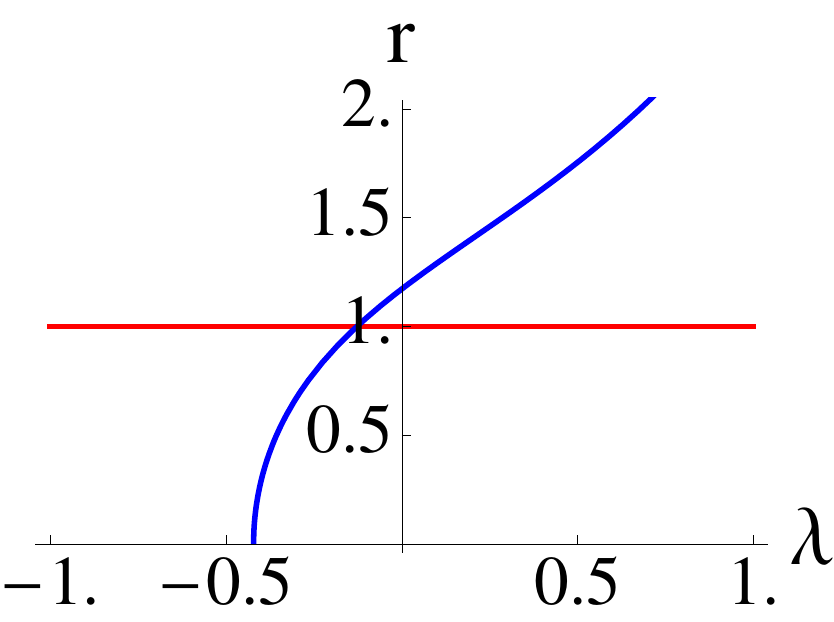}
&
\includegraphics[width=4cm, height=3cm,clip]{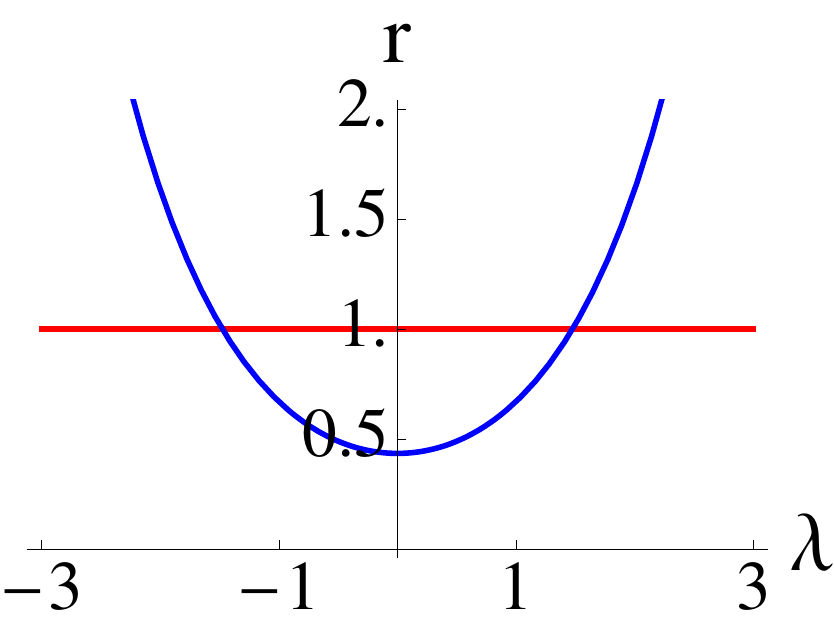}
&
\includegraphics[width=4cm,height=3cm,clip]{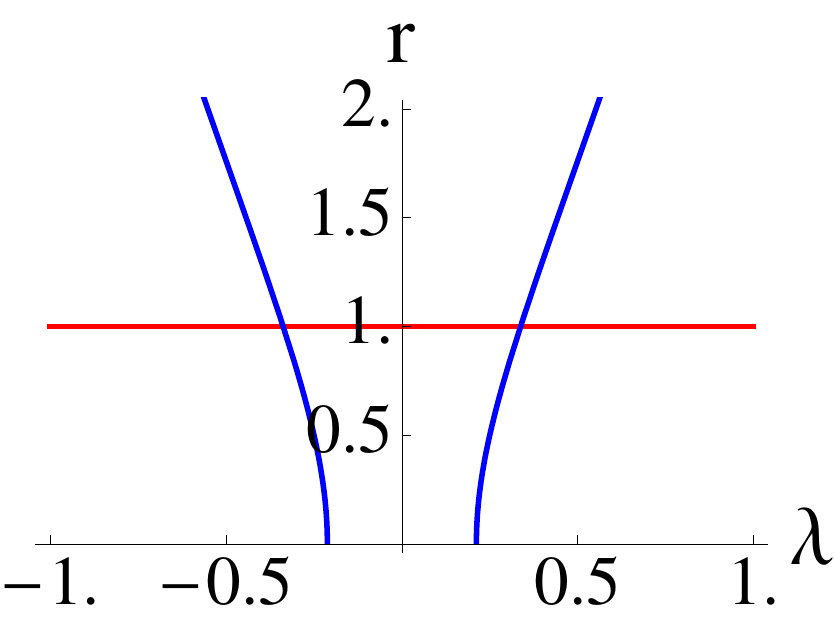}
\\
(i) $E=-0.1$, $J = 2$  & (ii) $E=-1.5$, $J = 2$ & (iii) $E=-0.9$, $J = 0.01$ & (iv) $E= - 5$, $J =3$
\\
\end{tabular}
\caption{Sample profiles of $r(\lambda)$ (in blue) in the AdS$_3$ black brane background. The event horizon $r_H =1$ is shown in red. The integration constant $\lambda_0$ appearing in \eqref{r_out(lambda)} has been set to $\lambda_0 =- 1/4 \ln (|A_+|/|A_-|)$.
}
\label{fig:r_out(lambda)}
\end{figure}

As we can see from \eqref{r_out(lambda)} and \eqref{r_out(lambda)2}, the
signs of $A_\pm,B_\pm$ determine the behavior of the geodesic.  The
relation between the values of the parameters $E,J$ and the signs of
$A_\pm,B_\pm$ is shown in Fig.~\ref{fig:A+A-B+B-}.  Note that not all
possible combinations of signs occur. We discuss different cases
(i)--(iv) shown in Fig.~\ref{fig:r_out(lambda)} in turn below.

(i) If $A_\pm,B_\pm>0$, the geodesic is entirely outside the horizon.
$r(\lambda)$ takes its minimum value
$r=r_H[1+(A_+^{1/2}+A_-^{1/2})^2/4]^{1/2}>r_H$ at $\lambda_{\text{min}}=(1/4)\ln(A_+A_-)$
while $r(\lambda)\to\infty$ as $\lambda\to\pm\infty$.
One sample profile of this case is shown in Fig.~\ref{fig:r_out(lambda)}(i).

(ii) If $A_+A_-<0$ and $B_+B_-<0$, the geodesic crosses
the horizon $r=r_H$ and reaches the singularity $r=0$.  Since in this
case
\be
\dot r(\lambda)
= \frac{r_H}{2} \frac{e^{2\lambda} - A_+A_-e^{- 2\lambda} }
{\sqrt{\left( e^{\lambda} +B_+ e^{-\lambda} \right)
\left( e^{\lambda} + B_- e^{-\lambda} \right) }} >0\,
\ee
(recall that we have set $\lambda_0=0$), $r(\lambda)$ is a monotonically
increasing function of $\lambda$ and the geodesic crosses the horizon
only once.  One sample plot of this behavior is presented in Fig.~\ref{fig:r_out(lambda)}(ii).

(iii) If $A_\pm<0$ and $B_\pm>0$, the geodesic crosses the horizon twice
at
\begin{align}
 \lambda_{H\pm}=\frac 1 2 \ln(-A_\pm)\label{eq:lambdaH}
\end{align}
but does not reach the singularity.  One sample plot of this behavior is
presented in Fig.~\ref{fig:r_out(lambda)}(iii).

(iv) If $A_\pm,B_\pm<0$, the geodesic crosses the horizon twice at
\eqref{eq:lambdaH} and furthermore hits the singularity twice at
\begin{align}
 \lambda_{O\pm}=\frac 1 2 \ln(-B_\pm)\label{eq:lambdaO}.
\end{align}
$r$ is pure imaginary between $\lambda_{O+}$  and  $\lambda_{O-}$.
One sample plot of this behavior is presented in Fig.~\ref{fig:r_out(lambda)}(iv).

If one is interested in geodesics that start and end on the boundary of
the BTZ black brane geometry \eqref{eq:BTZ}, the only relevant cases are (i) and
(iii).  However, if one is interested in geodesics in the Vaidya
geometry \eqref{eq:Vaidya} for which the BTZ geometry \eqref{eq:BTZ} is
only a part of the entire spacetime, then one should also consider cases
(ii) and (iv) since they can appear as a part of geodesics in the full
spacetime.

Substituting (\ref{r_out(lambda)}) into (\ref{GE1})
and (\ref{GE2}) and integrating the two equations, we find the expressions for $t$ and $x$:
\bea
t(\lambda) &= & t_0+ \frac{1}{2r_H} \ln  \Bigg | \frac{e^{ \lambda} + A_+e^{- \lambda} }{e^{ \lambda} + A_- e^{-\lambda} } \Bigg |  \,, \label{toutl}\\
x(\lambda) &= & x_0+ \frac{1}{2r_H} \ln \left( \frac{e^{\lambda} + B_- e^{-\lambda} }{e^{\lambda} + B_+ e^{-\lambda} }\right) \label{Xoutl}\,,
\eea
where $t_0,x_0$ are constants of integration corresponding the values of
$t,x$ at $\lambda=+\infty$.  Comparing (\ref{toutl}) with
\eqref{r_out(lambda)2}, it can be seen that the quantity inside the
absolute value of (\ref{toutl}) vanishes or diverges at the horizon
$r=r_H$ and thus $t(\lambda)$ diverges there.  This is due to the
well-known fact that the Schwarzschild time $t$ is not well-defined at
the horizon.  Note that $t(\lambda)\to\mp\infty$ as
$\lambda\to\lambda_{H\pm}$ and therefore $\lambda_{H+}$ ($\lambda_{H-}$)
corresponds to the past (future) horizon.

On the other hand, the Eddington-Finkelstein coordinate $v$ defined by
\be
\label{voutlambda}
v(\lambda) = t(\lambda) + \frac{1}{2r_H} \ln\frac{|r(\lambda)-r_H|}{r(\lambda)+ r_H}
\ee
(see (\ref{eq:BTZ})) is well-defined across the future horizon
$\lambda=\lambda_{H-}$, because the log divergence in $t(\lambda)$ gets
canceled by the second term of \eqref{voutlambda} and $v(\lambda)$ is
finite across the future horizon.  On the other hand, the divergence at
the past horizon $\lambda=\lambda_{H+}$ is not canceled and $v(\lambda)$
diverges across the past horizon.
The argument of the logarithm in (\ref{Xoutl}), on the other hand, is
always positive for $r>0$; see (\ref{r_out(lambda)}).

The above parametric expression of $r,t,x,v$ in terms of $\lambda$ is
useful for understanding the behavior of the geodesic, but for
computational purposes it is useful to eliminate $\lambda$ and write $t,x,v$ as
functions of $r$.  Inverting the relation (\ref{r_out(lambda)}), we
obtain the following two branches for $\lambda(r)$:
\bea\label{lambdapm}
\lambda_{\pm}(r)&=& \frac 1 2 \ln \left[ -1+ E^2  - J^2 + \frac{2 r^2}{r_H^2}
\pm \frac{2}{r_H^2} \sqrt{D(r)}\right],\qquad
D(r)\equiv r^4 + (-1+E^2 -J^2) r_H^2  r^2+ J^2 r_H^4.
\eea
Note that we have been setting $\lambda_0=0$.
Substituting (\ref{lambdapm}) into (\ref{toutl})--(\ref{voutlambda}) and setting $x_0 =0$, we get
\begin{align}
t(r)_{\pm} &= t_0 +\frac{1}{2r_H} \ln  \Bigg | \frac{r^2 - (E+1)r_H^2 \pm  \sqrt{D(r)}}{r^2 + (E-1)r_H^2 \pm  \sqrt{D(r)}} \Bigg |  \,,\label{t(r)pm}\\
x(r)_{\pm} &= \frac{1}{2r_H} \ln \left[ \frac{r^2 - J r_H^2 \pm  \sqrt{D(r)} }{r^2 + J r_H^2 \pm  \sqrt{D(r)} } \right]\,,\label{X(r)pm}\\
v(r)_{\pm}& = t_0 +\frac{1}{2r_H} \ln \left[ \frac{r-r_H}{r+r_H}\,\frac{r^2 - (E+1)r_H^2 \pm  \sqrt{D(r)} }{r^2 + (E-1)r_H^2 \pm  \sqrt{D(r)} }\right]\,.\label{v(r)pm}
\end{align}
We will refer to the branch given by $(t(r)_+,x(r)_+, v(r)_+)$ as branch
1 and the one given by $(t(r)_-,x(r)_-, v(r)_-)$ as branch 2.  Only by
combining both branches can we recover the full geodesic described by
\eqref{r_out(lambda)}, \eqref{toutl} and \eqref{Xoutl}.

From \eqref{toutl} and \eqref{Xoutl}, we can compute the coordinate difference
between the two boundary points on which the geodesics end:
\begin{align}
 \Delta t \equiv t(\lambda=-\infty)-t(\lambda=\infty) &={1\over 2r_H}\log{A_+\over A_-}={1\over 2r_H}\log{J^2-(1+E)^2\over J^2-(1-E)^2},
 \label{eq:DtBTZ}\\
 \ell     \equiv x(\lambda=-\infty)-x(\lambda=\infty) &={1\over 2r_H}\log{B_+\over B_-}={1\over 2r_H}\log{(J+1)^2-E^2\over (J-1)^2-E^2}.
 \label{eq:DXBTZ}
\end{align}
We see that geodesics connecting endpoints at the same time ($\Delta t=0$) correspond to taking
$E=0$.  If one is interested in such geodesics in the pure BTZ
geometry, then we have only to consider the $E=0$ case.  However, in section \ref{thermalization}, we will be interested in geodesics in the Vaidya geometry, which are composed of two pieces stretching respectively in AdS and BTZ, and glued together at the shell location.
In this case, one should also consider $E\neq 0$ geodesics since they can appear as
part of geodesics between equal-time endpoints in the full spacetime.

Now let us focus on the equal-time geodesics with $E=0$ and compute the
relation between the spatial boundary separation $\ell$ and the geodesic
length ${\cal L}_{\text{thermal}}$ in the black brane background.
By setting $E=0$ in  \eqref{eq:DXBTZ}, $\ell$ is computed as
\begin{align}
 \ell= {1\over r_H}\ln{J+1 \over  J-1},
 \qquad\text{or}\qquad J=\coth{r_H\ell\over 2}.
\label{eq:ellandJ-BTZ}
\end{align}
On the other hand, the geodesic length ${\cal L}_{\text{thermal}}$ is computed from \eqref{lambdapm}
as:
\begin{align}
 {\cal L}_{\text{thermal}}=\lambda_+(r=r_0) - \lambda_-(r=r_0)
 =2\ln{2r_0\over r_H\sqrt{|J^2-1|}}
 =2\ln{2r_0\sinh{r_H\ell\over 2}\over r_H}
 \label{LvsellBTZbare}
\end{align}
where $r_0 = 1/z_0$ is the bulk IR cutoff and we dropped the $\CO({1/r_0})$
quantity that vanishes as $r_0\to\infty$.  In the last equality, we used
\eqref{eq:ellandJ-BTZ}.

As discussed in Section \ref{geodesicapprox}, the geodesic length \eqref{LvsellBTZbare} is divergent in the $r_0\to
\infty$ limit and should be regularized by subtracting the corresponding divergent part of the
quantity in pure AdS geometry.  Because the pure AdS geometry can be
obtained by setting $M\to 0$ or $r_H\to 0$ in the BTZ geometry
\eqref{eq:BTZ}, we can simply send $r_H\to 0$ in various quantities to
obtain the corresponding quantities for pure AdS\@.  By sending $r_H\to 0$ in
\eqref{LvsellBTZbare}, we obtain the geodesic length in pure AdS:
\begin{align}
 {\cal L}_{\text{pure AdS}}=2\ln (2r_0)+\text{(finite)} .\label{LpureAdS}
\end{align}
Subtracting the divergent part of this from \eqref{LvsellBTZbare}, we
obtain
\begin{align}
 \delta {\cal L}_{\text{thermal}}(\ell)
 \equiv {\cal L}_{\text{thermal}} -2\ln (2r_0)
 =2\ln{\sinh{r_H\ell\over 2}\over r_H}\,,
 \label{LvsellBTZren}
\end{align}
which allows us to compute the thermal renormalized two-point function
defined in \eqref{ren_two-point}.  Observe that the two-point function
result computed in the geodesic approximation coincides with the weak
coupling result \eqref{weakcouplingscalar} for $\Delta=1$.  Actually
this is a consequence of conformal invariance, because the boundary CFT
lives in noncompact space (namely, the $x$ direction is not compactified
in our setting).

We can also obtain the expression for the equal-time geodesic in pure
AdS geometry by sending $r_H\to 0$ in \eqref{X(r)pm}.  In doing this, we
should send $J\to \infty$ at the same time because
\eqref{eq:ellandJ-BTZ} becomes $J=2/(r_H\ell)$.  In this
limit, \eqref{lambdapm} and
\eqref{X(r)pm} give
\begin{align}
 \lambda(r)_\pm&=
 \pm \cosh^{-1}{\ell r\over 2},
\label{lambda(r)pureAdS}\\
 x(r)_\pm
 &=\pm{\ell\over 2}\sqrt{1-\left({2\over\ell r}\right)^2}
 =\pm{\ell\over 2}\sqrt{1-\left({2z\over\ell}\right)^2},
\label{X(r)pureAdS}
\end{align}
where we dropped constants.  The second equation \eqref{X(r)pureAdS} says that the equal-time geodesic is a
semi-circle on the $z$-$x$ plane,
\begin{align}
 z^2+x^2=\left({\ell\over 2}\right)^2.\label{semicircle}
\end{align}

Note that the pure AdS results \eqref{LpureAdS}, \eqref{lambda(r)pureAdS}, \eqref{X(r)pureAdS} and
\eqref{semicircle} are correct not only for $d=2$ but for general
dimensions, because it is only the $z$-$x$ plane that is relevant for
such geodesics.

%%%%%%%%%%%%%%%%%%%%%%%%%%%%%%%%%%%%%%%%%%%%%%%%%%%%%%%%%%%%%%%%%%%%%%%

\subsubsection{Two-point function for \texorpdfstring{$d\geq 2$-}{d=2 or greater }dimensional field theories in equilibrium: numerical computation}

In any dimension we can consider  a Vaidya type background (\ref{eq:Vaidya}) with constant mass function $m(v)=M$. This is a ($d+1$)-dimensional black brane geometry, and in the special case of three dimensions it can be put in the form (\ref{eq:BTZ}).
We consider space-like geodesics connecting the boundary points $(t, x_1) = (t_0, -\ell/2)$ and $(t', x_1')=(t_0, \ell/2)$ of such black  brane geometries (for $d=3$, we also have $x_2 = x_2' $ and for $d=4$, $(x_2 ,x_3)= (x_2' ,x_3')$). If we parametrize the
geodesic in terms of the coordinate $x_1$, which we will simply denote $x$ in the rest of the Section, the solution will be given by the profiles $v=v(x)$ and $z=z(x)$. Inserting the IR bulk cut-off $z_0$, the boundary conditions read
\ba
z(-\ell/2) = z_0 = z(\ell/2)\,, \quad v(-\ell/2) = t_0 = v(\ell/2)\,,
\ea
and are symmetric with respect to the $z$- and $v$-axis. The geodesic length reads
\ba
{\cal L}_{\textrm{thermal}} = \int^{\ell/2}_{-\ell/2} dx \frac{\sqrt{1- 2 z'(x) v'(x) - (1 - M z(x)^d) v'(x)^2}}{z(x)},
\ea
with ${\ }^{\prime}\equiv d/dx$. The action is not explicitly dependent on $x$, which implies the existence of a conserved quantity. 
The conservation equation reads
\ba \label{conseqddim}
1 - 2 z' v' - (1- M z^d) v'^2 = \left(\frac{z_*}{z}\right)^2\,,
\ea
where $z_*$ is the value of $z(x)$ at the midpoint. Notice also that $v$ is a cyclic coordinate, implying a conserved momentum. From Eq. \eqref{eq:dt_and_dv}, it follows
\ba \label{tcritddim}
v  = t_0 - \int^z_{z_0} \frac{d \tilde z}{ 1- M {\tilde z}^d}\,.
\ea

\begin{figure}[h]
\centering
\includegraphics[width=3.in]{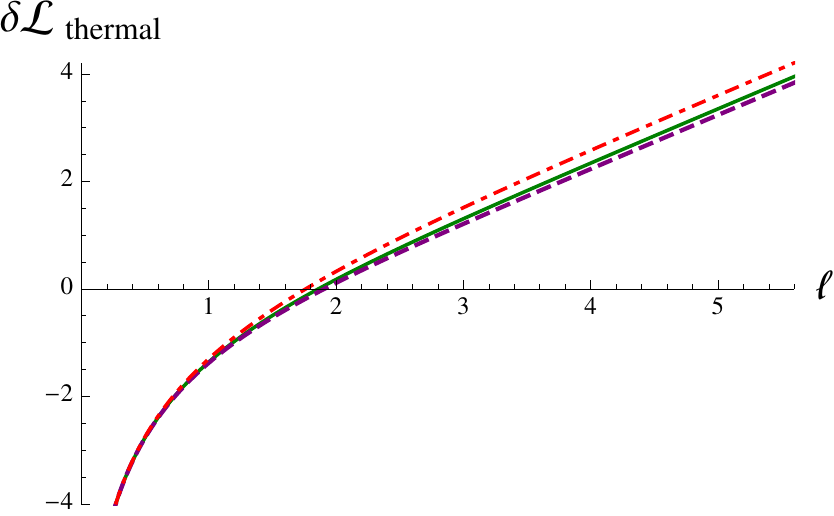}
\caption{$\delta {\cal L}_{\rm thermal}$ as a function of spatial scale $\ell$ for $d=2$ (red, dot dashed), $d=3$ (green) and $d=4$ (purple, dashed) for a black brane geometry with $M = 1$.  The results for $d=2$ agree with the analytical results of Section \ref{BTZanalytic} in the limit of a shell of zero thickness.
}
\label{fig:deltaLthermal}
\end{figure}

Inserting the relation between $v$ and $z$ of Eq.~(\ref{tcritddim}) into the conservation equation (\ref{conseqddim}) yields a first order differential equation in terms of $z(x)$ only. We take the derivative of  this equation with respect to $x$ and solve the resulting second order differential equation
\ba
z'' =-\frac{z_*^2}{z^3} + \frac{2-d}{2} M z^{d-3} z_*^2 + \frac{d}{2} M z^{d-1} ,
\ea
starting form the midpoint, where $z(0) = z_*$ and $z'(0) = 0$. We can then read off the boundary separation $\ell$ corresponding to a specific midpoint value $z_*$ via $z(\ell/2) = z_0$.
By substituting the conservation equation into the expression for the geodesic length, we can simplify the expression to
\ba
{\cal L}_{\textrm{thermal}} = 2 \int^{z_*}_{z_0} \frac{d z}{z} \frac{1}{\sqrt{(1-M z^d)(1-\frac{z^2}{z_*^2})}}\,.
\ea
The renormalized length $\delta {\cal L}_{\rm thermal}$ is
\ba
\delta {\cal L}_{\rm thermal} (\ell) \equiv {\cal L}_{\rm thermal} + 2 \ln\left(\frac{z_0}{2}\right)\,,
\ea
from which one obtains the renormalized equal-time two-point function through \eqref{ren_two-point}. We plot the renormalized length $\delta {\cal L}_{\rm thermal}$ as a function of the spatial scale $\ell$ in Fig.~\ref{fig:deltaLthermal}.

%%%%%%%%%%%%%%%%%%%%%%%%%%%%%%%%%%%%%%%%%%%%%%%%%%%%%%%%%%%%%%%%%%%%%%%

\subsection{Space-like Wilson loops}\label{WLs}

Above we examined the Wightman functions for field theories at finite temperature in order to get a sense of what
thermal equilibrium looks like according to this measure. A second non-local probe can be used to discuss the thermality of field theories is the (expectation value of the) Wilson loop. The Wilson loop is a gauge-invariant observable, constructed as the path-ordered contour integral over a closed loop $C$ of the gauge field
\ba
W(C) = \frac{1}{N} {\rm Tr}({\cal P} e^{ \oint_C A}).
\ea
Wilson loops contain useful information about the non-perturbative behavior of (non-abelian) gauge theories, but are in general hard to calculate.
In the AdS/CFT correspondence, the expectation value for the Wilson loop is related to the string partition function with a string worldsheet $\Sigma$ extending in the bulk and ending on the loop $C$ at the boundary
\ba
\langle W(C) \rangle = \int {\cal D} {\Sigma} \, e^{-A(\Sigma)},
\ea
where we integrate over all inequivalent string surfaces $\Sigma$ such that $\partial \Sigma = C$ at the AdS boundary and where $A(\Sigma)$ corresponds to the string action. In the strongly coupled limit, we can consider a saddle point approximation for the string partition function and reduce the calculation of the Wilson loop expectation value to determining the minimal area surface of the (classical) string worldsheet whose endpoints trace out the desired Wilson loop $C$ on the AdS boundary
\ba
\langle W(C) \rangle \sim e^{- \frac{1}{\alpha'} {\cal A}(\Sigma_0)}.
\ea
${\cal A}(\Sigma_0)$ represents the area of the minimal area surface $\Sigma_0$ with boundary $C$. The surface $\Sigma_0$ is a solution to the equations of motion arising from the bosonic action of the string. Thus, the expectation value of a Wilson loop in the AdS/CFT correspondence is given by the area of a minimal surface in AdS space which is bounded by the loop $C$ \cite{Maldacena:1998im}.

\begin{figure}[h]
\begin{center}{
\begin{tabular}{c}
 \includegraphics[width=12cm, height=12cm,clip]{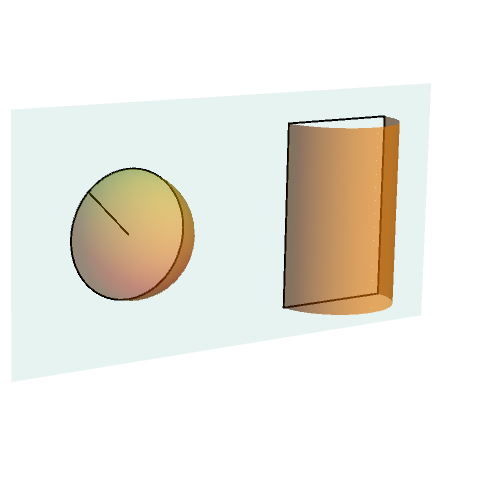}
 \begin{picture}(0,0)\put(-280,180){{\large$R$}}\put(-160,180){{\large $R$}}\put(-110,265){{\large $\ell$}}\end{picture}
  \end{tabular}
  \vspace{-1in}
  \textit{\caption{Two Wilson loop shapes with their minimal string surface: the circular Wilson loop (left) and the rectangular Wilson loop (right).\label{Wilsonsurfacessketch}}}}
\end{center}
\end{figure}
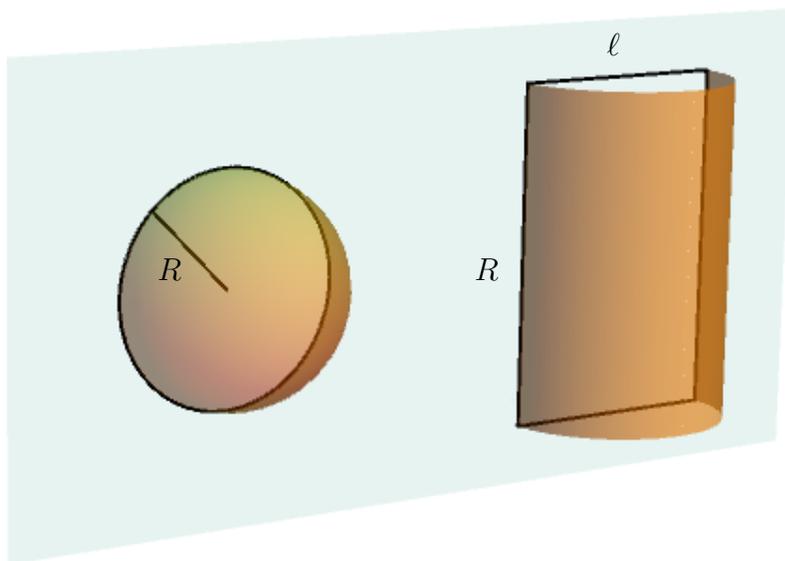

In this section, we will compute space-like Wilson loop expectation values in the vacuum and in the thermally equilibrated theory (the black brane background) for $d\geq 3$-dimensional field theories. In Sec.~\ref{thermalization}, we will compute Wilson loops in a strongly coupled theory following a quench and compare them to the thermal results to assess progress towards thermalization. We will study two loop shapes  - circles and strips as sketched in figure \ref{Wilsonsurfacessketch} - to test how and whether the shape of the loop affects the rate of thermalization.

%%%%%%%%%%%%%%%%%%%%%%%%%%%%%%%%%%%%%%%%%%%%%%%%%%%%%%%%%%%%%%%%%%%%%%%

\subsubsection{Circular Wilson loops}\label{s:CWL}

We first examine the circular space-like Wilson loop in a strongly coupled $d \geq 3$-dimensional field theory at finite temperature.
At zero temperature the dual geometry is pure AdS$_{d+1}$, and the associated minimal surface in AdS is simply a hemisphere \cite{Berenstein:1998ij}. This  surprisingly simple solution to a complicated second-order non-linear equation of motion  was obtained by taking an infinite straight Wilson line on the boundary, and applying a special conformal transformation to map this straight line to a circle. Since the conformal group of the boundary corresponds to reparameterizations of AdS we can apply the appropriate reparameterization to the AdS space, and it maps the original worldsheet  - which was flat and extended straight into the AdS - to the hemisphere.  This trick only works in pure AdS space where conformal invariance is preserved, so we are obliged to use numerical  methods of solving the equations of motion when we want to calculate the Wilson loop for thermal or out-of-equilibrium states.

For now, we want to study the minimal hemisphere surface in a thermal black brane background, with metric given by (\ref{eq:Vaidya}) with $m(v) = M$, where $M$ is the constant tension of the black brane. On the boundary we choose a two-dimensional plane $(x_1, x_2)$ in which the circular Wilson loop is set. To parameterize the Wilson loop we introduce polar coordinates $(\rho, \varphi)$. The minimal space-like surface with the circular Wilson loop as basis also has an azimuthal symmetry. The tip of the surface occurs at $(v,z,{\bf x} )= (v_*, z_*, {\bf 0})$. The cross section at fixed $z$ and $v$ is a circle, and thus the surface is conveniently parameterized in terms of the radii $\rho$ of these circles. The Nambu-Goto action for the string with circular symmetry in the ($d+1$)-dimensional AdS black brane background is
\ba\label{eq:ACWLBB}
A_{\rm NG}=\frac{1}{\alpha'}\int_{0}^{R} d \rho \frac{\rho}{z(\rho)^2} \sqrt{1-(1 - M z(\rho)^d ) v'(\rho)^2 - 2 z'(\rho) v'(\rho) },
\ea
where we integrated out the $\varphi$-factor as both $z$ and $v$ are independent of $\varphi$ due to the circular symmetry. This action has an explicit $\rho$-dependence, implying that the second order differential equation in $z(\rho)$ cannot be integrated to a first order differential equation. The coordinate $v$ is still a cyclic coordinate, which implies that the associated momentum is conserved. To solve for an equal-time Wilson loop, we  set the conserved energy to zero, which implies a relation between $v$ and $z$ as given in eq.~(\ref{tcritddim}).
Inserting this relation in the action,
\ba
A_{\rm NG} (R) = \frac{1}{\alpha'}\int_{0}^{R} d\rho  \frac{\rho}{z^2}\sqrt{1+\frac{z'^2}{1-M z^d}},
\ea
the resulting equation of motion reduces to a second order differential equation in $z(\rho)$
\ba\label{Wilsonloopdeom}
z'' = - \left( \frac{2}{z} + \frac{z'}{\rho} \frac{1}{1-M z^d} \right) (1-M z^d + z'^2) - \frac{d}{2} \frac{M z^{d-1} z'^2}{1-M z^d}.
\ea

One can easily check that the hemisphere is an analytic solution in pure AdS (obtained by setting $M=0$ in the above). It is given by
\ba\label{eq:hemi}
z(\rho) = \sqrt{z_*^2-\rho^2}.
\ea
In case of a black brane background ($M \neq 0$) we have to resort to numerical means. We impose boundary conditions at the midpoint assuming reflection symmetry along the $z$-axis
\ba
z(0) = z_*, \quad z'(0) = 0,
\ea
and read off the boundary radius $z(R) = z_0$.
Note that because the second term in the first bracket of (\ref{Wilsonloopdeom}) causes numerical issues at $\rho=0$ (the numerator and denominator both go to zero, but the ratio should take a fixed value), we cannot construct the solutions starting from the midpoint. Therefore we choose to solve (\ref{Wilsonloopdeom}) in the neighborhood of the midpoint by expanding around $\rho = 0$ to quadratic order (odd powers vanish by the symmetry)
\ba
z_p(\rho) = z_* - \frac{1- M z_*^d}{2 z_*} \rho^2 .
\ea
The boundary conditions for the numerical solution are obtained by matching at a point $\rho_p$ close to the midpoint; $z(\rho_p) = z_p(\rho_p)$ and $z'(\rho_p) = z_p'(\rho_p)$.

\begin{figure}[h]
\begin{center}
\includegraphics[width=3in]{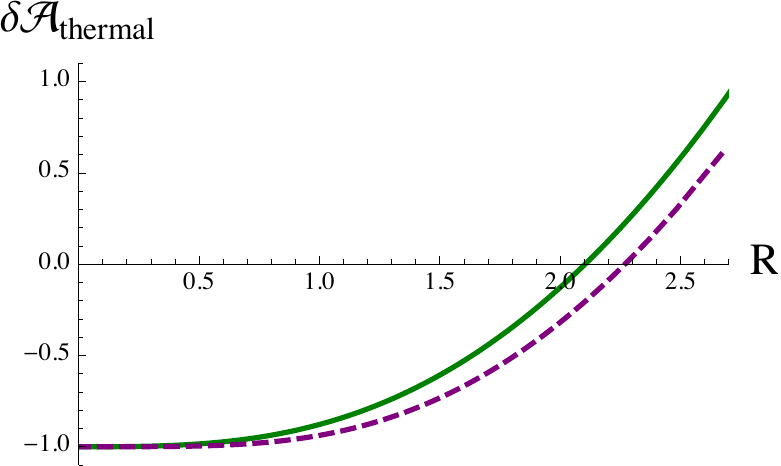}
\caption{$\delta {\cal A}_{\rm thermal}$ against the boundary radius $R$. The curves represent the string surface areas in a black brane background with unit mass for $d=3$ (green) and $d=4$ (purple, dashed).
}
\label{FigWilsonBraneArea}
\end{center}
\end{figure}

The logarithm of the expectation value of Wilson loop operators is approximated by the area (${\cal A}_{\rm thermal} = \alpha' A_{\rm NG}$) of the minimal surface. The largest contributions to the area are coming from the near-boundary region, because of the diverging AdS volume there. We subtract the divergent piece, which is  proportional to the UV-cutoff $1/z_0$.  This procedure is interpreted in the dual CFT as removing the UV divergence of the Coulomb self-energy of a point charge.  Subtraction of $R/z_0$ gives the regularized area
\ba\label{eq:WLAreg}
\delta {\cal A}_{\rm thermal} (R) = {\cal A}_{\rm thermal}(R) - \frac{R}{z_0}.
\ea
The regularized area is depicted for AdS$_4$ and AdS$_5$ in Fig.~\ref{FigWilsonBraneArea}. 

%%%%%%%%%%%%%%%%%%%%%%%%%%%%%%%%

\subsubsection{Infinite Rectangular Strips}\label{rectstripBB}

A less symmetric Wilson loop is the rectangular strip parametrized by the boundary coordinates ($x_1, x_2$). Assume that the strip is translationally invariant along the $x_2$-axis, such that  the profile of the associated minimal surface in AdS$_{d+1}$ is described by $z(x_1)$ and $v(x_1)$. In the following, we again denote $x \equiv x_1$. We impose the boundary conditions
\ba
z(-\ell/2) = z_0 = z(\ell/2), \quad v(-\ell/2) = t_0 = v(\ell/2).
\ea
As for the circular Wilson loop, the equation of motion for the minimal surface is obtained by minimizing the Nambu-Goto action.  In this case, because of the symmetries, the analysis closely resembles that of geodesics with some small differences in the action and the equations of motion. The area for a segment stretching between  $x_2 \in (0, R)$ in Eddington-Finkelstein coordinates is given by
\ba
{A}_{\rm NG}=\frac{R}{2 \pi \alpha'}\int^{\ell/2}_{\ell/2} d x \frac{ \sqrt{1-(1 - M z(x)^d ) v'(x)^2 - 2 z'(x) v'(x) }}{z(x)^2} .
\ea

\begin{figure}[h]
\begin{center}
\includegraphics[width=3in]{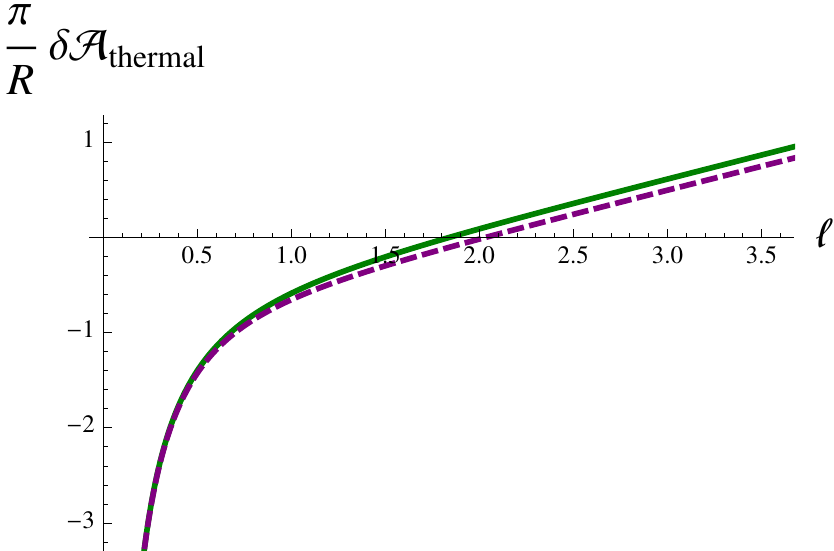}
\caption{ $\delta {\cal A}_{\rm thermal}/(R/\pi)$ as function of the spatial scale $\ell$. The curves represent the string surface areas in a black brane background with unit mass. The base of the surface is a rectangular strip at the AdS boundary. The green curve corresponds to d=3 and the purple, dashed curve to d=4.
}
\label{FigStripBBArea}
\end{center}
\end{figure}

The action does not exhibit an explicit $x$- or $v$-dependence, hence we can use the existence of two conserved quantities to simplify: these can be taken to be the midpoint of $z(x)$, $z_*$, and the momentum conjugate to $v(x)$. The $z_*$ conservation equation reads
\ba\label{conseqstrip}
1-(1 - M z^d ) v'^2 - 2 z' v' = \left(\frac{z_*}{z}\right)^4,
\ea
while the conservation of the momentum conjugate to $v(x)$ implies the same relation as in Eq.~(\ref{tcritddim}). Inserting this relation in the conservation equation gives a first-order differential equation in $z(x)$. We then take the derivative with respect to $x$ to obtain a second-order differential equation in $z(x)$
\ba\label{stripddimeom}
z'' = -2 \frac{z_*^4}{z^5} + \frac{4-d}{2} M z^{d-5} z_*^4 + \frac{d}{2} M z^{d-1}.
\ea
We impose the boundary conditions $z'(0) = 0$, $z(0) = z_*$ at the midpoint,
and read off the boundary separation $\ell$ via $z(\ell/2) = z_0$. Inserting the conservation equation in the Nambu-Goto action we get the following formula for the area  (${\cal A}_{\rm thermal} = \alpha' A_{\rm NG} $) of the minimal surface
\ba
{\cal A}_{\rm thermal}=\frac{R}{ \pi} \int^{z_*}_{z_0} \frac{dz}{z^2} \frac{1}{\sqrt{\left(1- \frac{z_*^4}{z^4}\right)(1- M z^d)}}.
\ea
The regularized area reads
\ba
\delta {\cal A}_{\rm thermal} (\ell,R) = {\cal A}_{\rm thermal} - \frac{1}{z_0} \frac{R}{\pi}\,,
\ea
and is plotted in Fig.~\ref{FigStripBBArea} for a four- and five-dimensional background. We have chosen to plot the results after dividing out $R/\pi$ as we are concerned purely with the dependence of the area on the boundary separation $\ell$.

%%%%%%%%%%%%%%%%%%%%%%%%%%%%%%%%

\subsection{Entanglement entropy}\label{s:3surf}

It is still an open question whether a suitable notion of ``local
entropy'' density exists which is valid out of equilibrium and which
satisfies some basics physical properties; in particular it must be a
non-decreasing function of time.  To define a local entropy density
using gauge/gravity duality, one typically employs horizons since their
area increases in time, and it has been suggested in
\cite{Figueras:2009iu, Chesler:2008hg} that apparent horizons provide
more compelling notions of local entropy density then global horizons
do. In our case, the entropy associated to the apparent horizon
instantaneously thermalizes, in the limit of a zero thickness shell, and does not provide a useful probe of
thermalization. Skepticism against the identification of entropy with apparent horizon area in full generality has also been raised in a number of other papers, including \cite{AbajoArrastia:2010yt,Figueras:2009iu,Hubeny:2011yk}.

Another way of assessing thermalization at different spatial scales in an out-of-equilibrium theory is by measuring the entanglement entropy associated with volumes of different sizes and shapes.
To review, consider dividing a quantum mechanical system in a state $|\Phi\rangle$  into two spatially disjoint parts $A$ and $B$.   The density matrix of the system restricted to $A$ and its entropy are computed by taking the traces
\be
\rho_A = \tr_B |\Phi\rangle\langle \Phi | \,; \qquad S(A) = - \tr_A ( \rho_A \ln \rho_A ) .
\ee
In a $(1+1)$-dimensional conformal field theory, where $A$ is an interval of length $\ell$, the entanglement entropy can be calculated analytically, yielding the universal result \cite{Calabrese:2009qy,Calabrese:2004eu}:
\be
{\rm vacuum:}\, S_{0}(\ell) =  {c \over 3} \ln\left({\ell \over a}\right)\,; \qquad
{\rm thermal \ equilibrium:}\, S_{T}(\ell) = {c \over 3} \ln\left( {\beta \over \pi a} \sinh {\pi \ell \over \beta} \right)\,.
\label{Sent-CC}
\ee
Here $a$ is the UV cutoff of the field theory, $c$ is the central charge, and $\beta=T^{-1}$ denotes the inverse temperature.

The entanglement entropy $S(A)$ describes the amount of information loss associated with the restriction of an observer to the volume $A$. In the vacuum state for $d>2$ dimensions, $S_0(A)$ is proportional to the surface area of $A$ \cite{Srednicki:1993im,Bombelli:1986rw}; in $d=2$ dimensions, as Eq.~(\ref{Sent-CC}) shows, $S_0$ depends logarithmically on the length of the interval $\ell$. At non-zero temperature, $S(A)$ receives an additional contribution, which can be interpreted as thermal entropy \cite{Ryu:2006bv,Nishioka:2009un}. In the limit $T\to\infty$, the thermal contribution is proportional to the volume of the region $A$, just like the statistically defined thermal equilibrium entropy. Computing the time dependent entanglement entropy as a function of spatial scale and studying its approach to $S_{T}(\ell)$ thus will provide a probe of scale-dependent thermalization.

There is a precise proposal for computing entanglement entropy in strongly coupled field theories with AdS duals \cite{Nishioka:2009un,Ryu:2006bv,Hubeny:2007xt}, where we think of the field theory as living on the boundary of an AdS space. Specifically, consider the boundary $\partial A$ of a connected region in the field theory whose entanglement entropy we wish to compute.  For a two-dimensional field theory, $\partial A$ is a pair of points, for a three-dimensional field theory $\partial A$ is a closed curve, and for a four-dimensional theory a surface. Now construct the minimal surface $\sigma_A$ in AdS space that meets $\partial A$ on the AdS boundary.   For a two-dimensional field theory, $\sigma_A$ is a geodesic in AdS$_3$ that approaches the boundary points $\partial A$, while for a three-dimensional field theory, $\sigma_A$ is a minimal surface in AdS$_4$ with boundary $\partial A$.    For a four-dimensional field theory, $\sigma_A$ is a minimal volume in AdS$_5$ which ends on the surface $\partial A$ on the AdS boundary. The entanglement entropy of the region $A$ of the field theory is then given by
\be
S_A = {{\rm Area}(\sigma_A) \over 4 G_N} ,
\ee
where $G_N$ denotes Newton's gravitational constant (and ``Area'' stands for the length of geodesics, the area of 2-surfaces, and the volume of 3-surfaces). The authors of \cite{Nishioka:2009un,Ryu:2006bv,Hubeny:2007xt} showed that this formula precisely reproduces the universal entropy formula of two-dimensional conformal field theories in thermal equilibrium \cite{Calabrese:2009qy,Calabrese:2004eu}. The restriction of the entanglement entropy to a finite spatial volume can be understood as a kind of coarse graining, by the discretization of the available momentum space modes, of the information contained in the quantum state \cite{Takayanagi:2010wp}.

Thus, to measure entanglement entropy in two dimensional field theories, we need to calculate space-like geodesics in the dual asymptotically AdS$_3$ background.  This means there is an intimate relation between two-dimensional entanglement entropy and the equal-time Wightman functions computed in the geodesic approximation. Indeed, the two quantities can be interpreted in terms of each other, as was also observed in \cite{AbajoArrastia:2010yt}. For three dimensional field theories, we need to compute space-like, two-dimensional  minimal surfaces in a dynamical, asymptotically AdS$_4$ background. These coincide precisely with classical space-like worldsheets of fundamental strings with a fixed boundary, as previously studied in \cite{Albash:2010mv}.   As we discussed above, the exponential of the action of such objects (essentially the area) is known to give the expectation value of Wilson loops in the field theory \cite{Maldacena:1998im}.   Thus, in these three-dimensional field theories, there is an intimate relation between entanglement entropy and Wilson loop expectation values, as we explore in this article.

\begin{figure}[h]
\begin{center}
\includegraphics[width=0.4 \textwidth]{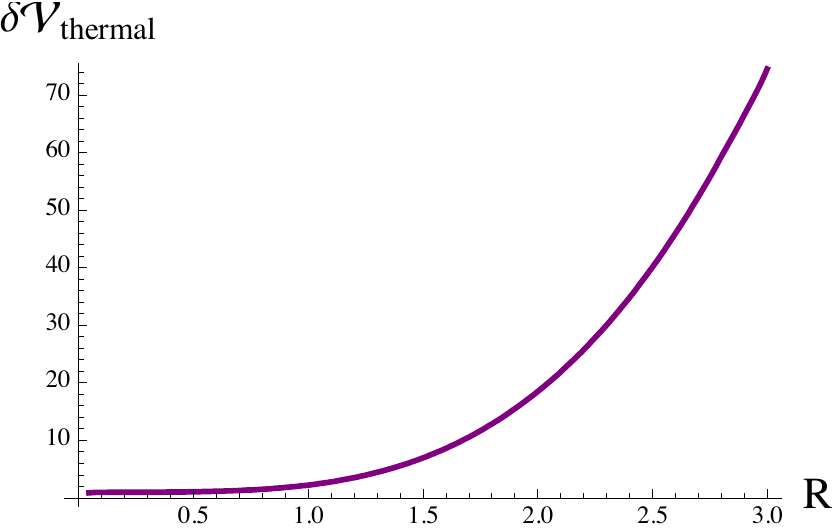}
\caption{Regularized volume $\delta {\cal V}_{\rm thermal}$ as a function of the spatial scale $R$ for d=4.}
\label{FigSphereBBArea}
\end{center}
\end{figure}

For four dimensional field theories, the proposed measure of entanglement entropy for spherical regions involves minimal volume hyper-surfaces whose cross-sections with respect to the AdS radial coordinate are spheres. In particular, in AdS$_5$ we can look for a 3-dimensional volume with 2-sphere cross-section which terminates on the AdS boundary.  This allows us to examine the thermalization of the entanglement entropy in the 4-dimensional field theory. 
The methods used to compute these volumes are very similar to those for the circular Wilson loop described in Section \ref{s:CWL}, with only some powers changing with the dimension. Briefly, the action for the general spherically symmetric $p$-volume in AdS$_d$ ($p\leq d-1$) black brane background is
\ba\label{eq:SphAct}
{ \mathcal V}_{\rm thermal}={\mathcal A}_{S^{p-1}}\int_{0}^{R} d \rho \frac{\rho^{p-1}}{z(\rho)^{p}} \sqrt{1-(1 - M z(\rho)^d ) v'(\rho)^2 - 2 z'(\rho) v'(\rho) },
\ea
where ${\mathcal A}_{S^{p-1}}$ is surface area of a unit $S^{p-1}$. Things continue to proceed as for the Wilson loops: the solution in the pure AdS ($M=0$) background is now a hemi-hypersphere and still obeys (\ref{eq:hemi}). What changes is the volume evaluated on the solution, which we evaluate with a cut-off near the boundary and subtract to renormalize as before.  The result in the vacuum ($M=0$) is
\ba\label{eq:Vsurf}
{ \mathcal V}_{AdS}=4\pi R\int_{0}^{\rho(z_0)} d \rho \frac{\rho^{2}}{(R^2-\rho^2)^2}=2\pi\left(\frac{R^2}{z_0^2}+\log{\frac{z_0}{\sqrt{2}R}}\right)+(\mbox{finite})\, .
\ea
The regularized volume is calculated analogously to previous cases by subtracting the divergent part of the volume of a pure AdS 3-surface with terminates on the same sphere on the boundary,
\be
\delta{ \mathcal V}_{\rm thermal}(R) ={ \mathcal V}_{\rm thermal}-2\pi\left(\frac{R^2}{z_0^2}+\log{\frac{z_0}{\sqrt{2}R}}\right)\, .
\ee
We plot the regularised volume $\delta{ \mathcal V}_{\rm thermal}$ as a function of the radius $R$ of the sphere on the boundary in Fig.~\ref{FigSphereBBArea}.

%%%%%%%%%%%%%%%%%%%%%%%%%%%%%%%%%%%%%%%%%%%%%%%%%%%%%%%%%%%%%%%%%%%%%%%

\section{Scale dependent thermalization from AdS/CFT}\label{thermalization}

In the previous section we discussed how the equal-time Wightman function (related to geodesics in AdS space) and Wilson loops (related to minimal surfaces in AdS space),  and entanglement entropy (related to minimal lengths, surfaces and volumes in different dimensions)  can probe thermal equilibrium.    In this section we will use the same probes to study the dynamics of equilibration in strongly coupled field theories.  The basic setup is to drop a shell of matter with vanishing rest mass (``null dust'') into AdS space to form a black hole.   This is dual to the homogeneous injection of energy into the dual field theory and its subsequent thermalization.   We want to probe the rates at which thermalization occurs on different spatial scales.
To this end we consider dynamical Vaidya-type backgrounds, again given by equation (\ref{eq:Vaidya}), but this time using the continuous mass function of equation (\ref{eq:mv}). This dynamical background continuously interpolates between pure AdS and AdS with a Schwarzschild black brane.
Recall that the parameter $v_0$ in the mass-function (\ref{eq:mv}) dictates how ``thick'' the shell is. $v_0\rightarrow 0$ is the step function limit corresponding to an infinitely thin shell.  In this limit, the infalling shell represents a shock wave.

%%%%%%%%%%%%%%%%%%%%%%%%%%%%%%%%%%%%%%%%%%%%%%%%%%%%%%%%%%%%%%%%%%%%%%%

\subsection{Equilibration of the two-point function}

%%%%%%%%%%%%%%%%%%%%%%%%%%%%%%%%%%%%%%%%%%%%%%%%%%%%%%%%%%%%%%%%%%%%%%%

\subsubsection{Two-dimensional field theories: analytic treatment}\label{BTZanalytic_Vaidya}

We first consider the $d=2$ case where analytic
computations are possible.  In this case, the Vaidya metric \eqref{eq:Vaidya} becomes
\be \label{AdS3Vaidya}
ds^2 = - [r^2 -m(v)]dv^2+ 2 dr dv + r^2 dx^2\,,\qquad r\equiv {1\over z}.
\ee
In order to have analytic control over the solution, let us consider the thin shell limit $v_0\to0$ of the mass profile \eqref{eq:mv}, which gives
\be \label{AdS3VaidyaMass}
m(v) = r_H^2 \theta(v)\,,\qquad r_H\equiv \sqrt{M},
\ee
where $\theta(v)$ is the step function.

Outside the shock wave, $v>0$, the metric is the standard planar black brane metric
\eqref{eq:BTZ}
\be \label{eq:outsideBTZ}
ds^2_{\text{out}} = - (r^2 - r_H^2)dt^2 + \frac{dr^2}{r^2 - r_H^2} + r^2 dx^2\,,\qquad
t= v- \frac{1}{2r_H} \ln  \frac{|r-r_H|}{r+ r_H}\,,
\ee
which we studied  in detail in the previous section. Inside the shock
wave, $v<0$, the metric is the Poincar\'e AdS$_3$:
\be \label{eq:insideAdS3}
ds^2_{\text{in}} = - r^2 dt^2 + \frac{dr^2}{r^2}+r^2 dx^2\,,\qquad
t= v+\frac 1 r\,.
\ee

We would like to study geodesics in the AdS$_3$ Vaidya geometry which is
\eqref{eq:outsideBTZ} and \eqref{eq:insideAdS3} glued together across
the infalling shell at $v=0$.  In particular, we focus on the equal-time
geodesic which starts and ends at the same time $t=t_0$ on the boundary
$r=\infty$.  There are two possible cases: (i) the geodesic
does not reach the shell and is entirely outside of it, and (ii) the
geodesic crosses the shell.  Because case (ii) is more involved than
case (i), let us first discuss case (i) briefly and then turn to the
discussion of case (ii).

In case (i), the geodesic is given by the equal-time geodesic in the
pure black brane geometry, namely by \eqref{t(r)pm}--\eqref{v(r)pm} with $E=0$.
Therefore, the relation between the (renormalized) geodesic length
$\delta {\cal L}_{\textrm{thermal}}$ and the boundary separation $\ell$ is given by
\eqref{LvsellBTZren}.

Now let us turn to case (ii).  In this case, the part of the geodesic
that is inside the shell is given by the equal-time geodesic in the pure
AdS geometry, namely by \eqref{X(r)pureAdS}, which we write in the
following form:
\begin{align}
 x(r)&
 =\frac{\sqrt{r^2-r_*^2}}{r_* r},
\label{X(r)pureAdS2}
\end{align}
where we set $r_*\equiv {2/\ell}$.
On the other hand, the part of the geodesic that is outside the shell is
given by the geodesic in the pure black brane geometry, namely by
\eqref{t(r)pm}--\eqref{v(r)pm}. In the present case, we should not set
the parameter $E$ to zero, because the geodesic gets refracted at the shell
and it does not have to be in a constant $t$ slice.

The two parts of the geodesics, the inside part and the outside part,
should be connected so that the total geodesic length is minimized.
Just like Snell's law, this can be stated as a refraction law for the angles
entering and exiting the shell, as follows.
Let us write the metric \eqref{eq:outsideBTZ}, \eqref{eq:insideAdS3} as
\begin{align}
 ds^2
 &= -f(r)^2dv^2+2drdv+r^2dx^2
\end{align}
where
\begin{align}
 f(r)^2&=
 \begin{cases}
  f_{\text{in}}(r)^2=r^2       & \qquad v<0 \\
  f_{\text{out}}(r)^2=r^2-r_H^2 & \qquad v>0
 \end{cases},
\end{align}
and focus on the region very close to the shell at $v=0$.  Consider a
point $P_{\text{in}}$ ($P_{\text{out}}$) just inside (outside) the shell, and let the
coordinate difference between $P_{\text{in}}$ and $P_{\text{out}}$ be $\Delta X=(\Delta
v,\Delta r,\Delta x)$.  Take another point $M$ on the shell
$v=0$, and let the coordinate difference between $P_{\text{in}}$ ($P_{\text{out}}$)
and $M$ be $\Delta X_{\text{in}}=(\Delta v_{\text{in}},\Delta r_{\text{in}},\Delta x_{\text{in}})$ ($\Delta X_{\text{out}}=(\Delta
v_{\text{out}},\Delta r_{\text{out}},\Delta x_{\text{out}})$),
so that $\Delta X_{\text{in}}+\Delta X_{\text{out}}=\Delta
X$.  Then the distance from $P_{\text{in}}$ to $P_{\text{out}}$ via $M$ is
\begin{align}
 \Delta s&=\sqrt{-f_{\text{in}}^2 \Delta v_{\text{in}}^2+2\Delta r_{\text{in}}\Delta v_{\text{in}}+r^2\Delta x_{\text{in}}^2}
 +\sqrt{-f_{\text{out}}^2 \Delta v_{\text{out}}^2+2(\Delta r-\Delta r_{\text{in}})\Delta v_{\text{out}}+r^2(\Delta x-\Delta x_{\text{in}})^2}.
\end{align}
We want to find the point $M$ that minimizes $\Delta s$.
Extremizing this with respect to $\Delta r_{\text{in}},\Delta x_{\text{in}}$, we find
\begin{align}
 {\Delta r_{\text{in}}\over \Delta v_{\text{in}}}
 &={\Delta r+{1\over 2}(f_{\text{in}}^2-f_{\text{out}}^2)\Delta v_{\text{out}}\over\Delta v},\qquad
 {\Delta x_{\text{in}}\over \Delta v_{\text{in}} }={\Delta x\over \Delta v}
\end{align}
and therefore
\begin{align}
 {\Delta r_{\text{out}}\over \Delta v_{\text{out}}}
 &={\Delta r+{1\over 2}(f_{\text{out}}^2-f_{\text{in}}^2)\Delta v_{\text{in}}\over\Delta v},\qquad
 {\Delta x_{\text{out}}\over \Delta v_{\text{out}} }={\Delta x\over \Delta v}
\end{align}
From these we obtain the following refraction law as required:
\be\label{RC}
\left.\frac{dr}{dv}\right|_{\text{in}} - \left.\frac{dr}{dv}\right|_{\text{out}}
= \frac{r_H^2}{2}\,, \qquad \left.\frac{dx}{dv}\right |_{\text{in}}=\left.\frac{dx}{dv} \right |_{\text{out}}\,.
\ee

By plugging the inside solution, \eqref{X(r)pureAdS2}, and the outside
solution, \eqref{t(r)pm}--\eqref{v(r)pm}, into the refraction condition
(\ref{RC}), we obtain the relation between the parameters of the geodesic:
\be\label{RC2}
E = \pm \frac{r_H \sqrt{r_{sw}^2 - r_*^2}}{2 r_{sw}^2} \quad \textrm{and} \quad J = \mp \frac{r_*}{r_H}\,,
\ee
where $r_{sw}$ is the value of $r$ at which the geodesic intersects the
shell, and how to determine it will be explained below.  The first sign
combination applies to branch 1 for $r_{sw} \le r_H/\sqrt 2$ and
branch 2 for $r_{sw} \ge r_H/\sqrt 2$, while the second sign applies to
branch 1 for $r_{sw} \ge r_H/\sqrt 2$ and branch 2 for $r_{sw} \le
r_H/\sqrt 2$. Recall that branches 1 and 2 were defined in section \ref{BTZanalytic},
below eq. \eqref{v(r)pm}.

It is possible that the geodesic crosses the horizon ($r=r_H$) before
reaching the shell ($v=0$).  When this happens, let us require that the
geodesic goes to the boundary $r=\infty$ at $\lambda=\infty$ and, as we
decrease $\lambda$, it enters the horizon with $v$ staying finite.  In the
previous section, we saw that the geodesic crosses the horizon at
$\lambda=\lambda_{H\pm}$ and, below \eqref{voutlambda}, we observed that
$v$ is finite across $\lambda=\lambda_{H-}$ but not across
$\lambda=\lambda_{H+}$.  Therefore, we need that
$\lambda_{H+}<\lambda_{H-}$ which means that $E<0$.  So, we choose the
second sign combination in \eqref{RC2}:
\be \label{RCE&J}
E = - \frac{r_H \sqrt{r_{sw}^2 - r_*^2}}{2 r_{sw}^2} \quad \textrm{and} \quad J = \frac{r_*}{r_H}\,.
\ee
This leaves branch 1 for $r_{sw} \ge {r_H/ \sqrt 2}$ and branch 2 for  $r_{sw} \le {r_H/ \sqrt 2}$.

The equation \eqref{RCE&J} depends on $r_{sw}$ which is where the
geodesic hits the shell.  Because the shell is at $v=0$, the value of
$r_{sw}$ is determined by
\begin{align}
 0=v(r_{sw})
 &={1\over 2r_H}\log\left[
 {r_{sw}-r_H\over r_{sw}+r_H}
 \,
 {
 r_{sw}^2-(1+E)r_H^2\pm\sqrt{r_{sw}^4-(1+J^2-E^2)r_H^2r_{sw}^2+J^2r_H^4}
 \over
 r_{sw}^2-(1-E)r_H^2\pm\sqrt{r_{sw}^4-(1+J^2-E^2)r_H^2r_{sw}^2+J^2r_H^4}
 }
 \right]+t_0
\label{rsweq0}
\end{align}
where in the second equality we used \eqref{v(r)pm} and the $\pm$ signs
correspond to branch 1 and 2, respectively.  If we rewrite $E,J$ in
favor of $r_{sw},r_*$ using the relation \eqref{RCE&J}, this becomes
\begin{align}
 0&=
 {1\over 2r_H}\log\left[
 {2r_{sw}^2(r_{sw}-r_H)+\sqrt{r_{sw}^2-r_*^2\,}\,(2r_{sw}^2-2r_{sw} r_H+r_H^2)
 \over
 2r_{sw}^2(r_{sw}+r_H)+\sqrt{r_{sw}^2-r_*^2\,}\,(2r_{sw}^2+2r_{sw} r_H+r_H^2)}
 \right]
 +t_0.
\label{rsweq}
\end{align}
Note that the two possible signs in
\eqref{rsweq0} lead to the same  condition \eqref{rsweq}.
To solve this, let us make the convenient definitions
\be \label{def}
\rho \equiv \frac{r_{sw}}{r_H}\,, \quad \rho \sin\theta \equiv \frac{r_*}{r_H}\,, \quad a \equiv e^{2 r_H t_0}\,.
\ee
The range of the parameters is $\theta \in [0, \frac{\pi}{2}],$ $\rho\ge 0$, and $a\ge 1$.  In terms of these,  the relation (\ref{rsweq}) becomes
\be
\frac{4 \rho (1+c)}{c+ 2\rho (1+ \rho)(1+c)}=1- \frac 1 a\,,
\ee
with $c \equiv \cos\theta$. This can be solved for $\rho$, giving two branches of solutions:
\be
\rho(a,c)_{\pm} = \frac{a+1}{2(a-1)}
\pm{1\over 2}\sqrt{\left({a+1\over a-1}\right)^2-{2c\over c+1}}\,.
\label{eq:rhopm}
\ee

We can see that only $\rho_+$ is allowed, as follows.  First, we can show
that, for fixed $c$, the derivative of $\rho(a,c)_{\pm}$
with respect to $a$ is always negative (positive) in the range
$a\in(1,\infty)$, $c\in(0,1]$ \footnote{ ${\partial \rho(a,c)_{-}}/{\partial a}$
actually vanishes when $c=0$, while ${\partial \rho(a,c)_{+}}/{\partial a}<0$ also
for $c=0$.  }, so that $\rho(a,c)_{\pm}$ monotonically
decreases (increases) as we increase $a$. Furthermore
\be
\rho(a,c)_{\pm} \rightarrow  \frac 1 2 \left(1 \pm  \sqrt{\frac{1-c}{1+c}} \right) \quad \textrm{as} \quad
a \to \infty \quad (\text{or}\quad t_0\to\infty)\,.
\ee
Therefore, for any finite values of $t_0$,
\be
\rho(a,c)_{-} <  \frac 1 2 \left(1 -  \sqrt{\frac{1-c}{1+c}} \right)  \le \frac 1 2
\le \frac 1 2 \left(1 +  \sqrt{\frac{1-c}{1+c}} \right)
< \rho(a,c)_{+}.
\ee
For $r_{sw} \ge {r_H/ \sqrt 2}$ (i.e.\ $\rho>1/\sqrt{2}$), this means
that we should take the $\rho_{+}$ branch.  The case with $r_{sw} <
{r_H/ \sqrt 2}$ (i.e.\ $\rho<1/\sqrt{2}$) needs some more work.  We only
need to consider geodesics that do not reach $r=0$, since otherwise we cannot
glue branch 1 to branch 2 at $\lambda_{min}$ before the geodesic
reaches the singularity. Thus, when $r_{sw} < {r_H/\sqrt 2}$, we
need $A_{\pm} <0 $ and $B_{\pm}>0$, that is $J< 1+E$, for $E<0$ and $0<
J< 1$. Using \eqref{RCE&J}, the last condition  translates into the
following condition for $\rho$:
\bea
\begin{cases}
\frac 1 2 \left( 1- \sqrt{\frac{1 + c}{1 - c}} \right) <
-\frac 1 2 \left( 1- \sqrt{\frac{1 + c}{1 - c}} \right)
<  \rho <   \frac {1}{\sqrt{ 2}}
 & \text{for}\quad c \in [0,\frac{1}{\sqrt 2}]\,,\\[2ex]
\frac 1 2 \left( 1+ \sqrt{\frac{1-c}{1+c}} \right) <  \rho <  \frac {1}{\sqrt{ 2}}
 & \text{for}\quad  c \in ( \frac{1}{\sqrt 2}, 1]\,.
\end{cases}
\eea
This means that we should take $\rho_{+}$ in (\ref{eq:rhopm}) for all $c \in [0,1]$.

\begin{figure}[h]
\begin{minipage}[h]{7.5cm}
\begin{center}
\includegraphics[width=7.5cm, height=5cm,clip]{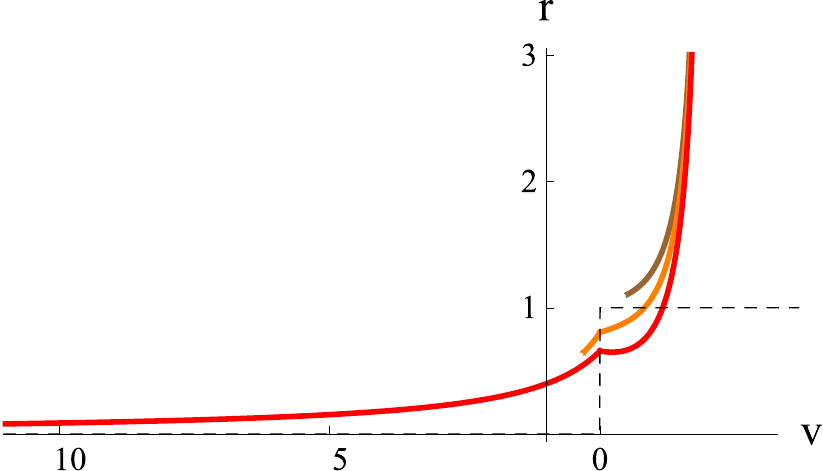}
\caption{Geodesics in the $(v,r)$ plane for fixed $t_0 =2$ and $\ell \approx 3.0$ (brown, top), $\ell \approx 4.6$ (orange, middle),
$\ell \approx 68.2$ (red, bottom). In black dashed, the apparent horizon.}
\label{fig:r(v)Shellnew.pdf}
\end{center}
\end{minipage}
~~~~~
\begin{minipage}[h]{7.5cm}
\begin{center}
\includegraphics[width=7.5cm,height=5cm,clip]{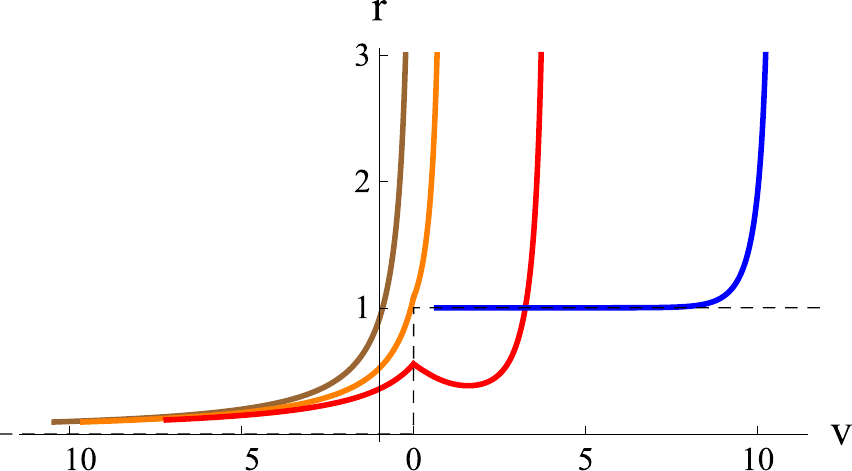}
\caption{Geodesics in the $(v,r)$ plane for fixed $\ell \approx 21.3$ and, from left to right, $t_0 = 0.1$ (brown), $t_0 =1$ (orange),
$t_0 =4$ (red), $t_0 \approx 10.6$ (blue). In black dashed, the apparent horizon.}
\label{fig:r(v)ell=21new.pdf}
\end{center}
\end{minipage}
\end{figure}

In Figures \ref{fig:r(v)Shellnew.pdf} and \ref{fig:r(v)ell=21new.pdf}, geodesics in the $(v,r)$ plane are plotted for different values of $t_0$ and $\ell$.
The geodesics start from the boundary at $(r,v) = (\infty, t_0)$, plunge in the bulk in the planar BTZ geometry, eventually refract at the shell at $v=0$ and propagate in pure AdS\@.
All profiles are symmetric under $x \to -x$. The curve in black dashed is the apparent horizon.

% CP diagram

\begin{figure}[h]
\begin{minipage}[h]{7.5cm}
\begin{center}
\includegraphics[width=0.5 \textwidth,clip]{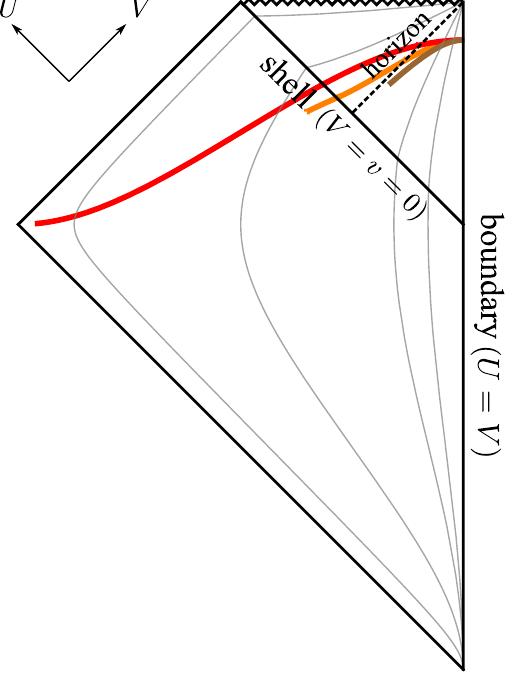}
\caption{Geodesics in the spacetime Carter-Penrose diagram for fixed $t_0 =2$ and $\ell \approx 3.0$ (brown, bottom), $\ell \approx 4.6$ (orange, middle),
$\ell \approx 68.2$ (red, top). Surfaces of constant $r$ are plotted in gray.}
\label{fig:CPdiagram_t02}
\end{center}
\end{minipage}
~~~~~
\begin{minipage}[h]{7.5cm}
\begin{center}
\includegraphics[width=0.5 \textwidth,clip]{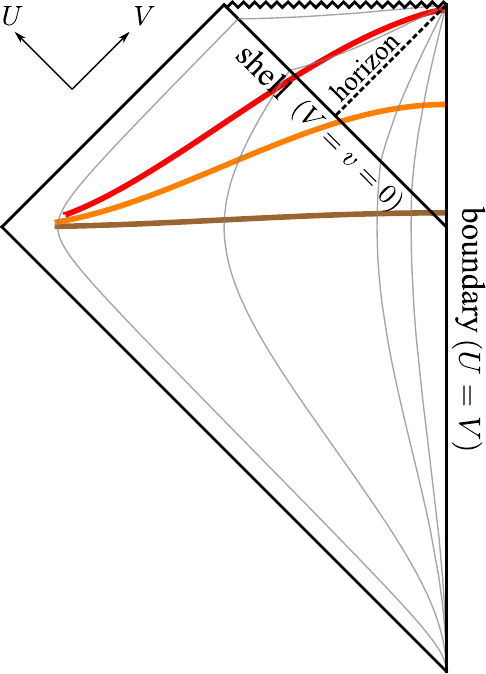}
\caption{Geodesics in the spacetime Carter-Penrose diagram for fixed $\ell \approx 21.3$ and $t_0 = 0.1$ (brown, bottom), $t_0 =1$ (orange, middle),
$t_0 =4$ (red, top).  Surfaces of constant $r$ are plotted in gray. We do not display in the diagram the geodesic with $t_0 \approx 10.6$ (in blue in Fig.~\ref{fig:r(v)ell=21new.pdf}) because it would cover the horizon. }
\label{fig:CPdiagram_ell}
\end{center}
\end{minipage}
\end{figure}

In Figures \ref{fig:CPdiagram_t02} and \ref{fig:CPdiagram_ell}, we display the boundary time and spatial separation dependence of the geodesics in the Carter-Penrose diagram of Vaidya spacetime.
The transformations to the $(U,V)$ coordinates are given by
\be
U \equiv \frac{2}{\pi} \tan^{-1} \left[ \frac{r_H}{2} \left(t + \frac 1 r \right)\right]\,, \qquad V \equiv \frac{2}{\pi} \tan^{-1} \left[ \frac{r_H}{2} \left(t - \frac 1 r \right)\right] \,,
\ee
for $v<0$,
\be
U \equiv \frac{2}{\pi} \tan^{-1} \left[ \tanh \left( \frac{r_H}{2} \left(t + \frac{1}{r_H} \coth^{-1} \frac{r}{r_H}\right) \right)\right]\,, \qquad  V \equiv \frac{2}{\pi} \tan^{-1} \left[ \tanh \left( \frac{r_H}{2} \left(t - \frac{1}{ r_H} \coth^{-1} \frac{r}{r_H}\right) \right)\right]\,,
\ee
for $v>0$ and $r>r_H$, and
\be
U \equiv \frac{2}{\pi} \tan^{-1} \left[ \coth \left( \frac{r_H}{2} \left(t + \frac{ 1}{ r_H} \tanh^{-1} \frac{r}{r_H}\right) \right)\right]\,, \qquad  V \equiv \frac{2}{\pi} \tan^{-1} \left[ \tanh \left( \frac{r_H}{2} \left(t - \frac{ 1}{ r_H} \tanh^{-1} \frac{r}{r_H}\right) \right)\right]\,,
\ee
for $v>0$ and $r<r_H$. In these coordinates, the ``singularity''  is located
at $U+V=1$ and is depicted by a wavy line in
Fig.~\ref{fig:CPdiagram_t02} and \ref{fig:CPdiagram_ell}, the shell is
at $V=0$ and the horizon is at $U=1/2$ (dashed line).

%Geodesic length

With all the relations above in hand, we can compute the geodesic length
and spatial boundary separation in terms of the parameters of the geodesic.  The geodesic length is the sum of the
geodesic length for the inside part, which can be computed from
\eqref{lambda(r)pureAdS}, and the one for the outside part, which can be
computed from \eqref{lambdapm}.  The result is
\begin{align}
 \delta {\cal L}
 = 2 \ln \frac{r_{sw}+\sqrt{r_{sw}^2 - r_*^2}}{r_*}  \label{Lb}
 - \ln \left[ \frac{4 r_{sw}^4 (2 r_{sw}^2 -r_*^2 - r_H^2) + r_H^4 (r_{sw}^2 -r_*^2) + 4 r_{sw}^3 (2r_{sw}^2 - r_H^2) \sqrt{r_{sw}^2 -r_*^2}}{4 r_{sw}^4}\right].
\end{align}
Here, just as in \eqref{LvsellBTZren}, $\delta {\cal L}$ has been renormalized
by subtracting an IR divergent quantity $2\ln (2r_0)$.  The spatial
boundary separation can also be computed by summing the contribution
from the inside part \eqref{X(r)pureAdS2} and the outside part
\eqref{X(r)pm}, the result being
\begin{align}
\ell & = 2 \frac{\sqrt{r_{sw}^2 -r_*^2}}{r_* r_{sw}} + \frac{1}{r_H} \ln \left[  \frac{2 r_{sw} (r_{sw}^2 +r_* r_H) +(2 r_{sw}^2 - r_H^2) \sqrt{r_{sw}^2 - r_*^2}}{2 r_{sw} (r_{sw}^2 - r_* r_H) +(2 r_{sw}^2 - r_H^2) \sqrt{r_{sw}^2 - r_*^2}} \right]\,. \label{Xb}
\end{align}

Although $\delta {\cal L}$ and $\ell$ are written in terms of $r_{sw}$ and $r_*$ in
the above expressions, we can write them in terms of $t_0$ and $s \equiv \sqrt{1-c^2}$ using
the definitions (\ref{def}) and plugging in the solution
$\rho=\rho(a,c)_{+} $ of Eq.~\eqref{eq:rhopm}
\be
2 \rho = \coth (r_H t_0) + \sqrt{\coth^2(r_H t_0) -\frac{2c}{c+1}}\,.
\ee
Explicitly,
\begin{align}
\label{Lren}
\delta {\cal L}(t_0, \ell)
& = 2 \ln \left[ \frac{\sinh(r_H t_0)}{r_H s(\ell, t_0)} \right]\,,
\end{align}
where $s(\ell, t_0) \in [0,1]$ is parametrically defined by
\begin{align}
 \ell
 =
 {1\over r_H}\left[{2c\over s\rho}+\ln\left({2(1+c)\rho^2+2s\rho-c \over 2(1+c)\rho^2-2s\rho-c}\right)\right]\,.
\end{align}

%%%%%%%%%%%%%%%%%%%%%%%%%%%%%%%%%%%%%%%%%%%%%%%%%%%%%%%%%%%%%%%%%%%%%%%

\subsubsection{\texorpdfstring{$d\geq 2$}{d>1}-dimensional field theories: numerical analysis}\label{sec:dyngeo}

As in the AdS$_3$ case, we consider a four- or five-dimensional dynamical metric which interpolates between pure planar AdS at early times and a Schwarzschild black brane at late times. The transition is induced by an infalling shell of null dust. The metric in ($d+1$) dimensions was given in Eq.~\eqref{eq:Vaidya}. 
We again consider the mass function (\ref{eq:mv}), where the profile parameter is set to $v_0 = 0.01$ in the remainder. We consider geodesics with a boundary separation along $x_1$, denoted $x$ in the following, while the ($d-2$) coordinates ($x_2, \dots ,x_{d-2}$) of both endpoints are the same.
This implies that $z$ and $v$ only depend on $x$, i.e.\ $z=z(x)$ and $v=v(x)$. We assume that $x$ runs between $-\ell/2$ and $+\ell/2$. Then the length of the geodesic is given by:
\ba
{\cal L} =  \int_{-\ell/2}^{\ell/2} dx\, \frac{\sqrt{1- (1-m(v)z^d)v'^2 - 2 z' v'}}{z(x)}\,,
\label{geodlength4dim}.
\ea
where  ${\ }^{\prime}\equiv d/dx$. 
We notice that the integrand has no explicit  $x$-dependence, implying the existence of a conserved quantity, similar to the black brane case. The conservation equation reads:
\ba
1-(1-m(v) z^d) v'^2 - 2 z' v' = \left(\frac{z_*}{z}\right)^2\,.
\label{geod4dimcons}
\ea
The two equations of motion following from (\ref{geodlength4dim}) are
\ba
z v'' + 2 z' v' - 1 +  v'^2 + \frac{d-2}{2} m(v) z^d v'^2 &=& 0,
\label{geod4dimeom1}
\\
z'' + (1- m(v) z^d) v'' - \frac{\dot m (v)}{2} z^d v'^2 - d\, m(v) z^{d-1} z' v' &=& 0,
\label{geod4dimeom2}
\ea
where $\dot m (v) = d m(v)/dv$. 
It is possible to show that (\ref{geod4dimcons}) and (\ref{geod4dimeom1}) imply (\ref{geod4dimeom2}) after taking the derivative of (\ref{geod4dimcons}) with respect to $x$. Therefore we restrict ourselves to solving (\ref{geod4dimcons}) and (\ref{geod4dimeom1}). We construct solutions $z(x)$ and $v(x)$ with the $x\rightarrow -x$ symmetry which is already present in the equations of motion, and which satisfy the boundary conditions
\ba
z(0)=z_*, \quad v(0)=v_*, \quad v'(0)=0=z'(0).
\ea
After constructing a geodesic for a specific choice of values $(z_*, v_*)$, information about the boundary separation and boundary time at which the geodesic is inserted follows from
\ba
z(\ell/2) = z_0, \quad v(\ell/2) = t_0.
\ea
The on-shell length is obtained from Eq.~(\ref{geodlength4dim}), upon use of the conservation equation (\ref{geod4dimcons}),
\ba
{\cal L}(\ell,t_0) = 2 \int^{\ell/2}_{0} dx \frac{z_*}{z(x)^2}.
\ea
This length should be regularized by subtracting the cut-off dependent part $2 \ln (2/z_0)$, yielding $\delta {\cal L}(\ell,t_0)$.

%%%%%%%%%%%%%%%%%%%%%%%%%%%%%%%%%%%%%%%%%%%%%%%%%%%%%%%%%%%%%%%%%%%%%%%

\subsubsection{Thermalization}\label{Them2pt}

\begin{figure}[ht]
\centering
\centering
\includegraphics[width=0.25\linewidth]{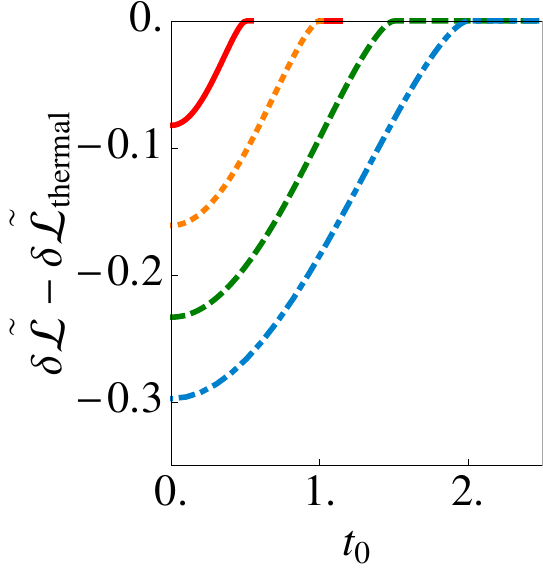}
\includegraphics[width=0.25\linewidth]{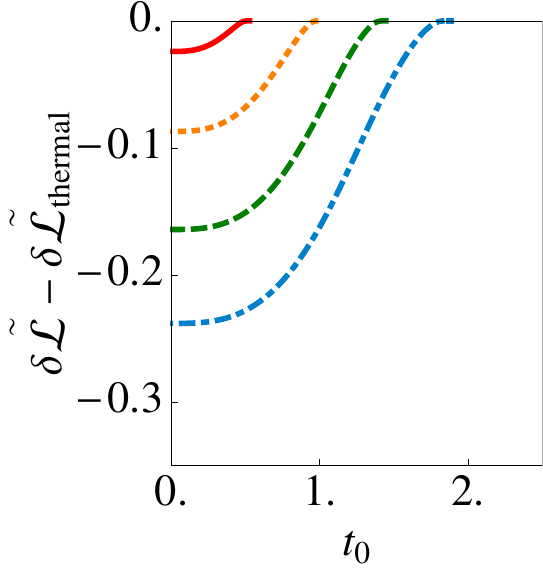}
\includegraphics[width=0.25\linewidth]{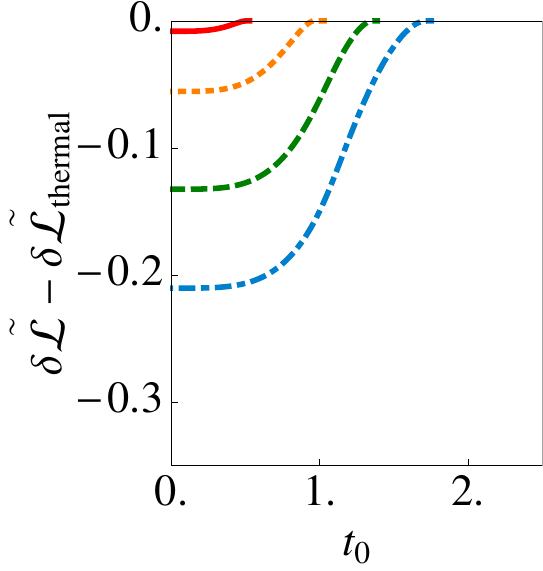}
\caption{$\delta{\tilde {\cal L}} - \delta{\tilde {\cal L}}_{{\rm thermal}}$ ($\tilde{\cal L} \equiv \cal L / \ell $) as a function of  boundary time $t_0$ for $d=2$ (left), $d=3$ (middle), and $d=4$ (right) for a thin shell ($v_0 = 0.01$).   The boundary separations were taken to be $\ell = 1,2,3,4$ (from top to bottom curve).  All quantities are given in units of $M$.  These numerical results coincide with our analytical expressions in AdS$_3$ ($d=2$) in the limit $v_0 \to 0$.}
\label{fig:deltaL}
\end{figure}

Having collected the geodesic lengths in the infalling shell backgrounds in AdS$_{3,4,5}$, we can use these to describe the process of thermalization following a quench. 
To this end, we measure the approach to thermal equilibrium by comparing $\delta {\cal L}$ at any given time with the late time result $\delta {\cal L}_{\rm thermal}$.  In any dimension, this compares the logarithm of the two-point correlator at different spatial scales with the logarithm of the thermal correlator. However, we find it more revealing to consider  $\tilde{{\cal L}} \equiv \delta {\cal L}/\ell$, where we divide by the spatial separation on the boundary (we will discuss the significance of this later in Section \ref{s:dynahyper} where we deal with entanglement and  Kolmogorov-Sina\"i entropies). In figure \ref{fig:deltaL} we plot this measure for two-, three- and four-dimensional field theories.  In all three cases we observe a delay in the onset of thermalization. The reason for this ``delay'' is simply that the effect of the medium only becomes fully apparent at distances of the order of the thermal screening length $\ell_{\rm D} \sim (\pi T)^{-1}$. Although a very small volume of linear dimension
$\ell \ll \ell_{\rm D}$ would appear fully thermalized after a time $t_0=\ell/2$, its contribution to the entropy of a large volume would be disproportionately small, because it does not support the momentum modes that constitute the thermal medium at large. As a consequence, the rapid linear increase of the logarithm of the two-point function seen in Fig.~\ref{fig:deltaL} thus only sets in after some delay. Obviously, this effect is more pronounced in higher dimensions.

\begin{figure}[t]
\centering
\includegraphics[width=0.25\linewidth]{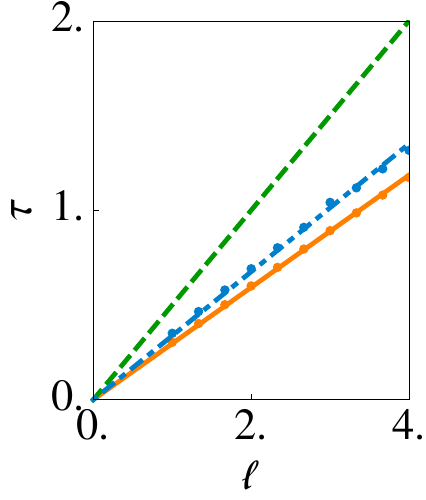}
\includegraphics[width=0.25\linewidth]{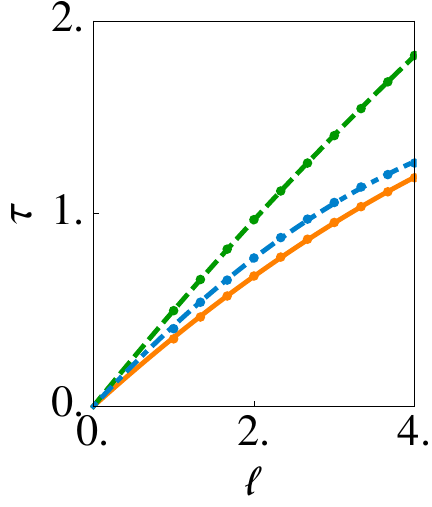}
\includegraphics[width=0.25\linewidth]{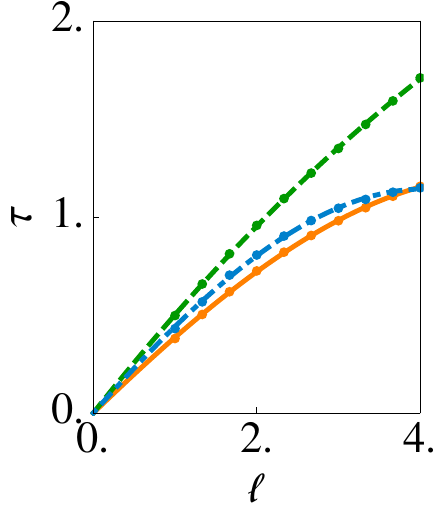}
\caption{Thermalization times ($\tau_{{\rm crit}}$,  top line; $\tau_{{\rm max}}$, middle line; $\tau_{1/2}$, bottom line) as a function of spatial scale for $d=2$ (left), $d=3$ (middle) and $d=4$ (right) for a thin shell ($v_0 = 0.01$).   All thermalization time scales are linear in $\ell$ in two dimensions, and deviate from linearity in $\ell$ in three and four dimensions.
}
\label{fig:thermtime}
\end{figure}

From the curves in Fig.~\ref{fig:deltaL} we can extract different thermalization times for any spatial scale:
\begin{itemize}
\item[(1)] The critical time $\tau_{\rm crit}$ at which the tip of the geodesic grazes the middle of the shell at $v=0$.  This can be computed by asking when a geodesic with a given boundary separation in the black brane geometry outside the infalling shell just grazes the latter: $\tau_{\rm crit} (\ell) =  \int^{z_*}_{z_0} \frac{d z}{ 1- M z^d}$, where $z_*$ is determined by the boundary separation $\ell$.
\item[(2)] The half-thermalization time $\tau_{1/2}$:  time that measures the duration for the curves to reach half of their equilibrium value.
\item[(3)]  The time $\tau_{\rm max}$: time at which thermalization proceeds most rapidly, thus for which the curves in Fig.~\ref{fig:deltaL} are steepest.
\end{itemize}
For two, three and four dimensional field theories all these times are plotted in Fig.~\ref{fig:thermtime}.   In $d=2$ we can analytically derive the linear relation $\tau_{{\rm crit}} \equiv \ell/ 2$, as also observed in  \cite{AbajoArrastia:2010yt}.

The linearity of $\tau_{{\rm crit}}(\ell)$ in two dimensions is expected from general arguments in conformal field theory \cite{Calabrese:2009qy}, and the coefficient is as small as possible under the constraints of causality following a quantum quench.    The thermalization time scales $\tau_{1/2}$ and $\tau_{{\rm max}}$ for three- and four-dimensional field theories (Fig.~\ref{fig:thermtime}, middle and right) are sublinear in the spatial scale.  In the range we study, the complete thermalization time $\tau_{{\rm crit}}$ deviates slightly from linearity, and is somewhat shorter than $\ell/2$.  These observations pose the question whether a rigorous causality bound for thermalization processes exists or not. We will come back to this question below, after examining other probes than two-point functions.    One of the issues here is that the initial conditions are homogeneous -- thus different domains will be independently coming to equilibrium at the same temperature possibly leading to apparent violations of causality.

General arguments for a sharp quench in a 2-dimensional field theory \cite{Calabrese:2009qy,AbajoArrastia:2010yt} predict a nonanalytic feature where thermalization at a spatial scale $\ell$ is completed abruptly at $\tau_{{\rm crit}}(\ell)$.  This is evident in Fig.~\ref{fig:deltaL} (left) as a sudden change in the slope at $\tau_{{\rm crit}}$, smoothed out only by the small non-zero thickness of the shell, or equivalently, by the intrinsic duration of the quench.    We find a similar (higher-order) non-analyticity for $d=3,4$ (Fig.~\ref{fig:deltaL}, middle and right) and expect this to be a general feature for quantum quenches in all dimensions.

Fig.~\ref{fig:thermtime} shows that complete thermalization of the equal-time correlator is first observed at short length scales, or large momentum scales (see also \cite{Lin:2008rw}),  in contrast with the behavior of weakly coupled gauge theories. In the ``bottom-up'' scenario \cite{Baier:2000sb} applicable to that case, hard quanta of the gauge field do not equilibrate directly by randomizing their momenta, but  gradually degrade their energy by radiating soft quanta, which fill up the thermal phase space and equilibrate by collisions among themselves. This bottom-up scenario is  linked to the infrared divergence of the splitting functions of gauge bosons and fermions in perturbative gauge theory. It contrasts with the ``democratic'' splitting properties of excitations in strongly coupled SYM theory that favor an approximately equal sharing of energy and momentum \cite{Hatta:2008tx}. One might, therefore, have expected that thermalization  proceeds fundamentally differently (``top-down'') in strongly coupled gauge theories, and this is evident within the AdS/CFT paradigm with the natural initial conditions used here.

%%%%%%%%%%%%%%%%%%%%%%%%%%%%%%%%%%%%%%%%%%%%%%%%%%%%%%%%%%%%%%%%%%%%%%%

\subsection{Equilibration of Wilson loops}

Another nonlocal probe of thermalization is the Wilson loop expectation value.  As we discussed in Sec.~\ref{WLs}, this quantity is related in the AdS/CFT correspondence to the area of a minimal surface in AdS space that bounds the desired loop on the AdS boundary.  In this section we compute such minimal areas in the infalling shell background.

\subsubsection{Circular Wilson loops}\label{s:dynoWL}

We first consider minimal surfaces corresponding to circular Wilson loops in the Vaidya background. The corresponding minimal surfaces were analyzed for the vacuum (pure AdS) and finite temperature equilibrium (AdS black brane) situations in Sect. \ref{s:CWL}.  This set-up was also analyzed in \cite{Albash:2010mv} in the context of studying entanglement entropy. The parameterization of the surface and boundary conditions are as given in Section \ref{s:CWL}, the only difference being that the mass function $m(v)$ is now given by (\ref{eq:mv}). The area functional is a slight modification of (\ref{eq:ACWLBB}) (${\cal A} = \alpha' A_{\rm NG}$) given by
\ba
{\cal A}(t_0, R) =\int_{0}^{R} d\rho \frac{\rho}{z^2}\sqrt{1- \left(1-m(v)z^d\right)v'^2 - 2 z' v' }\,,
\label{loopaction}
\ea
where  ${\ }^{\prime}\equiv d/d\rho$.
The explicit $\rho$ dependence again means there is no conservation equation. The equations of motion become quite involved and we omit them here.

We solve the equations of motion numerically, and although we want our output as a function of the boundary radius $R$, for practical purposes we have to input boundary conditions at the tip of the surface at $\rho=0$ where $z=z_*$ and $v=v_*$. For each value of $z_*$ we consider, we find numerical solutions for the functions $z(\rho)$ and $v(\rho)$ for various values of $v_*$, each of which reaches the boundary (which is cut-off at $z_0$) at  $\rho_0(v_*)$. We then make an interpolation of $\rho_0(v_*)$, which allows us to find the exact value of $v_*$ such that $\rho_0(v_*)=R$. For each value of $z_*$, we now know the value of $v_*$ giving the surface with boundary radius $R$, and we calculate the area (using the functional above) and the boundary time $t_0=v(R)$. These are the quantities we finally plot. There are further numerical challenges with a divergence as $\rho\rightarrow 0$ and with integrating our solutions near the AdS boundary, where contributions to the area are weighted much higher. Thus we used an expansion around $\rho=0$, and different techniques to deal with a separate integral near the boundary, as necessary.

\begin{figure}[h]
\centering
\includegraphics[width=0.3\linewidth]{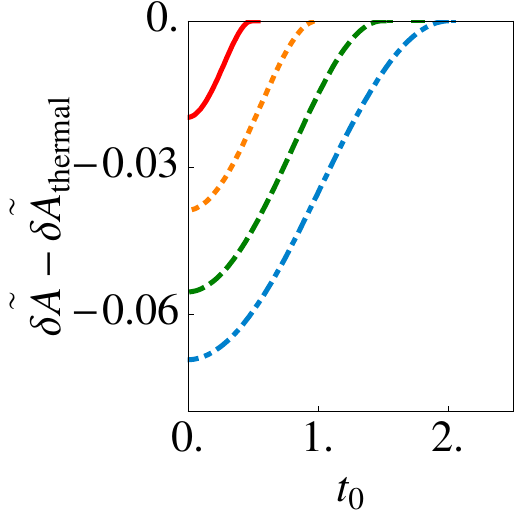}
\includegraphics[width=0.3\linewidth]{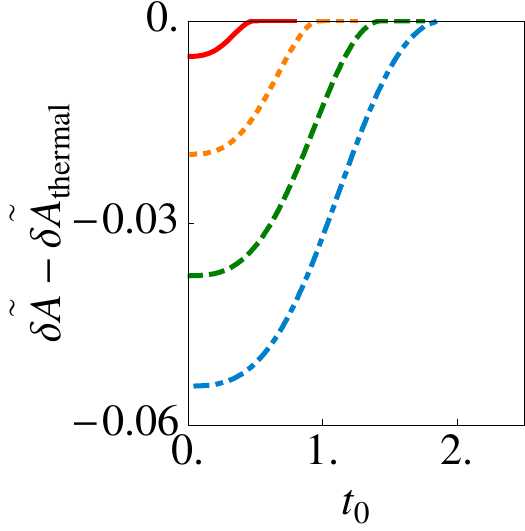}
\caption{$\delta\tilde{{\cal A}} - \delta\tilde{{\cal A}}_{{\rm thermal}}$ ($\tilde {\cal A} \equiv {\cal A}/ (/\pi R^2)$) as a function of  $t_0$ for circular Wilson loop radii $R = 0.5, 1, 1.5, 2$ (top curve to bottom curve) and mass shell parameters $v_0 = 0.01$, $M=1$, in three-dimensional (left panel) and four-dimensional (right panel) field theories.
}
\label{DL-Avt-2}
\end{figure}

As in (\ref{eq:WLAreg}) we regulate the boundary by subtracting the cut-off dependent piece of the AdS area,
\be
\delta {\cal A} (t_0,R) = {\cal A}(t_0,R)-\frac{R}{z_0} \,.
\ee
We define $ \delta \tilde{{\cal A}} \equiv \delta {\cal A}/ (\pi R^2)$ dividing by the area of the region on the boundary bounded by the loop.
We then plot $\delta \tilde{{\cal A}}- \delta \tilde{{\cal A}}_{\rm thermal}$ as a function of the boundary time, subtracting the black brane value to show the approach to thermalization. This is shown in Fig.~\ref{DL-Avt-2} for AdS$_4$ and AdS$_5$ for a series of values of $R$.

\begin{figure}[h]
\centering
\includegraphics[width=0.25\linewidth]{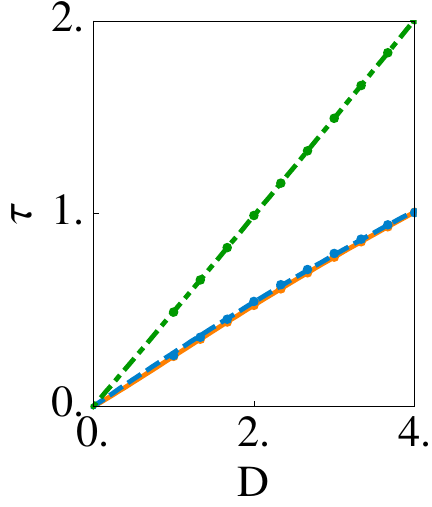}
\hspace{0.05\linewidth}
\includegraphics[width=0.25\linewidth]{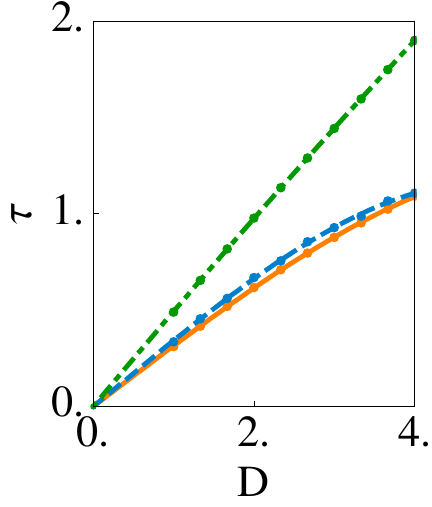}
\caption{Wilson loop thermalization times ($\tau_{{\rm  crit}}$,  top line; $\tau_{{\rm max}}$, middle line; $\tau_{1/2}$, bottom line) as a function of the diameter for circular Wilson loop operators in  three-dimensional (left) and four-dimensional (right) field theories.
}
\label{DL-Avt-3}
\end{figure}

Repeating the analysis performed for Wightman functions, we calculate the three thermalization times defined earlier (see Sec.~\ref{Them2pt}) as a function of the loop diameter (Fig.~\ref{DL-Avt-3}). We use the diameter rather than the radius here as it is the analogue of the separation $\ell$ that we plotted for the geodesics.   The complete thermalization time $\tau_{{\rm crit}}(D)$ is close to being a straight line of slope $1/2$ for three dimensional theories  over the range of scales that we study  (also see \cite{Albash:2010mv}) (it would be unit slope as a function of the radius).  But in four dimensions $\tau_{{\rm crit}}(D)$  deviates somewhat from linearity and is shorter than $D/2$.   Overall, our thermalization times for Wilson loop averages are remarkably similar to those for two-point correlators .  This suggests that higher-order correlators thermalize similarly to the basic Green function in these strongly coupled theories, so that any of these non-local probes fairly assesses progress towards thermalization.

%%%%%%%%%%%%%%%%%%%%%%%%%%%%%%%%

\subsubsection{Infinite Rectangular Strips}\label{rectstripDyn}

As a second example we consider an infinite strip, similarly to \cite{Albash:2010mv}.    A key difference between the strip and the Wilson circle is that the size of the latter is set by a single scale, the radius.  The strip could be regarded as a limit of a ellipsoidal loop with a highly elongated semi-major axis.  Thus there are in effect two scales involved in the Wilson strip -- the width and the (infinite) length.  As before, at early times the background is AdS$_4$, but evolves to a black brane at late times due to an infalling shell of null dust. The metric is at all times given by (\ref{eq:Vaidya}) with $d=3$ or 4. Again, the mass function $m(v)$ determining the evolution of the metric as a function of the bulk light-cone time $v$ is modeled by the expression (\ref{eq:mv}). We set the parameter $v_0 = 0.01$. At the AdS$_4$ boundary we consider a rectangular strip parametrized by the coordinates $x_1$ and $x_2$, such that
\ba
x_1 \in (-\ell/2, \ell/2 ), \quad x_2 \in (0, R),
\ea
where $R$ will be taken to infinity \footnote{This is exactly the same configuration as in \cite{Albash:2010mv}.}. The coordinate $x_1$ is denoted $x$ in the rest of the section. The area of the string surface with this rectangular base is given by
\ba
{\cal A}(t_0, \ell, R) = \frac{R}{2 \pi} \int_{-\ell/2}^{\ell/2} dx \frac{\sqrt{1- (1-m(v)z^d)v'^2 - 2 z' v' }}{z^2},
\label{stripaction}
\ea
when we consider the following embedding profile for the string:
\ba
v \equiv v(x), \quad z\equiv z(x), \quad z(\pm \ell /2) =z_0, \quad v(\pm \ell/2) = t_0 .
\ea
We notice that $z$ and $v$ only depend on $x$ and that $x$ itself does not appear in the action, similarly to the geodesic length in a dynamical Vaidya background. There exists a conserved quantity corresponding to the maximum value of $z$ denoted as $z_*$. The conservation equation then reads:
\ba
1-(1-m(v) z^d) v'^2 - 2 z' v' = \left(\frac{z_*}{z}\right)^4.
\label{stripcons}
\ea
From the action (\ref{stripaction}) we obtain the following two equations of motion:
\ba
z v'' + 4 z' v' - 2 + 2 v'^2 + \frac{d-4}{2} m(v)z^d v'^2 &=& 0,
\label{stripeom1}
\\
z'' + (1- m(v) z^d) v'' - \frac{\dot m(v)}{2} z^d v'^2 - d\, m(v) z^{d-1} z' v' &=& 0.
\label{stripeom2}
\ea
One can show that taking the derivative of (\ref{stripcons}) with respect to $x$ and combinig with (\ref{stripeom1}) leads to (\ref{stripeom2}). This it is sufficient to solve (\ref{stripcons}) and (\ref{stripeom1}). In order to do this numerically we use the symmetry  of the string surface, i.e.\ we construct one half of the solution $z(x)$ and $v(x)$ and then use the reflection symmetry to construct the other half. Our boundary conditions are $z(0) = z_*,\quad v(0)= v_*,\quad v'(0) = 0 = z'(0)$.
After constructing a string surface satisfying these boundary conditions, we read off the boundary separation and the boundary time using $z(\ell/2) = z_0$, $v(\ell/2) = t_0$.

\begin{figure}[ht]
\centering
\centering
\includegraphics[width=0.25\linewidth]{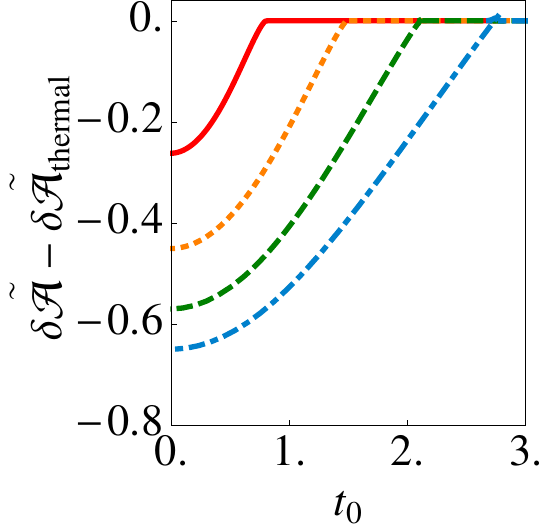} \hfil
\includegraphics[width=0.25\linewidth]{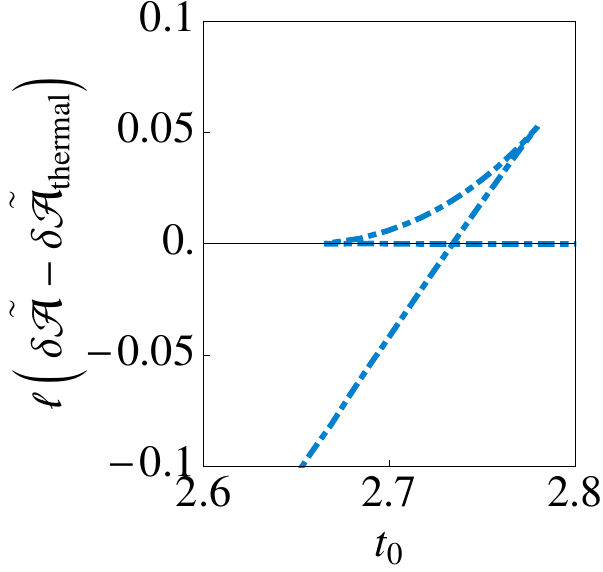} \hfil
\includegraphics[width=0.25\linewidth]{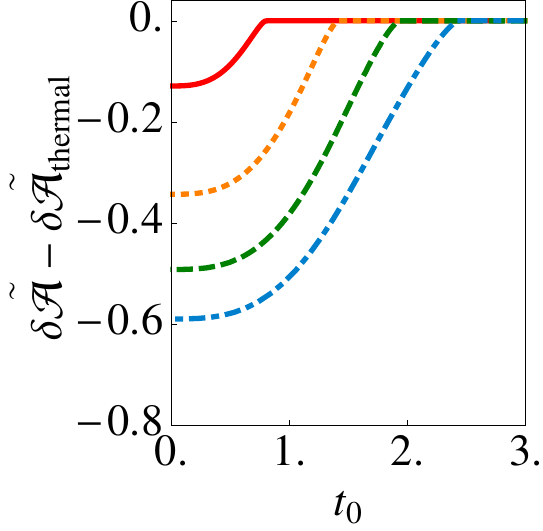}
\caption{$\delta{ \cal \tilde A} - \delta{ \cal \tilde A}_{{\rm thermal}}$ (${\cal \tilde A} \equiv {\cal A} / (R \ell/ \pi )$) as a function of  boundary time $t_0$ for $d=3$ (left) and $d=4$ (right) for a thin shell ($v_0 = 0.01$). The boundary separations were taken to be $\ell = 1,2,3,4$ (from top to bottom curve).  All quantities are given in units of $M$. The middle panel shows a zoomed-in version of the swallow tail for $\ell$ = 4, where we plot $ \ell (\delta{ \cal \tilde A} - \delta{ \cal \tilde  A}_{{\rm thermal}})$ to amplify the effect. }
\label{fig:deltaAstrip}
\end{figure}

Using the symmetries of the embedded string surface  and the conservation equation (\ref{stripcons}), we obtain the on-shell area as:
\ba
{\cal A}(t_0, \ell, R)  = \frac{R}{\pi} \int^{\ell/2}_{0} dx \frac{z_*^2}{z^4}.
\ea
We regulate the area by subtracting the cut-off dependent part
\ba
\delta {\cal A} (t_0, \ell, R) = {\cal A}(t_0, \ell, R)  - \frac{1}{z_0} \frac{R}{\pi}.
\ea
We then compare the area in the shell background with the area in a thermalized background at different times for fixed scales in Fig.~\ref{fig:deltaAstrip}, where we defined $\delta{\cal \tilde A} \equiv \delta{\cal A} / (R \ell/ \pi )$. In four dimensions ($d=3$) it turns out that there is a range of times $t_0$ and Wilson strip widths $\ell$ for which there are three different minimal surfaces in AdS$_4$.   Thus, in this case, we observe a ``swallow tail'' in the Wilson loop thermalization curve for large enough boundary separations, as was also noted in \cite{Albash:2010mv}. However, in five dimensions ($d=4$) the thermalization curves do not exhibit the swallow tail.

\begin{figure}[h]
\centering
\centering
\includegraphics[width=0.25\linewidth]{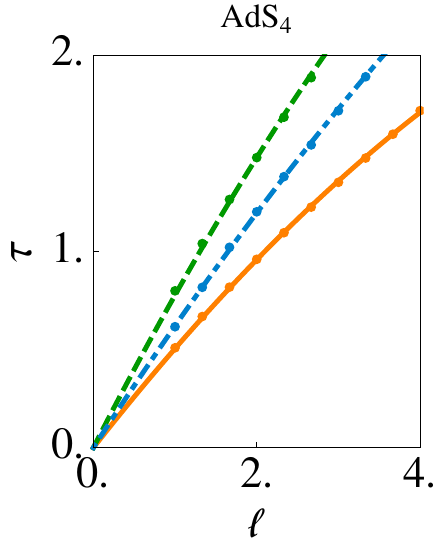} \hfil
\includegraphics[width=0.25\linewidth]{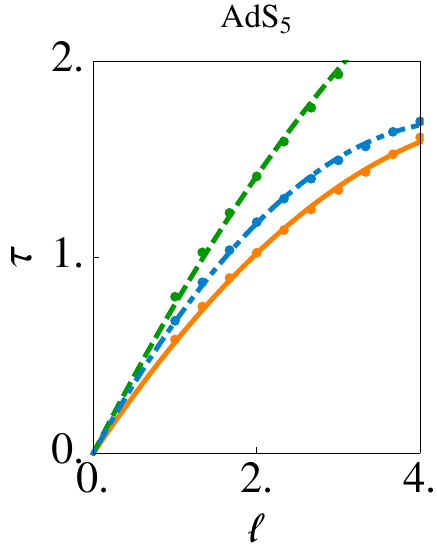}
\caption{Wilson strip thermalization times ($\tau_{{\rm crit}}$,  top line; $\tau_{{\rm max}}$, middle line; $\tau_{1/2}$, bottom line) as a function of spatial scale for $d=3$ (left) and $d=4$ (right) for a thin shell ($v_0 = 0.01$).   All thermalization times deviate from linearity in the spatial scale.
}
\label{fig:timesstrip}
\end{figure}

In Fig.~\ref{fig:timesstrip} we plot the three different thermalization times $\tau_{\rm crit}$, $\tau_{1/2}$ and $\tau_{\rm max}$ for $d = 3$ and $d = 4$ dimensional field theories. In three and four dimensions, the time for complete thermalization, $\tau_{\rm crit}$ (dashed green curve), exceeds the linear relation $\ell/2$.  The half-time $\tau_{1/2}$ (orange curve) is sub-linear. The time $\tau_{\rm max}$ (dot-dashed blue curve) exhibits a different behavior in three and four dimensions.   The time for complete thermalization for a strip of width $\ell$ is slower than the thermalization time of a circular loop of  radius $R$.  This may be because the long direction of the strip provides a second, larger scale.

%%%%%%%%%%%%%%%%%%%%%%%%%%%%%%%%

\subsection{Equilibration of the entanglement entropy}
\label{s:dynahyper}

Finally we consider how the entanglement entropy equilibrates following a quench.  Because it measures all contributions to the information loss caused by the restriction of the field theory to a finite volume, this quantity provides for a more comprehensive measure of equilibration than either the two-point function or the Wilson loop expectation value. The entanglement entropy can be considered as a special case of the standard coarse grained entropy of equilibrating quantum systems \cite{Takayanagi:2010wp}.   As discussed in Sec.~\ref{s:3surf}, the entanglement entropy in a two-dimensional field theory is related to geodesic lengths in AdS$_3$, while in three-dimensional field theory entanglement is related to minimal surface areas in AdS$_4$.  Both of these quantities have been computed above in the infalling shell background in the process of studying two-point functions and Wilson loops.  It is interesting that the Wightman function in 2 dimensions and the Wilson loop in 3 dimensions are so closely related to entanglement.

\begin{figure}[h]
\begin{center}
\includegraphics[width=0.28 \textwidth]{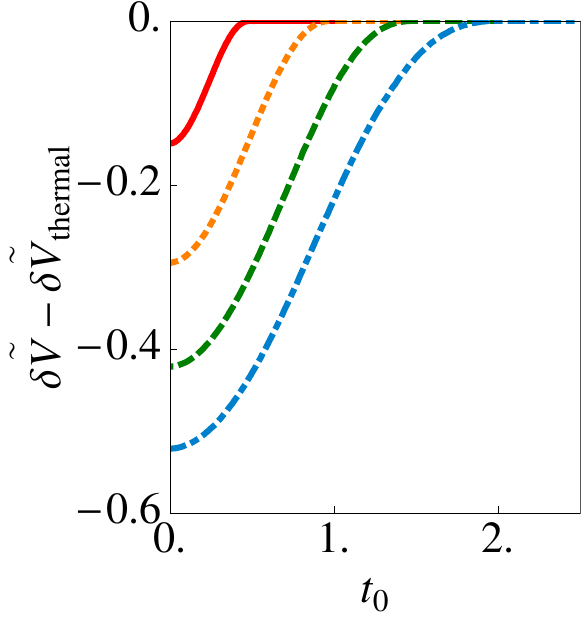}\hfil
\includegraphics[width=0.25 \textwidth]{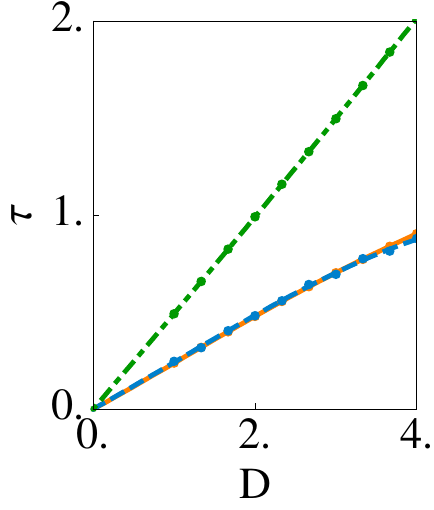} \\
\caption{The left panel shows $\delta\tilde{{\cal V}} - \delta\tilde{{\cal V}}_{\rm thermal}$ ($\tilde {\cal V} \equiv {\cal V}/(4\pi R^3/3)$) as a function of boundary time $t_0$ for a 3-surface in $d=4$ for a thin shell ($v_0=0.01$) and 4 equally spaced boundary radii $R=0.5,1,1.5,2$.  The right figure shows the thermalization times ($\tau_{crit}$, top line; $\tau_{max}$, middle line; $\tau_{1/2}$, bottom line) as a function of spatial scale for the same 3-surface. \label{fig:Sph}
}
\end{center}
\end{figure}

For 4-dimensional theories, to study the entanglement entropy in spherical regions, we need to compute the volume of minimal three-surfaces  with $S^2$ cross-section in the asymptotically AdS$_5$ infalling shell geometry.   As with the pure AdS and black brane cases we discussed in Section \ref{s:3surf}, the method is a straightforward generalization of the circular Wilson loop case, so we do not repeat the details of Section \ref{s:dynoWL}. We replace the black brane tension $M$ in equation (\ref{eq:SphAct}) with the dynamical mass function (Eq.~\eqref{eq:mv}). We can regulate the resulting volumes by subtracting the divergent part of the 3-volume of the 3-surface of the same boundary radius in pure AdS, which is given analytically in (\ref{eq:Vsurf}). We call this regulated volume $\delta {\cal V}$.  However, as in previous cases, to illustrate the thermalization process we find it more instructive to subtract the renormalized 3-volume of the solution of the same radius in the black brane background, $\delta {\cal V}_{thermal}$. We ultimately plot $\delta\tilde{{\cal V}}-\delta \tilde{{\cal V}}_{thermal}$ against the boundary time in the left panel of Fig.~\ref{fig:Sph}. $\tilde{{\cal V}}$ is defined as ${\cal V}$ divided by the volume of the region that it bounds on the boundary.
As in previous cases we can also calculate the three thermalization times we defined earlier for the 3-volume, at different values of the diameter $D=2R$. These are shown in the right panel of Fig.~(\ref{fig:Sph}).

Our thermalization times for Wilson loop averages and entanglement entropy seem remarkably similar to those for two-point correlators.  Slightly ``faster-than-causal'' thermalization, possibly due to the homogeneity of the initial configuration, seems to occur for the probes that do not correspond to entanglement entropy in each dimension.  For the latter, the thermalization time is linear in the spatial scale and saturates the causality bound. As the actual thermalization rate of a system is set by the slowest observable, our results suggest that in strongly coupled theories with a gravity dual, thermalization occurs ``as fast as possible'' at each scale, subject to the constraint of causality.

\begin{figure}[t]
\centering
\includegraphics[width=2in,height=2in]{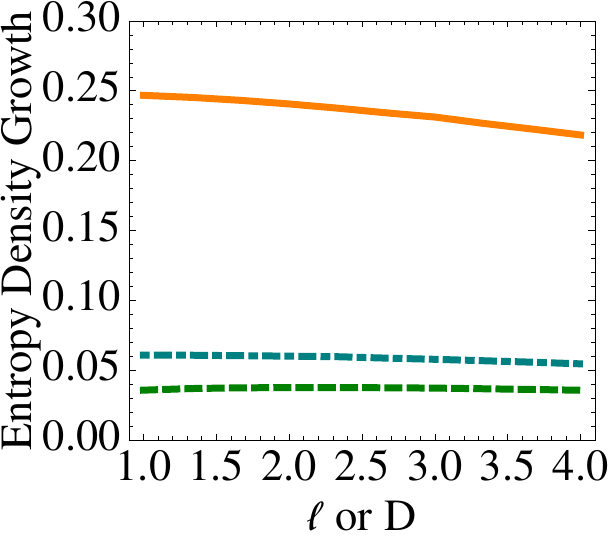}
\includegraphics[width=2in,height=2in]{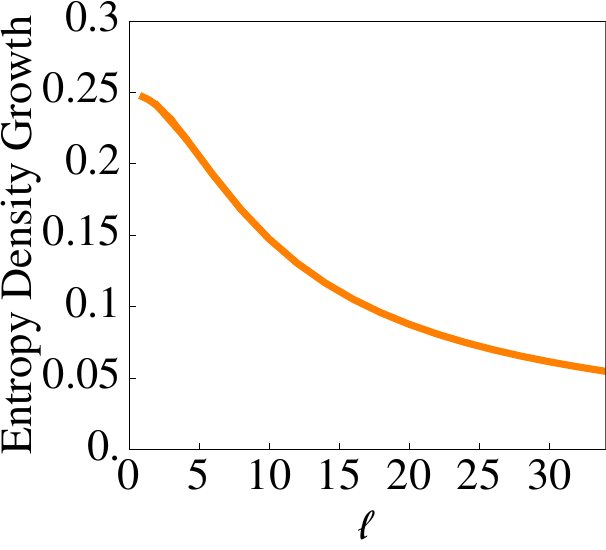}
\includegraphics[width=2in,height=2in]{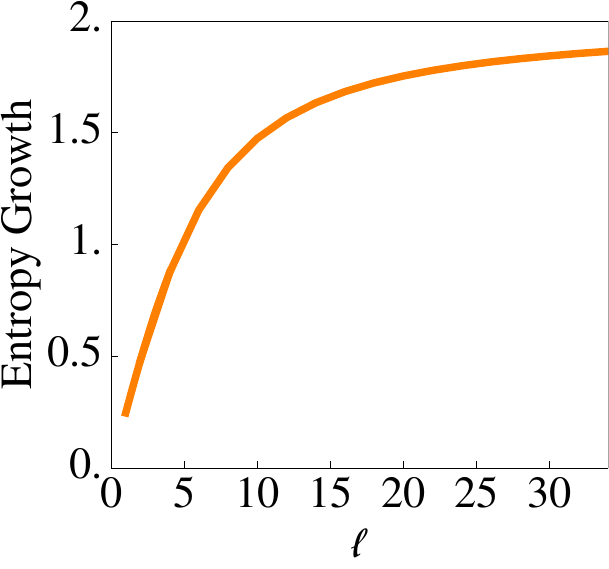}
\caption{(Left) Maximal growth rate of entanglement entropy {\it density} vs. diameter of entangled region for $d=2,3,4$ (top to bottom).  (Middle) Same plot for $d=2$, larger range of $\ell$.  (Right) Maximal entropy growth rate for $d=2$.}
\label{fig:KSAdS3}
\vspace{-0.1in}
\end{figure}

The average growth rate of the {\it coarse grained} entropy in nonlinear dynamical systems is measured by the Kolmogorov-Sina\"i (KS) entropy rate $h_{\rm KS}$ \cite{Sinai}, which is given by the sum of all positive Lyapunov exponents.  For a classical SU(2) lattice gauge theory in 4 dimensions, $h_{\rm KS}$  has been shown to be proportional to the volume \cite{Bolte:1999th}. For a system starting far from equilibrium, the KS entropy rate generally describes the rate of growth of the coarse grained entropy during a period of linear growth after an initial dephasing period and before the close approach to equilibrium \cite{Latora:1999xx}.   Here we observe similar linear growth of {\it entanglement} entropy density in $d=2,3,4$ (leftmost panels Figs.~\ref{fig:deltaL}, \ref{DL-Avt-2}, \ref{fig:Sph}).  For small boundary volumes, the growth rate of entropy {\it density} is nearly independent of the boundary volume (almost parallel slopes in the leftmost panels of Figs.~\ref{fig:deltaL}, \ref{DL-Avt-2}, \ref{fig:Sph} and nearly  constant maximal growth rate in Fig.~\ref{fig:KSAdS3} left). Equivalently, the growth rate of the entropy is proportional to the volume -- suggesting that entropy growth is a local phenomenon.    However, in $d=2$ where our analytic results enable study of large boundary volumes $\ell$, we find that the growth rate of the entanglement entropy density changes for large $\ell$, falling asymptotically as $1/\ell$ (Fig.~\ref{fig:KSAdS3} middle panel).  Equivalently, the entropy has a growth rate that approaches a constant limiting value for large $\ell$ (Fig.~\ref{fig:KSAdS3} right panel), and thus cannot arise from a local phenomenon.  This behavior suggests that entanglement entropy and coarse grained entropy have different dynamical properties.

In summary, we have investigated the scale dependence of thermalization following a quench in 2-, 3-, and 4-dimensional strongly coupled field theories with gravity duals. We found that the entanglement entropy sets a time scale for equilibration that saturates a causality bound.  Our results raise interesting questions about the relationship between the entanglement entropy growth rate and the KS entropy growth rate defined by coarse graining of the phase space distribution.

%%%%%%%%%%%%%%%%%%%%%%%%%%%%%%%%%%%%%%%%%%%%%%%%%%%%%%%%%%%%%%%%%%%%%%%

\section{Discussion}\label{Discussion}

In this work, we explored the approach to thermal equilibrium of simple nonlocal observables in strongly coupled conformal field theories with a gravity dual. In particular, we investigated the behavior during thermalization of equal-time two-point functions, Wilson loops and entanglement entropy in boundary field theories holographically represented by asymptotically AdS geometries. The thermalization process was modeled in the bulk by the collapse of a spatially homogeneous thin shell of null dust that eventually forms a black hole. The thickness $v_0$ of the shell is related to the duration of the process of energy deposition in the boundary field theory. In the limit $v_0 \ll z_0$, where $z_0$ parametrizes the effective UV cut-off of the field theory, the energy deposition can be considered as instantaneous.

Our observables were calculated in a semi-classical approximation appropriate for high-dimension operators, in which they correspond to geodesics or minimal surfaces in the bulk. For the AdS$_3$ case, where the thermal limit is represented by the planar BTZ black hole, we were able to solve the problem analytically. Not only did this provide a check on our numerical approach, which is required for higher dimensional cases, it also allowed us to obtain results in regimes beyond reach of our numerical analysis.

In all cases we found that the thermal limit is reached after a finite time $\tau_{{\rm crit}}$, which is a function of the geometric size of the probe in the boundary field theory, {\em e.g.}\ the separation $\ell$ of the two points of the equal-time Wightman function or the radius $R$ of the circular Wilson loop. For those cases where the logarithm of the correlation function is proportional to the entanglement entropy of the enclosed area on the boundary, {\em i.e.}\ for geodesics in AdS$_3$, minimal surfaces bounded by a circle in AdS$_4$, and minimal co-dimension two surfaces bounded by a spherical shell in AdS$_5$, we found that $\tau_{{\rm crit}}=\ell/2$ (AdS$_3$) and $\tau_{{\rm crit}}=D/2$ (AdS$_{4,5}$), in the regimes we were able to study. Our result confirms a general rule for two-dimensional conformal field theories implied by causality (as also observed in \cite{AbajoArrastia:2010yt}) and generalizes it to higher dimensions (as also done in \cite{Albash:2010mv} for AdS$_4$).

The conclusion we can draw from these results is that the decoherence and equilibration of instantaneously deposited energy generally propagates at the speed of light in the conformal field theory. This makes sense, because the process can be qualitatively understood as a cascade of gauge boson splittings, which transfers energy from all momentum scales into the infrared until all energy is distributed thermally. The speed by which this process evolves is determined by the speed of propagation of the gauge quanta, {\em i.e.}\ the speed of light. Alternatively, if the initial energy deposition were spatially inhomogeneous, one would expect the dissipation of the local inhomogeneities in the initial energy density to be constrained by the speed of sound, which is equal to $c/\sqrt{3}$ in a conformal field theory. This argument invites the conjecture that local equilibration of energy, in the sense of hydrodynamics, is controlled by the speed of light in conformal field theories, whereas global equilibration requiring hydrodynamic transport of energy is governed by the speed of sound. On the other hand, we found that the na\"ive causality argument does not hold  for the two-point function in 3d and 4d field theories and for Wilson loops in 4d field theories, which do not have an interpretation as entanglement entropy. In this case we obtained $\tau_{{\rm crit}} < \ell/2$. This is not as surprising as it might sound: we already know of other observables that thermalize faster than the entanglement entropy, for instance one-point functions.

We also found that the transition to full thermal equilibrium is abrupt and non-analytic, as expected from the causality argument. One might have objected that the non-analyticity of the transition to
full thermal equilibrium is probably due to the geodesic approximation
and will be smoothed out once we consider the exact Wightman
function. Corrections to the geodesic approximation can be worked out by
considering the exact first-quantized path integral representation of
the propagator, whose saddle points are given by geodesics. Possible
corrections are then due to fluctuations around the saddle
point. However, if the geodesic saddle-point lies entirely within the
AdS Schwarzschild geometry and does not graze the shell, such higher
order corrections will not be sensitive to the presence of the shell,
since the shell is invisible in perturbation theory around the geodesic.
This remains valid when we couple the heavy bulk scalar field to other
bulk degrees of freedom.  At best, there could possibly be
non-perturbative corrections due to other complex saddle points which
are sensitive to the existence of the shell, but is unclear how to
detect those in time-dependent situations. Therefore, for sufficiently
small separations the two-point function appears to be exactly given by
the thermal answer, which will then abruptly change to a non-thermal
answer once the geodesic starts crossing the shell. As a result, we
expect the non-analyticity to be valid beyond the geodesic approximation.
The non-analyticity is probably a consequence of our treatment of the endpoints of the Wightman function, which are both located at the same bulk variable $z_0$. The origin of a similar problem has been studied in the theory of boundary critical phenomena, where one typically introduces an extrapolation length $\tau_0$ to tame the sharp UV cut-off of the conformal field theory \cite{Calabrese:2006xx,Calabrese:2007rg}. This corresponds to the introduction of a smooth UV cut-off of the form $\exp(-E\tau_0)$, where $E$ is the energy. The sharp cusp in the asymptotic result is then rounded over a region $|t - \ell/2| \sim \tau_0$. In our holographic setting, a similar approach would smear the two endpoints of the correlation function independently over some range of values $z_0$. We have not studied the consequences of such a prescription. 

One may wonder whether our results are in tension with those of \cite{Ebrahim:2010ra}, where instantaneous thermalization was found for two-point functions related to Brownian motion of a ``quark'' represented by a string stretching from the boundary to the horizon of an AdS$_3$-Vaidya spacetime. While the two-point functions studied in \cite{Ebrahim:2010ra} did not involve high-dimension operators, so that there is no reason to expect a geodesic approximation to be valid, one can get some intuition by computing geodesics in the induced geometry on the string worldsheet. We have done so, and find that geodesics connecting two points on the string outside the shell never cross the shell. So within a geodesic approximation, this would explain why the observable considered in \cite{Ebrahim:2010ra} thermalizes instantaneously, in contrast to the observables studied in the present paper.

The fact that correlation functions of a small geometric size thermalize earlier than those of a large size implies that high momentum modes in the boundary field theory approach thermal equilibrium faster than long wavelength modes. In other words, the boundary field theory thermalizes first in its ultraviolet domain and later in the infrared or, as one might say, thermalization proceeds from the top down. In part, this is a consequence of the fact that the specific energy injection mechanism considered here has support in the UV: the energy shell is injected into the geometry near the boundary and then falls into the deeper regions of the AdS space, which represent IR modes of the boundary field theory. While this appears natural when viewed from the gravity side, it represents a radical deviation of the thermalization behavior of the dual field theory from that known in weakly coupled non-Abelian gauge theories, where thermalization occurs from the bottom up, independent of how the energy is injected into the field theory. For the reader unfamiliar with heavy ion collision phenomenology, we point out that the scale at which energy is deposited in the collision -- the parton saturation scale $Q_s$ of the colliding nuclei -- is higher by an inverse power of the coupling constant $\alpha_s$ than the thermal scale $T$ after thermalization. Thus, while the energy is not injected in the {\em extreme} UV in a nuclear collision, it is certainly not injected in the infrared. We also note that the difference in thermalization behavior is closely related to the observation that highly localized excitations in the strongly coupled gauge theory do not evolve into jets as is commonly found in weakly coupled, asymptotically free gauge theories that permit a perturbative treatment.

It is tempting to speculate to which extent our results may apply to the thermalization of QCD matter that is produced in relativistic heavy ion collisions. Phenomenologically, it is known that the thermalization has to occur fast, allowing the matter to expand according to the laws of nearly ideal hydrodynamics. If the relevant length scale for thermalization is given by the thermal scale $\ell \sim \hbar/T$, our results suggest that $\tau_{{\rm crit}} \sim 0.5\hbar/T$ if the matter is strongly coupled. For initial temperature value $T\approx 300-400$ MeV at heavy ion collider energies, we obtain the estimate $\tau_{{\rm crit}} \sim 0.3$ fm/$c$, comfortably short enough to account for the experimental observations. We know that such a short thermalization time applies to all momentum scales, because jets are created in the nuclear collisions, albeit at suppressed levels. However, as already mentioned, the relevant momentum scale for energy deposition is the parton saturation scale $Q_s$ of the colliding nuclei, {\em i.e.}\ the transverse momentum below which most partons in the colliding nuclei are found. For achievable energies and large nuclei, $Q_s \leq 3$ GeV; in this range of momenta QCD may well be considered as a strongly coupled gauge theory.

Though this was not the emphasis in this paper, the rapid thermalization studied in this paper may also be of relevance for black hole physics. It has been suggested that black hole creation is the fastest possible form of thermalization that exists in nature \cite{Sekino:2008he} and it would therefore be of particular interest to more directly link the causality bound to this conjecture.

We also encountered phenomena that do not appear to have a simple explanation. For example, for long rectangular Wilson loops in the boundary gauge theory of AdS$_4$, the area of the minimal surface exhibits a ``swallow tail'' behavior: the transition to the thermal limit proceeds via a discontinuous succession of shapes of the minimal surface. It is unclear why this phenomenon occurs for the rectangular Wilson loop but not for the two-point function, as the differential equation for the stationary surface is quite similar to that for the geodesic curve.

Finally, we briefly comment on the difference between dynamically collapsing shells and quasistatic shells that adiabatically approach the event horizon. As we show in the Appendix, the swallow tail behavior is ubiquitous for quasistatic shells close to the horizon, leading to significant differences between dynamical and quasistatic shells for large boundary separations.   

%%%%%%%%%%%%%%%%%%%%%%%%%%%%%%%%%%%%%%%%%%%%%%%%%%%%%%%%%%%%%%%%%%%%%%%

\section*{Acknowledgments}

We thank V.~Hubeny, M.~Rangamani and S.~Ross for helpful discussions. This research is supported  by the Belgian Federal Science Policy Office through the Interuniversity Attraction Pole IAP VI/11, by FWO-Vlaanderen through project G011410N, by the Foundation of Fundamental Research on Matter (FOM), by DOE grant DE-FG02-05ER-41367, by the BMBF, and by Academy of Finland grant 1127482. AB and WS are Aspirant FWO.

%%%%%%%%%%%%%%%%%%%%%%%%%%%%%%%%%%%%%%%%%%%%%%%%%%%%%%%%%%%%%%%%%%%%%%%

\appendix

%%%%%%%%%%%%%%%%%%%%%%%%%%%%%%%%%%%%%%%%%%%%%%%%%%%%%%%%%%%%%%%%%%%%%%%

\section{The quasistatic approximation}

In the main text we considered a dynamical shell of matter falling into
AdS space as a model of thermalization.  An adiabatic approximation to
the process of thermalization can be constructed by treating the
computation of correlation functions and Wilson loops in a quasistatic
manner.  To do this we observe that the dynamical shell is falling into
AdS space on a null trajectory.  Then, at any given time, we imagine a
shell located statically at the corresponding location on the null
trajectory and compute Wightman functions, Wilson loops and entanglement entropy via the
geodesic and minimal surface methods described in the previous section.
In this quasistatic approximation the geodesics and minimal surfaces
with endpoints at a given boundary time remain localized on the equal
time surface in the bulk of AdS space (Fig.~\ref{FigCrossingTheShell}).

As we will see, in the quasistatic approximation, there are generically
multiple geodesics connecting a pair of endpoints on the AdS boundary.
We demonstrate this explicitly for AdS$_3$ and have found similar
results for AdS$_4$ (not shown).  There can also be multiple minimal
surfaces that trace out a given loop on the AdS$_4$ boundary.  This
``swallow tail phenomenon'' was also described in the full dynamical
setting for the strip Wilson loop (see \cite{Albash:2010mv} and above).
When multiple geodesics or minimal surfaces are present in a dynamical
setting, it will generally be necessary to analyze which of these
saddlepoints lie within steepest descent integration contours in the
path integral.  Here, for completeness, we document the situations we
have found where multiple saddlepoints exist (see below and
Table~\ref{Overview}).  Note that in a quasistatic setting the saddlepoint
of minimal action will dominate.

\begin{table}[h]
\begin{center}
\begin{tabular}{|c|c|c|}
\hline
\multirow{2}{*}{holographic probe} &\multirow{2}{*}{quasistatic} &\multirow{2}{*}{ dynamical}\\
&&\\
 \hline \multirow{2}{*}{ geodesics in AdS$_3$} &\multirow{2}{*}{$\surd$} &\multirow{2}{*}{$\times$}\\
 &&\\
\multirow{2}{*}{ geodesics in AdS$_4$}&\multirow{2}{*}{ $\surd$} & \multirow{2}{*}{ $\times$}\\
&&\\
\multirow{2}{*}{ strip in AdS$_4$} &\multirow{2}{*}{  $\surd$} &\multirow{2}{*}{  $\surd$}\\
&&\\
\multirow{2}{*}{ circular loop in  AdS$_4$}& \multirow{2}{*}{ $\surd$} &\multirow{2}{*}{ $\times$}\\
&&\\
 \hline
\end{tabular}
\end{center}
\caption{Overview: swallow tail ($\surd$);  no swallow tail  ($\times$)}
\label{Overview}
\end{table}%

%%%%%%%%%%%%%%%%%%%%%%%%%%%%%%%%%%%%%%%%%%%%%%%%%%%%%%%%%%%%%%%%%%%%%%%

\subsection{Geodesics in the quasistatic approximation}

We study the equal-time geodesic in the shell background in the
quasistatic approximation for $d=2$.  In the quasi-static approximation,
the AdS$_3$ geometry with a shell at a fixed radial coordinate $r=r_s$
can be written as
\begin{align}
 ds^2=-f(r)^2dt^2+{dr^2\over f(r)^2}+r^2dX^2
 \quad;\qquad
 f(r)^2=
\begin{cases}
 f_{\text{in}}(r)^2 \,\, =r^2        & r<r_s\\
 f_{\text{out}}(r)^2 =r^2-r_H^2  & r>r_s
\end{cases}
\label{eq:QSAdS3}
\end{align}
where $r_H$ ($<r_s$) is the position of the would-be horizon.  Outside
the shell, $r>r_s$, the metric \eqref{eq:QSAdS3} is the planar black brane
 metric that we studied in the text in \ref{BTZanalytic}, while
inside the shell, $r>r_s$, it is the empty AdS$_3$ we discussed
around \eqref{lambda(r)pureAdS}--\eqref{semicircle}.
Therefore, for studying the equal-time geodesic in the shell spacetime
\eqref{eq:QSAdS3}, we can use our previous results. Outside the shell,
the relevant expressions are found in \eqref{r_out(lambda)},
\eqref{r_out(lambda)2}, \eqref{Xoutl}, and \eqref{X(r)pm}.  The outside
geodesics are parametrized by two parameters $E$ and $J$ which
correspond to the energy and angular momentum, respectively. We take
$E=0$ so that they describe equal-time geodesics (see below
\eqref{eq:DXBTZ}). Also, we can take $J\ge 0$ without loss of
generality, because flipping the sign of $J$ just corresponds to the
reverse parametrization of the same geodesic.  Inside the shell, the
expressions for geodesics are found in \eqref{lambda(r)pureAdS},
\eqref{X(r)pureAdS}, and \eqref{X(r)pureAdS2}.  The inside geodesics are
parametrized by $r_*\equiv 2/\ell$. As is clear from
\eqref{X(r)pureAdS2}, $r_*$ is the minimum value of $r$ on the geodesic.

The geodesic in the entire shell background spacetime \eqref{eq:QSAdS3} is
obtained by gluing outside and inside geodesics across the shell using a
refraction law so that the geodesic length is extremized.  
In the current situation the refraction law can be derived in very much the same way as it was in the shockwave background around \eqref{RC}, and the result is
\begin{align}
 \left. f_{\text{in}}{dX\over dr}\right|_{r=r_s}
 =
 \left. f_{\text{out}}{dX\over dr}\right|_{r=r_s}.\label{RCshell}
\end{align}
By substituting the expression for the outside geodesic \eqref{X(r)pm}
and the inside one \eqref{X(r)pureAdS2} into \eqref{RCshell}, we see
that the parameters in the inner and outer regions are related by
\begin{align}
 r_*=r_HJ.\label{r*andJ}
\end{align}

Just as we did in the main text for the shockwave geometry, we can compute
the renormalized geodesic length $\delta\cal L$ and spatial boundary separation
$\ell$ in terms of the parameters $J,r_*$ of the geodesic.  The geodesic
length is the sum of the geodesic length for the inside part, which can
be computed from \eqref{lambda(r)pureAdS}, and the length for the outside
part, which can be computed from \eqref{lambdapm} with $E=0$.  The
result is
\begin{align}
 \label{ell,L_QAdS3}
\ell&
 =
{2\sqrt{r_s^2-r_H^2J^2}\over r_Hr_sJ }
 +
 {2\over r_H}\ln\left[{r_s(J+1)\over J\sqrt{r_s^2-r_H^2}+\sqrt{r_s^2-r_H^2J^2}}\right],
\quad
 \delta{\cal L}
 =
2\ln\left[ {r_s+\sqrt{r_s^2-r_H^2J^2}\over
 r_HJ\bigl({\sqrt{r_s^2-r_H^2}+\sqrt{r_s^2-r_H^2J^2}}\,\bigr)} \right] .
\end{align}
Here, we regularized ${\cal L}$ by subtracting the divergent
quantity $2\ln(2r_0)$ just as we did in the main text.  Note that
\eqref{ell,L_QAdS3} is valid if the geodesic is partly inside the shell,
which is the case if $J\le {r_s/r_H}$. In this case, \eqref{ell,L_QAdS3}
gives implicitly the relation between $\ell$ and $\delta{\cal L}$
through the parameter $J$ (note that $J$ and $r_*$ are related by
\eqref{r*andJ}).

If $J\ge{r_s/r_H}$, instead, the entire geodesic is outside the shell
and the relation between $\ell$ and $\delta{\cal L}$ is given by
\eqref{LvsellBTZren}.

\begin{figure}[htb]
\begin{center}
\begin{tabular}{c@{~~~~~~}c}
\includegraphics[width=0.45 \textwidth]{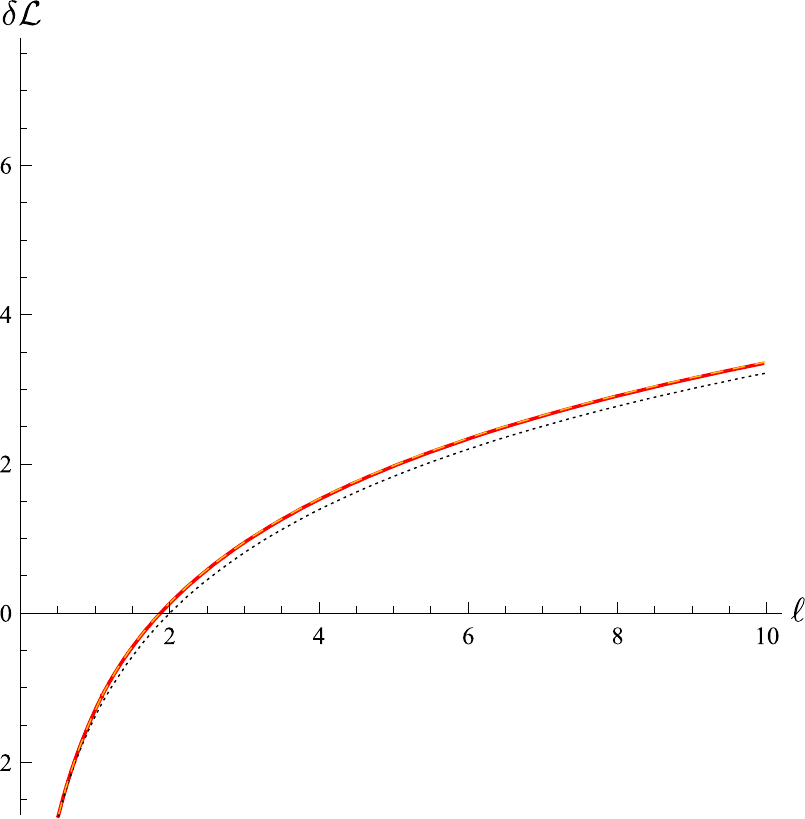} &
\includegraphics[width=0.45 \textwidth]{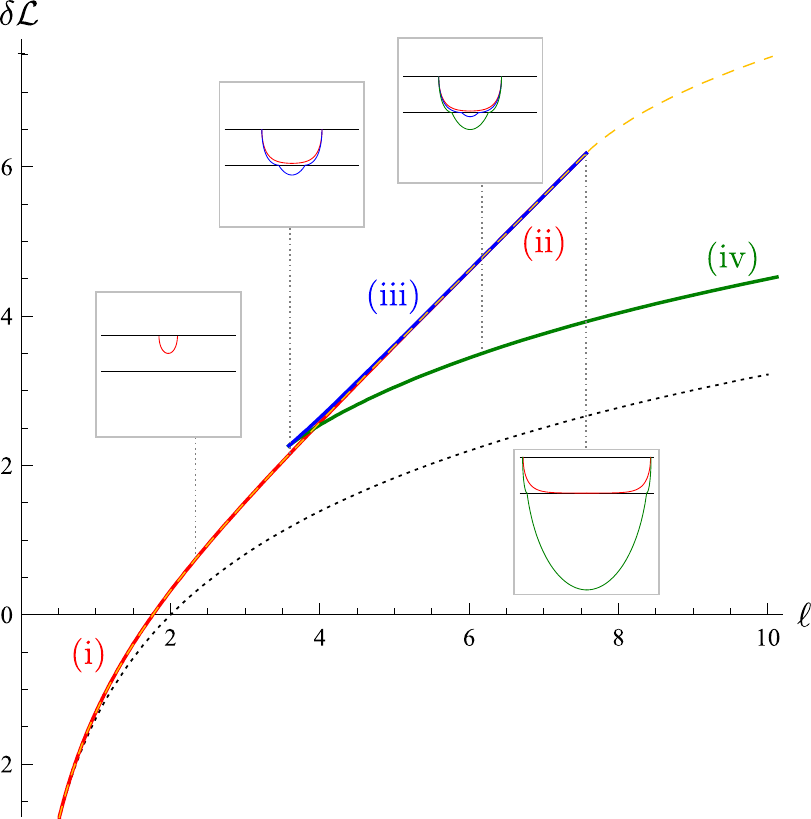}
\\
 {\bf (A)} ${r_s/ r_H}=2$ & {\bf (B)} ${r_s/ r_H}=1.001$
\end{tabular}
\caption{ The $\delta {\cal L}$ versus $\ell$ curves (solid lines) for
the values of $r_s/ r_H$ (A) sufficiently larger than $1$ and (B)
sufficiently close to $1$.  We set $r_H=1$. For comparison, we plotted
$\delta{\cal L}(\ell)$ for pure AdS$_3$ in dotted black lines and for
AdS$_3$ Vaidya in dashed yellow lines. Note that most of the Vaidya
curves (dashed yellow) are overlapping with quasi-static curves (solid,
in various colors).  For the values of $t_0$ used for the Vaidya curves,
see text (Eq.\ \eqref{t0andrs}).  In (B), we used (partly) different
colors for the different regimes (i)--(iv) explained in the text in
Eq.\ \eqref{eq:dLellregimes}. In the insets are the actual shapes of the
geodesics for some selected values of $\ell$, in different colors
corresponding to different branches, in the $(X,z=1/r)$ coordinates
(actual values of the coordinates not shown).  The two horizontal lines
in each inset represent the boundary ($z=0$) and the shell
($z_s=1/r_s$). } \label{fig:shellgeodesic}
\end{center}
\end{figure}

In Fig.~\ref{fig:shellgeodesic}, we plotted $\delta{\cal L}$ versus
$\ell$ for different values of $r_s/r_H$. We see that the curve has more
nontrivial structure for ${r_s/ r_H}\approx 1$.  So, let us focus on
this case.  If $r_s/r_H\approx 1$, then from \eqref{ell,L_QAdS3} one can
derive that, depending on the value of $J$, the relation between $\ell$
and $\delta{\cal L}$ has the following regimes:
\begin{align}
 \label{eq:dLellregimes}
 \begin{array}{rlllll}
 \text{(i)}
  & J\gg 1
  &:&\ell\ll {1\over r_H}
  &,& \delta{\cal L}\approx  2\ln{\ell\over 2}
  \\[1ex]
  \text{(ii)}
  & J\ge {r_s\over r_H}~\text{and}~ J\sim 1
  &:&\frac{1}{r_H}\ll \ell \le \frac{1}{r_H}\ln\frac{r_s+r_H}{r_s-r_H}
  &,& \delta{\cal L} \approx r_H \ell -2\ln(2r_H)
  \\[1ex]
  \text{(iii)}
  & J\le {r_s\over r_H}~\text{and}~ J\sim 1
  &:& {1\over r_H}\ll \ell \le {1\over r_H}\ln{r_s+r_H\over r_s-r_H}
  &,& \delta{\cal L} \approx r_H \ell -2\ln(2r_H)
  \\[1ex]
  \text{(iv)} &
  {1\over r_H}\gg J >0
  &:& \ell \gg {\sqrt{r_s^2-r_H^2}\over r_H^2}
  &,& \delta{\cal L} \approx
  2\ln {\ell\over 2}
  +2\ln {2r_s \over {r_s+\sqrt{r_s^2-r_H^2}}} .
\end{array}
\end{align}

The geodesic does not cross the shell in regimes (i) and (ii) while it
crosses the shell in regimes (iii) and (iv).
In regime (i), the $\ell$-$\delta{\cal L}$ relation is the same as the
one for empty AdS$_3$, because the geodesic is near the boundary of the
AdS space where the background metric is same as the empty AdS metric.
In regime (ii) the geodesic length $\delta{\cal L}$ grows linearly with
the boundary separation $\ell$, because the geodesic extends almost
parallel to the shell just above it.  This happens because the shell at
$r=r_s$ is very close to the would-be horizon at $r=r_H$.  Regime (iii)
is similar to regime (ii) except that a small middle part of the
geodesic dips into the inside geometry. In regime (iv), the
$\ell$-$\delta{\cal L}$ relation is the same as the one for empty AdS,
up to a constant shift. This is because most of the geodesic is inside
the shell where the metric is equal to the empty AdS metric. However,
the part of geodesic outside the shell still sees the BTZ metric and
leads to the constant shift that survives even for $\ell\to \infty$.

So, if $r_s/r_H\approx 1$, within a certain range of the spatial
separation $\ell$, there are three possible geodesics and three
corresponding values of $\delta {\cal L} (\ell)$; see
Fig.~\ref{fig:shellgeodesic}.  The upper bound of the range is
$\ell_{max}=(1/r_H)\ln[(r_s+r_H)/(r_s-r_H)]$ while the lower bound can
be computed from the condition $\partial\ell/\partial J=0$ and is given
in the $r_s\to r_H$ limit by $\ell_{min}\approx 3.7/r_H$.  As $r_s\to
r_H$, we can make the upper bound $\ell_{max}$ arbitrarily large and the
``spike'' in Fig.~\ref{fig:shellgeodesic} arbitrarily long.

In general, in the presence of multiple geodesics in Lorentzian
signature, which geodesic makes the dominant contribution to the
correlation function is a subtle issue \cite{Festuccia:2005pi,
Fidkowski:2003nf, Louko:2000tp}.  However, in the present approximation
where the spacetime is static, it is simply the shortest geodesic that
contributes most to the correlation function.  Therefore, in spite of
the existence of the peculiar ``linear'' regimes (ii) and (iii), it is
(i) and (iv) that determine the correlation function.

For comparison, in Fig.~\ref{fig:shellgeodesic}, we also plotted
$\delta{\cal L}(\ell)$ for the Vaidya geometry we studied in
\ref{BTZanalytic_Vaidya}.  The relevant expressions are found in
\eqref{eq:rhopm} and \eqref{Lren}. The Vaidya geodesics are parametrized
by $t_0$, the time at which the boundary operators are inserted.  In the
plot, we determined the value of $t_0$ in terms of $r_s$ so that the
shock wave in the Vadiya geometry is at $r=r_s$ at time $t_0$ just as in
the quasi-static metric \eqref{eq:QSAdS3}.  Namely, because the shock wave is at
$v=0$, by setting $v=0$ in \eqref{eq:BTZ}, we obtain the relation between $t_0$ and $r_s$:
\begin{align}
 t_0={1\over 2r_H}\ln{r_s+r_H\over r_s-r_H}.
\label{t0andrs}
\end{align}
Note that, in the large $\ell$ limit with $r_s$ fixed, the quasi-static computation
\eqref{ell,L_QAdS3} gives the asymptotic behavior:
\begin{align}
 \delta{\cal L} \approx
  2\ln {\ell\over 2}  +2\ln {2r_s \over {r_s+\sqrt{r_s^2-r_H^2}}}
  =2\ln {\ell\over 2}  +2\ln \left[{\cosh(r_Ht_0)\over \cosh^2\left({r_Ht_0\over 2}\right)} \right]
 \qquad \text{(quasi-static),}
\end{align}
where in the second expression we used
\eqref{t0andrs}.
In the same limit (which corresponds in \eqref{Lren} to $c \to 1$, $r_*/r_H \to 0$ with fixed $t_0$), the Vaidya
result  \eqref{Lren} gives the following asymptotic behavior:
\be
\delta {\cal L}
\approx 2 \ln \frac{\ell}{2} - \ln \frac{16 a}{(\sqrt a +1)^4}
=2 \ln \frac{\ell}{2} +4 \ln \left(\cosh{r_Ht_0\over 2}\right)
 \qquad \text{(Vaidya)}\,,
\ee
where we used $a=e^{2r_Ht_0}$.
These lead to the difference in the $\ell\to\infty$ behavior of the quasi-static and Vaidya curves
that we can observe in
Fig.~\ref{fig:shellgeodesic}.

%%%%%%%%%%%%%%%%%%%%%%%%%%%%%%%%%%%%%%%%%%%%%%%%%%%%%%%%%%%%%%%%%%%%%%%

\subsection{Wilson loops in the quasistatic approximation}

In four dimensions the entanglement entropy is related to the area of the space-like string surface whose base is a Wilson loop at the AdS boundary.  In \cite{Albash:2010mv} it was noticed that there can exist three different string surfaces for the same boundary time and the same boundary separation in a dynamical background with a thermal quench (see also \ref{rectstripDyn}). In this subsection we will investigate whether the quasi-static approximation for Wilson loops and Wilson strips exhibits multiple string surfaces for the same boundary.

\subsubsection{Circular loops}\label{quasiWilso}

Similar to the geodesic analysis, the four dimensional background can be divided into a black brane part (outside the shell) and a pure AdS part (inside the shell). However, we cannot solve the differential equations analytically and instead we rely on numerical methods. Outside the shell the solution satisfies the differential equation (\ref{Wilsonloopdeom}) in the black brane background (with $d=3$), while inside the shell we must find a minimal surface in pure AdS (i.e.\ we set the mass $M$ to zero in eq.~(\ref{Wilsonloopdeom}). The minimal surface in pure AdS consists of a hemisphere. In the thin-shell limit, at the position of the shell $z=z_s$, the minimal surface should satisfy the refraction condition
\be
\left.\frac{dz_{\text{out}}}{d\rho}\right|_{z_s} = \sqrt{1- M z_s^3}\left.\frac{dz_{\text{in}}}{d\rho}\right|_{z_s}\,.
\ee
This is derived analogously to the geodesic case (\ref{RCshell}) in which case it agrees with numerical results obtained by minimizing the path of the geodesic. Note that in the dynamical case we do not use such a condition and solve the equations of motion in the whole space with mass function (\ref{eq:mv}). However, here, taking advantage of this simple relation allows to use much less computationally intensive numerics in the limit of a thin shell (where the mass function is a step-function). We start by constructing a hemispherical string surface whose tip is given by $z(0)=z_* > z_s$ and $z'(0) = 0$, thus lying inside the shell. For this solution we can determine the intersection point $\rho_1$ with the shell and the gradient at that point analytically.
We can then use the refraction condition to generate the rest of the solution in the black brane background, using the boundary condition $z(\rho_1)=z_s$ and $z'(\rho_1) = z'_{\rm out}$. Where this part of the solution intersects the boundary (cut-off at $z_0$) gives us the boundary radius $R$ of the Wilson loop. We can also construct surfaces wholly outside the shell, which are just the thermal Wilson loops from section \ref{s:CWL}. Thus we get $R(z_*)$ for any shell position $z_s$.

\begin{figure}[h]
\begin{center}
\includegraphics[width=0.3 \textwidth]{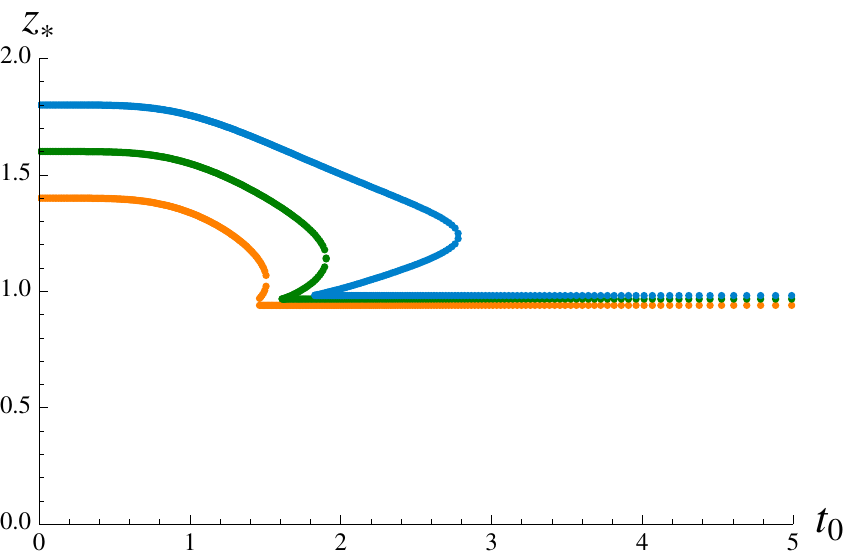}
\hfil
\includegraphics[width=0.3 \textwidth]{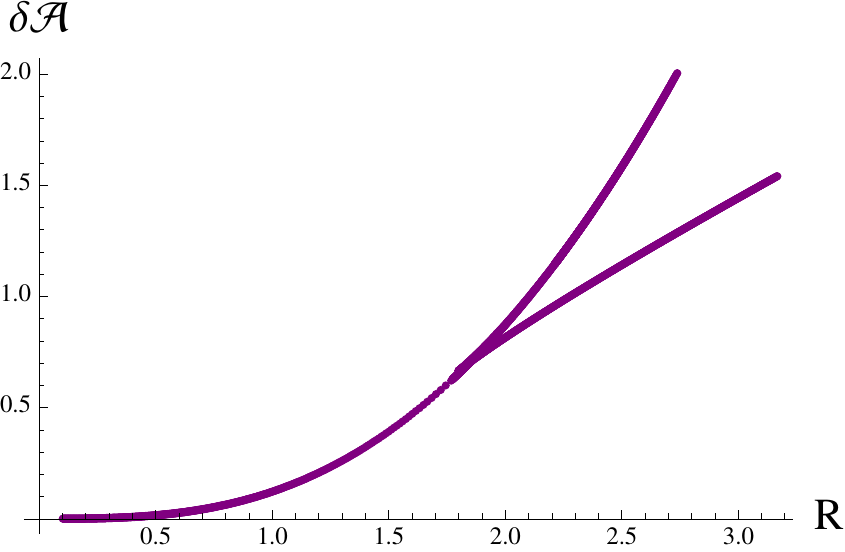}
\hfil
\includegraphics[width=0.3 \textwidth]{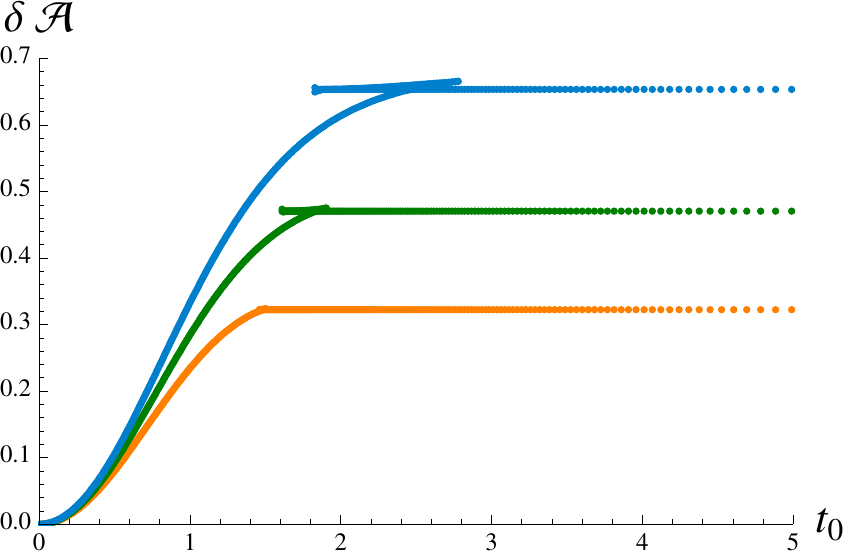}
\\
({\bf A})
\hfil ({\bf B}) \hfil ({\bf C})
 \\
\caption{({\bf A})$z_*$ against $t_0$ for  $R=1.4,1.6,1.8$. Three $z_*$-values for a certain range of $t_0$ makes it  clear we have multiple branches.
({\bf B}) $\delta {\cal A}$ against $R$ for  $M^{1/3} z_s=1/ 1.001$. We have multiple solutions for the same radius in the region with the spike.
({\bf C}) $\delta {\cal A}$ against $t_0$ for $R=1.4,1.6,1.8$, we have three branches and the behavior of the curve there is complicated.}
\label{QLZ}
\end{center}
\end{figure}

As with the geodesics case we can get multiple surfaces with the same boundary radius and shell position (such as in Fig.~\ref{fig:shellgeodesic}). An easy way to check whether these multiple solutions exist before calculating any areas (see also \cite{Albash:2010mv}) is to invert our numerical data appropriately so that we can plot it on a graph of  $z_*$ against boundary time $t_0$, the boundary time being obtained as a function of the shell position $z_s$:
\ba
t_0 = \int^{z_s}_{z_0}  \frac{d z}{ 1- M z^3}.
\ea
The multiple solutions will show up as multiple possible values of $z_*$ for the same $t_0$.  We see this in Fig.~\ref{QLZ}(A) where we have plotted for three different boundary radii.

The area of the minimal surface consists of three parts: we add the areas of the piece in empty AdS (inside the shell) and of the piece in the black brane background (outside the shell) and subtract the area of the solution in empty AdS with the same boundary radius,
\ba
\delta {\cal A}(R, t_0) = \int_{0}^{\rho_1(t_0)} d\rho  \frac{\rho}{z^2}\sqrt{1+z'^2} + \int_{\rho_1(t_0)}^{R} d\rho  \frac{\rho}{z^2}\sqrt{1+\frac{z'^2}{1-M z^3}} - \left(\frac{R}{z_0} -1\right).
\ea
In Fig.~\ref{QLZ}(B), we plot $\delta {\cal A}(R, t_0)$ as a function of $R$ for a fixed shell position. The three branches appear in a very similar fashion to those for geodesics in AdS$_3$ shown in Fig.~\ref{fig:shellgeodesic}. Fig.~\ref{QLZ}(C) then shows $\delta {\cal A}(R, t_0)$ as a function of the boundary time $t_0$ for three different boundary radii. The appearance of the three branches gives a `swallow tail' as found in \cite{Albash:2010mv} in the dynamical case for the infinite rectangular strip, and as we also find in Section \ref{rectstripDyn}. However, there are no such multiple solutions in the dynamical Wilson loop case as shown in \cite{Albash:2010mv} and Section \ref{s:dynoWL}. This can be taken as an indication of the limits of the quasistatic approximation in accurately capturing the behavior as the shell gets closer to the horizon (recall that to get the three branches in Figure \ref{QLZ}(B) we have to wait till the shell approaches the would be horizon, here $M^{1/3} z_s=1/ 1.001$). The appearance of three branches does not, however, signify that something is unphysical. As mentioned, they also appear in the dynamical case for the infinite strip. What it does mean is that since we are using a saddle point approximation to calculate field theory quantities and our space time is Lorentzian, when there are multiple geodesics we should carefully follow a steepest descent procedure.

The numerics are hard to pursue around the earliest time where the three branches occur in figure \ref{QLZ}(C) and we see some noise around one corner of the `swallow tail'. Here the tip of the sphere at $z_*$, which is our initial condition, is very close to the shell, and very small variations in $z_*$ give rise to large variations in $R$ and hence $\delta {\cal A}$.

%%%%%%%%%%%%%%%%%%%%%%%%%%%%%%%%%%%%%%%%%%%%%%%%%%%%%%%%%%%%%%%%%%%%%%%

\subsubsection{Strip}\label{quasistrip}

We can perform a similar analysis for the Wilson strip in a quasi-static background. The equation (\ref{stripddimeom}) is now the differential equation we solve numerically in the black brane background (with $d=3$). Considering a similar boundary setup for the strip as in section \ref{rectstripBB}, we follow a  similar procedure to construct solutions as for the quasi-static Wilson loops. Also the refraction condition through the shell (in the thin shell approximation) is the same as for circular Wilson loops. Hence, we will not work this case out explicitly will immediately give the results.  The string surface area consists of three different pieces: the part of the solution inside the shell (empty AdS) ${\cal A}_{AdS}$, the part of the solution outside the shell (black brane background) ${\cal A}_{BB}$, and a subtracting part corresponding to the full solution in empty AdS with the same boundary separation $\ell$. The full area in the shell background is given by
\ba
\delta {\cal A} (\ell, t_0)= {\cal A}_{AdS}(z_s(t_0),  \tilde x) + {\cal A}_{BB}( \tilde x, \ell) - {\cal A}_{AdS}(z_0, \ell),
\ea
where $\tilde x$ represents the space-like separation of the part inside the shell. 
The area formula in empty AdS is given by the relation
\be
{\cal A}_{AdS} (z, y)\equiv \frac{R}{\pi}\left(\frac{1}{z}+\frac{1}{2 y}\frac{\pi \Gamma(-1/4)\Gamma(3/4)}{\Gamma(1/4)^2}\right),
\ee
where $R$ corresponds to the height of the Wilson strip at the boundary (see \ref{rectstripBB}), 
and the black brane area is given by
\ba
{\cal A}_{BB}(y_1, y_2 ) \equiv \frac{R}{\pi} \int_{y_1}^{y_2} dx \frac{1}{z(x)^2} \sqrt{1 + \frac{z'(x)^2}{1-M z^3}}.
\ea

\begin{figure}[h]
\begin{center}
\includegraphics[width=0.3 \textwidth]{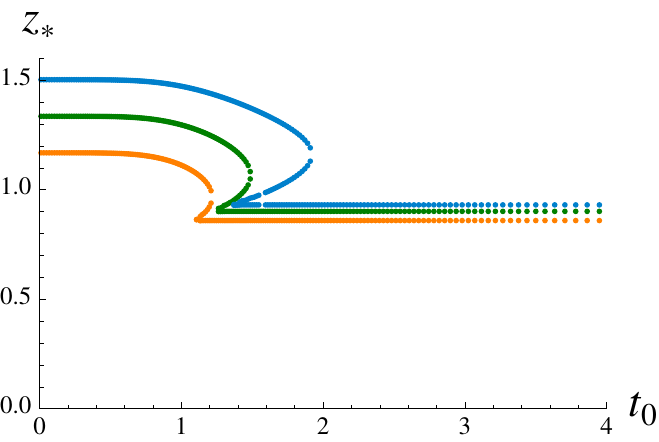}
\hfil
\includegraphics[width=0.3 \textwidth]{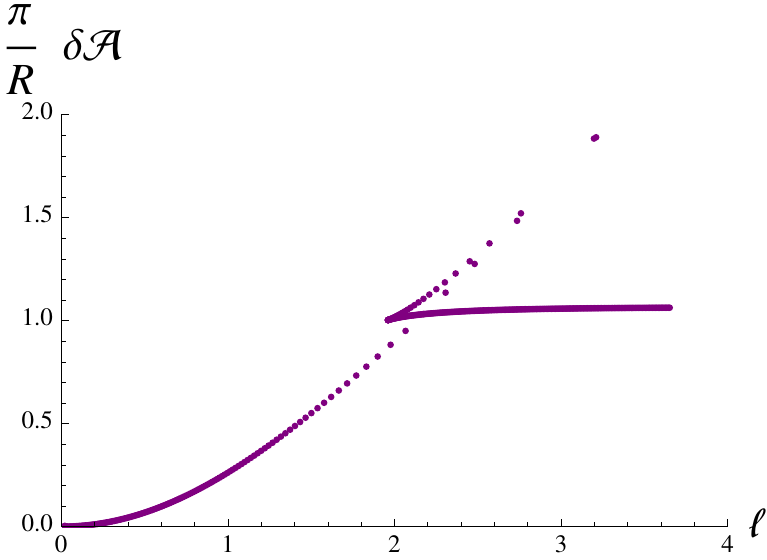}
\hfil
\includegraphics[width=0.3 \textwidth]{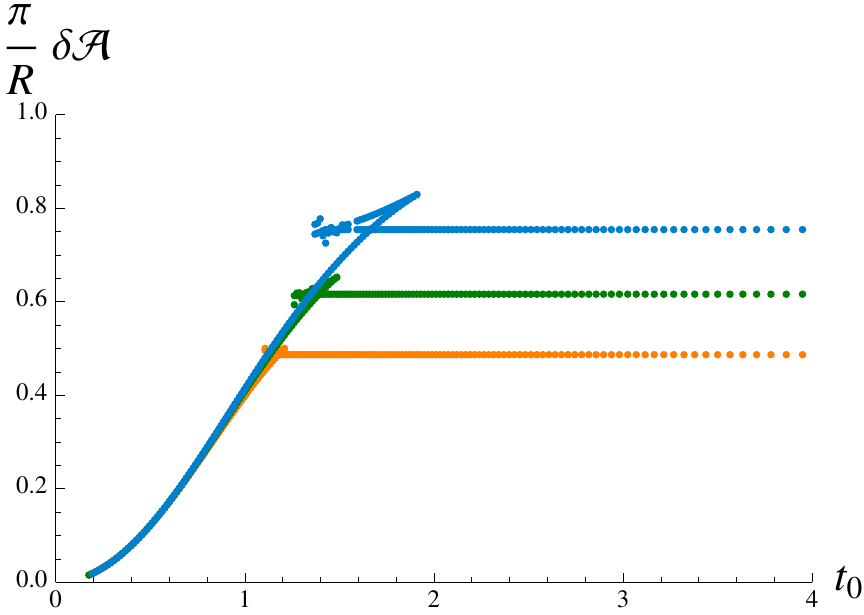}
\\
({\bf A})
\hfil ({\bf B})
\hfil ({\bf C})
 \\
\caption{({\bf A}) $z_*$ against $t_0$ for $\ell$ = 1.4, 1.6, 1.8. We see indeed the existence of three different solutions for the same boundary time $t_0$.
({\bf B}) $\delta{\cal A}/(R/\pi)$ against $\ell$ for a fixed position of the shell: $M^{1/3} z_s=1/ 1.001$.
({\bf C}) $\delta{\cal A}/(R/\pi)$ against $t_0$ for $\ell$ = 1.4, 1.6, 1.8.
}
\label{StripAvsx}
\end{center}
\end{figure}

In Fig.~\ref{StripAvsx}(A), we plot the midpoint of $z(x)$, namely $z_*$, as a function of the boundary time $t_0$. In the middle panel of Fig.~\ref{StripAvsx}, we plot the area $\delta {\cal A} (\ell, t_0)$ (with $R/ \pi$ divided out) as a function the boundary separation $\ell$ for a fixed shell position $z_s$ (thus for a fixed boundary time $t_0$). On the right panel of Fig.~\ref{StripAvsx} we plot the area $\delta {\cal A} (\ell, t_0)$ (with $R/\pi$ divided out) as a function of the boundary time $t_0$ for three fixed boundary separations.
As in the dynamical case, we have three branches of solutions appearing, but here they show up already for smaller values of the radius.

%%%%%%%%%%%%%%%%%%%%%%%%%%%%%%%%%%%%%%%%%%%%%%%%%%%%%%


\begin{thebibliography}{99}

\bibitem{Whitepapers:2005}
%\bibitem{Arsene:2004fa}
  I.~Arsene {\it et al.}  [BRAHMS Collaboration],
  %``Quark Gluon Plasma an Color Glass Condensate at RHIC? The perspective from
  %the BRAHMS experiment,''
  Nucl.\ Phys.\  A {\bf 757}, 1 (2005);
  %[arXiv:nucl-ex/0410020].
  %%CITATION = NUPHA,A757,1;%%
%\bibitem{Adcox:2004mh}
  K.~Adcox {\it et al.}  [PHENIX Collaboration],
  %``Formation of dense partonic matter in relativistic nucleus nucleus
  %collisions at RHIC: Experimental evaluation by the PHENIX  collaboration,''
  Nucl.\ Phys.\  A {\bf 757}, 184 (2005);
  %[arXiv:nucl-ex/0410003].
  %%CITATION = NUPHA,A757,184;%%
%\bibitem{Back:2004je}
  B.~B.~Back {\it et al.} [PHOBOS Collaboration],
  %``The PHOBOS perspective on discoveries at RHIC,''
  Nucl.\ Phys.\  A {\bf 757}, 28 (2005);
  %[arXiv:nucl-ex/0410022].
  %%CITATION = NUPHA,A757,28;%%
%\bibitem{Adams:2005dq}
  J.~Adams {\it et al.}  [STAR Collaboration],
  %``Experimental and theoretical challenges in the search for the quark  gluon
  %plasma: The STAR collaboration's critical assessment of the  evidence from
  %RHIC collisions,''
  Nucl.\ Phys.\  A {\bf 757}, 102 (2005).
  %[arXiv:nucl-ex/0501009].
  %%CITATION = NUPHA,A757,102;%%

\bibitem{Gyulassy:2004zy}
  M.~Gyulassy and L.~McLerran,
  %``New forms of QCD matter discovered at RHIC,''
  Nucl.\ Phys.\  A {\bf 750}, 30 (2005).
  %[arXiv:nucl-th/0405013].
  %%CITATION = NUPHA,A750,30;%%

\bibitem{Harris:1996zx}
  J.~W.~Harris and B.~M\"uller,
  %``The search for the quark-gluon plasma,''
  Ann.\ Rev.\ Nucl.\ Part.\ Sci.\  {\bf 46}, 71 (1996)
  %[arXiv:hep-ph/9602235].
  %%CITATION = ARNUA,46,71;%%

%\cite{Maldacena:1997re}
\bibitem{Maldacena:1997re}
  J.~M.~Maldacena,
  %``The large N limit of superconformal field theories and supergravity,''
  Adv.\ Theor.\ Math.\ Phys.\  {\bf 2}, 231 (1998)
  [Int.\ J.\ Theor.\ Phys.\  {\bf 38}, 1113 (1999)]
  [arXiv:hep-th/9711200];
  %%CITATION = IJTPB,38,1113;%%
%\cite{Gubser:1998bc}
%\bibitem{Gubser:1998bc}
  S.~S.~Gubser, I.~R.~Klebanov, A.~M.~Polyakov,
  %``Gauge theory correlators from noncritical string theory,''
  Phys.\ Lett.\  {\bf B428 } (1998)  105-114.
  [hep-th/9802109];
%\cite{Witten:1998qj}
%\bibitem{Witten:1998qj}
  E.~Witten,
  %``Anti-de Sitter space and holography,''
  Adv.\ Theor.\ Math.\ Phys.\  {\bf 2 } (1998)  253-291.
  [hep-th/9802150].

%\cite{Policastro:2001yc}
\bibitem{Policastro:2001yc}
  G.~Policastro, D.~T.~Son and A.~O.~Starinets,
  %``The shear viscosity of strongly coupled N = 4 supersymmetric Yang-Mills
  %plasma,''
  Phys.\ Rev.\ Lett.\  {\bf 87}, 081601 (2001)
  [arXiv:hep-th/0104066].
  %%CITATION = PRLTA,87,081601;%%

%\cite{Janik:2005zt}
\bibitem{Janik:2005zt}
  R.~A.~Janik and R.~B.~Peschanski,
  %``Asymptotic perfect fluid dynamics as a consequence of AdS/CFT,''
  Phys.\ Rev.\  D {\bf 73}, 045013 (2006)
  [arXiv:hep-th/0512162].
  %%CITATION = PHRVA,D73,045013;%%

\bibitem{CasalderreySolana:2011us}
  J.~Casalderrey-Solana, H.~Liu, D.~Mateos, K.~Rajagopal and
  U.~A.~Wiedemann,
  %``Gauge/String Duality, Hot QCD and Heavy Ion Collisions,''
  arXiv:1101.0618 [hep-th].

%\cite{Baier:2000sb}
\bibitem{Baier:2000sb}
  R.~Baier, A.~H.~Mueller, D.~Schiff and D.~T.~Son,
  %``'Bottom-up' thermalization in heavy ion collisions,''
  Phys.\ Lett.\  B {\bf 502}, 51 (2001)
  [arXiv:hep-ph/0009237].
  %%CITATION = PHLTA,B502,51;%%

%\cite{Mueller:2005un}
\bibitem{Mueller:2005un}
  A.~H.~Mueller, A.~I.~Shoshi and S.~M.~H.~Wong,
  %``A possible modified 'bottom-up' thermalization in heavy ion collisions,''
  Phys.\ Lett.\  B {\bf 632}, 257 (2006)
  [arXiv:hep-ph/0505164].
  %%CITATION = PHLTA,B632,257;%%

%boost invariance

%\cite{Kovchegov:2007pq}
\bibitem{Kovchegov:2007pq}
  Y.~V.~Kovchegov, A.~Taliotis,
  %``Early Time Dynamics in Heavy Ion Collisions from AdS/CFT Correspondence,''
  Phys.\ Rev.\  {\bf C76 } (2007)  014905.
  [arXiv:0705.1234 [hep-ph]].

%\cite{Chesler:2008hg}
\bibitem{Chesler:2008hg}
  P.~M.~Chesler and L.~G.~Yaffe,
  %``Horizon formation and far-from-equilibrium isotropization in supersymmetric
  %Yang-Mills plasma,''
  Phys.\ Rev.\ Lett.\  {\bf 102} (2009) 211601
  [arXiv:0812.2053 [hep-th]];
  %%CITATION = PRLTA,102,211601;%%
%\cite{Chesler:2009cy}
%\bibitem{Chesler:2009cy}
  P.~M.~Chesler, L.~G.~Yaffe,
  %``Boost invariant flow, black hole formation, and far-from-equilibrium dynamics in N = 4 supersymmetric Yang-Mills theory,''
  Phys.\ Rev.\  {\bf D82 } (2010)  026006.
  [arXiv:0906.4426 [hep-th]].

%\cite{Beuf:2009cx}
\bibitem{Beuf:2009cx}
  G.~Beuf, M.~P.~Heller, R.~A.~Janik and R.~Peschanski,
  %``Boost-invariant early time dynamics from AdS/CFT,''
  arXiv:0906.4423 [hep-th].
  %%CITATION = ARXIV:0906.4423;%%

%\cite{Catterall:2010fx}
\bibitem{Catterall:2010fx}
  S.~Catterall, A.~Joseph and T.~Wiseman,
  %``Thermal phases of D1-branes on a circle from lattice super Yang-Mills,''
  JHEP {\bf 1012}, 022 (2010)
  [arXiv:1008.4964 [hep-th]].
  %%CITATION = JHEPA,1012,022;%%

%\cite{Hanada:2010gs}
\bibitem{Hanada:2010gs}
  M.~Hanada,
  %``A proposal of a fine tuning free formulation of 4d N=4 super Yang-Mills,''
  JHEP {\bf 1011}, 112 (2010)
  [arXiv:1009.0901 [hep-lat]].
  %%CITATION = JHEPA,1011,112;%%

% quasinormal modes

\bibitem{Chandrasekhar:1975zza}
  S.~Chandrasekhar and S.~L.~Detweiler,
  %``The quasi-normal modes of the Schwarzschild black hole,''
  Proc.\ Roy.\ Soc.\ Lond.\  A {\bf 344}, 441 (1975);
  %%CITATION = PRSLA,A344,441;%%
%
%\bibitem{Horowitz:1999jd}
  G.~T.~Horowitz and V.~E.~Hubeny,
  %``Quasinormal modes of AdS black holes and the approach to thermal
  %equilibrium,''
  Phys.\ Rev.\  D {\bf 62}, 024027 (2000)
  [arXiv:hep-th/9909056];
  %%CITATION = PHRVA,D62,024027;%%
%
%\cite{Starinets:2002br}
%\bibitem{Starinets:2002br}
  A.~O.~Starinets,
  %``Quasinormal modes of near extremal black branes,''
  Phys.\ Rev.\  {\bf D66 } (2002)  124013.
  [hep-th/0207133];
%
%\bibitem{Kovtun:2005ev}
  P.~K.~Kovtun and A.~O.~Starinets,
  %``Quasinormal modes and holography,''
  Phys.\ Rev.\  D {\bf 72}, 086009 (2005)
  [arXiv:hep-th/0506184];
  %%CITATION = PHRVA,D72,086009;%%
  %\cite{Janik:2006gp}
%\bibitem{Janik:2006gp}
  R.~A.~Janik and R.~B.~Peschanski,
  %``Gauge / gravity duality and thermalization of a boost-invariant perfect
  %fluid,''
  Phys.\ Rev.\  D {\bf 74}, 046007 (2006)
  [arXiv:hep-th/0606149];
  %%CITATION = PHRVA,D74,046007;%%
%
 %\cite{Friess:2006kw}
%\bibitem{Friess:2006kw}
  J.~J.~Friess, S.~S.~Gubser, G.~Michalogiorgakis, S.~S.~Pufu,
  %``Expanding plasmas and quasinormal modes of anti-de Sitter black holes,''
  JHEP {\bf 0704 } (2007)  080.
  [hep-th/0611005];
%\cite{Iqbal:2008by}
%\bibitem{Iqbal:2008by}
  N.~Iqbal, H.~Liu,
  %``Universality of the hydrodynamic limit in AdS/CFT and the membrane paradigm,''
  Phys.\ Rev.\  {\bf D79 } (2009)  025023.
  [arXiv:0809.3808 [hep-th]].

%\cite{Hubeny:2010ry}
\bibitem{Hubeny:2010ry}
  V.~E.~Hubeny and M.~Rangamani,
  %``A holographic view on physics out of equilibrium,''
  Adv.\ High Energy Phys.\  {\bf 2010}, 297916 (2010)
  [arXiv:1006.3675 [hep-th]].
  %%CITATION = 00642,2010,297916;%%

%\cite{Danielsson:1999zt}
\bibitem{Danielsson:1999zt}
  U.~H.~Danielsson, E.~Keski-Vakkuri and M.~Kruczenski,
  %``Spherically collapsing matter in AdS, holography, and shellons,''
  Nucl.\ Phys.\  B {\bf 563}, 279 (1999)
  [arXiv:hep-th/9905227].
  %%CITATION = NUPHA,B563,279;%%

  %\cite{Giddings:1999zu}
\bibitem{Giddings:1999zu}
  S.~B.~Giddings and S.~F.~Ross,
  %``D3-brane shells to black branes on the Coulomb branch,''
  Phys.\ Rev.\  D {\bf 61}, 024036 (2000)
  [arXiv:hep-th/9907204].
  %%CITATION = PHRVA,D61,024036;%%

%\cite{Danielsson:1999fa}
\bibitem{Danielsson:1999fa}
  U.~H.~Danielsson, E.~Keski-Vakkuri and M.~Kruczenski,
  %``Black hole formation in AdS and thermalization on the boundary,''
  JHEP {\bf 0002}, 039 (2000)
  [arXiv:hep-th/9912209].
  %%CITATION = JHEPA,0002,039;%%

%\cite{Giddings:2001ii}
\bibitem{Giddings:2001ii}
  S.~B.~Giddings and A.~Nudelman,
  %``Gravitational collapse and its boundary description in AdS,''
  JHEP {\bf 0202}, 003 (2002)
  [arXiv:hep-th/0112099].
  %%CITATION = JHEPA,0202,003;%%

% shock waves collision

%\cite{Kang:2004jd}
\bibitem{Kang:2004jd}
  K.~Kang, H.~Nastase,
  %``High energy QCD from Planckian scattering in AdS and the Froissart bound,''
  Phys.\ Rev.\  {\bf D72 } (2005)  106003.
  [hep-th/0410173];
%\cite{Gubser:2007zr}
%\bibitem{Gubser:2007zr}
  S.~S.~Gubser, S.~S.~Pufu, A.~Yarom,
  %``Shock waves from heavy-quark mesons in AdS/CFT,''
  JHEP {\bf 0807 } (2008)  108.
  [arXiv:0711.1415 [hep-th]];
%\cite{Grumiller:2008va}
%\bibitem{Grumiller:2008va}
  D.~Grumiller, P.~Romatschke,
  %``On the collision of two shock waves in AdS(5),''
  JHEP {\bf 0808 } (2008)  027.
  [arXiv:0803.3226 [hep-th]];
%\cite{Gubser:2008pc}
%\bibitem{Gubser:2008pc}
  S.~S.~Gubser, S.~S.~Pufu and A.~Yarom,
  %``Entropy production in collisions of gravitational shock waves and of heavy
  %ions,''
  Phys.\ Rev.\  D {\bf 78}, 066014 (2008)
  [arXiv:0805.1551 [hep-th]];
  %%CITATION = PHRVA,D78,066014;%%
%\cite{Albacete:2008vs}
%\bibitem{Albacete:2008vs}
  J.~L.~Albacete, Y.~V.~Kovchegov, A.~Taliotis,
  %``Modeling Heavy Ion Collisions in AdS/CFT,''
  JHEP {\bf 0807 } (2008)  100.
  [arXiv:0805.2927 [hep-th]];
%\cite{AlvarezGaume:2008fx}
%\bibitem{AlvarezGaume:2008fx}
  L.~Alvarez-Gaume, C.~Gomez, A.~Sabio Vera, A.~Tavanfar, M.~A.~Vazquez-Mozo,
  %``Critical formation of trapped surfaces in the collision of gravitational shock waves,''
  JHEP {\bf 0902 } (2009)  009.
  [arXiv:0811.3969 [hep-th]];
%\cite{Albacete:2009ji}
%\bibitem{Albacete:2009ji}
  J.~L.~Albacete, Y.~V.~Kovchegov, A.~Taliotis,
  %``Asymmetric Collision of Two Shock Waves in AdS(5),''
  JHEP {\bf 0905 } (2009)  060.
  [arXiv:0902.3046 [hep-th]];
%\cite{Lin:2009pn}
%\bibitem{Lin:2009pn}
  S.~Lin, E.~Shuryak,
  %``Grazing Collisions of Gravitational Shock Waves and Entropy Production in Heavy Ion Collision,''
  Phys.\ Rev.\  {\bf D79 } (2009)  124015.
  [arXiv:0902.1508 [hep-th]];
%\cite{Gubser:2009sx}
%\bibitem{Gubser:2009sx}
  S.~S.~Gubser, S.~S.~Pufu, A.~Yarom,
  %``Off-center collisions in AdS(5) with applications to multiplicity estimates in heavy-ion collisions,''
  JHEP {\bf 0911 } (2009)  050.
  [arXiv:0902.4062 [hep-th]];
%\cite{Kovchegov:2009du}
%\bibitem{Kovchegov:2009du}
  Y.~V.~Kovchegov, S.~Lin,
  %``Toward Thermalization in Heavy Ion Collisions at Strong Coupling,''
  JHEP {\bf 1003 } (2010)  057.
  [arXiv:0911.4707 [hep-th]];
%
%\bibitem{Kovchegov:2010zg}
  Y.~V.~Kovchegov,
  %``Shock Wave Collisions and Thermalization in AdS_5,''
  arXiv:1011.0711 [hep-th];
  %%CITATION = ARXIV:1011.0711;%%
%\cite{Chesler:2010bi}
%\bibitem{Chesler:2010bi}
  P.~M.~Chesler, L.~G.~Yaffe,
  %``Holography and colliding gravitational shock waves in asymptotically AdS_5 spacetime,''
  Phys.\ Rev.\ Lett.\  {\bf 106 } (2011)  021601.
  [arXiv:1011.3562 [hep-th]].

%\cite{Lin:2006rf}
\bibitem{Lin:2006rf}
  S.~Lin and E.~Shuryak,
  %``Toward the AdS/CFT gravity dual for High Energy Collisions: I.Falling into
  %the AdS,''
  Phys.\ Rev.\  D {\bf 77}, 085013 (2008)
  [arXiv:hep-ph/0610168];
  %%CITATION = PHRVA,D77,085013;%%
%\cite{Lin:2007fa}
%\bibitem{Lin:2007fa}
  S.~Lin and E.~Shuryak,
  %``Toward the AdS/CFT Gravity Dual for High Energy Collisions: II. The Stress
  %Tensor on the Boundary,''
  Phys.\ Rev.\  D {\bf 77}, 085014 (2008)
  [arXiv:0711.0736 [hep-th]].
  %%CITATION = PHRVA,D77,085014;%%

%\cite{Bhattacharyya:2009uu}
\bibitem{Bhattacharyya:2009uu}
  S.~Bhattacharyya and S.~Minwalla,
  %``Weak Field Black Hole Formation in Asymptotically AdS Spacetimes,''
  arXiv:0904.0464 [hep-th].
  %%CITATION = ARXIV:0904.0464;%%

%\cite{Janik:2006ft}
\bibitem{Janik:2006ft}
  R.~A.~Janik,
  %``Viscous plasma evolution from gravity using AdS/CFT,''
  Phys.\ Rev.\ Lett.\  {\bf 98 } (2007)  022302.
  [hep-th/0610144].

%\cite{Lin:2008rw}
\bibitem{Lin:2008rw}
  S.~Lin and E.~Shuryak,
  %``Toward the AdS/CFT Gravity Dual for High Energy Collisions.
  %3.Gravitationally Collapsing Shell and Quasiequilibrium,''
  Phys.\ Rev.\  D {\bf 78}, 125018 (2008)
  [arXiv:0808.0910 [hep-th]].
  %%CITATION = PHRVA,D78,125018;%%

%\cite{Hubeny:2007xt}
\bibitem{Hubeny:2007xt}
  V.~E.~Hubeny, M.~Rangamani, T.~Takayanagi,
  %``A Covariant holographic entanglement entropy proposal,''
  JHEP {\bf 0707 } (2007)  062.
  [arXiv:0705.0016 [hep-th]].

\bibitem{AbajoArrastia:2010yt}
  J.~Abajo-Arrastia, J.~Aparicio and E.~Lopez,
  %``Holographic Evolution of Entanglement Entropy,''
  arXiv:1006.4090 [hep-th].

%\cite{Albash:2010mv}
\bibitem{Albash:2010mv}
  T.~Albash and C.~V.~Johnson,
  %``Evolution of Holographic Entanglement Entropy after Thermal and
  %Electromagnetic Quenches,''
  arXiv:1008.3027 [hep-th].
  %%CITATION = ARXIV:1008.3027;%%

 %\cite{Erdmenger:2011jb}
\bibitem{Erdmenger:2011jb}
  J.~Erdmenger, S.~Lin and T.~H.~Ngo,
  %``A moving mirror in AdS space as a toy model for holographic
  %thermalization,''
  arXiv:1101.5505 [hep-th].
  %%CITATION = ARXIV:1101.5505;%%

%\cite{CaronHuot:2011dr}
\bibitem{CaronHuot:2011dr}
  S.~Caron-Huot, P.~M.~Chesler and D.~Teaney,
  %``Fluctuation, dissipation, and thermalization in non-equilibrium AdS_5 black
  %hole geometries,''
  arXiv:1102.1073 [hep-th].
  %%CITATION = ARXIV:1102.1073;%%

%\cite{Skenderis:2008dh}
\bibitem{Skenderis:2008dh}
  K.~Skenderis and B.~C.~van Rees,
  %``Real-time gauge/gravity duality,''
  Phys.\ Rev.\ Lett.\  {\bf 101}, 081601 (2008)
  [arXiv:0805.0150 [hep-th]];
  %%CITATION = PRLTA,101,081601;%%
%\cite{Skenderis:2008dg}
%\bibitem{Skenderis:2008dg}
  K.~Skenderis and B.~C.~van Rees,
  %``Real-time gauge/gravity duality: Prescription, Renormalization and
  %Examples,''
  JHEP {\bf 0905}, 085 (2009)
  [arXiv:0812.2909 [hep-th]].
  %%CITATION = JHEPA,0905,085;%%

%\cite{Arsiwalla:2010bt}
\bibitem{Arsiwalla:2010bt}
  X.~Arsiwalla, J.~de Boer, K.~Papadodimas and E.~Verlinde,
  %``Degenerate Stars and Gravitational Collapse in AdS/CFT,''
  JHEP {\bf 1101}, 144 (2011)
  [arXiv:1010.5784 [hep-th]].
  %%CITATION = JHEPA,1101,144;%%
  
  %\cite{Das:2010yw}
\bibitem{Das:2010yw}
 S.~R.~Das, T.~Nishioka and T.~Takayanagi,
 %``Probe Branes, Time-dependent Couplings and Thermalization in AdS/CFT,''
 JHEP {\bf 1007}, 071 (2010)
 [arXiv:1005.3348 [hep-th]].
 %%CITATION = JHEPA,1007,071;%%

%\cite{Hashimoto:2010wv}
\bibitem{Hashimoto:2010wv}
 K.~Hashimoto, N.~Iizuka and T.~Oka,
 %``Rapid Thermalization by Baryon Injection in Gauge/Gravity Duality,''
 arXiv:1012.4463 [hep-th].
 %%CITATION = ARXIV:1012.4463;%%

%\cite{Kovner:2001vi}
\bibitem{Kovner:2001vi}
  A.~Kovner and U.~A.~Wiedemann,
  %``Eikonal evolution and gluon radiation,''
  Phys.\ Rev.\  D {\bf 64}, 114002 (2001);
  %[arXiv:hep-ph/0106240].
  %%CITATION = PHRVA,D64,114002;%%

%\cite{Liu:2006he}
\bibitem{Liu:2006he}
  H.~Liu, K.~Rajagopal and U.~A.~Wiedemann,
  %``Wilson loops in heavy ion collisions and their calculation in AdS/CFT,''
  JHEP {\bf 0703}, 066 (2007).
  %[arXiv:hep-ph/0612168].
  %%CITATION = JHEPA,0703,066;%%

%\cite{Kovtun:2006pf}
\bibitem{Kovtun:2006pf}
  P.~Kovtun and A.~Starinets,
  %``Thermal spectral functions of strongly coupled N = 4 supersymmetric
  %Yang-Mills theory,''
  Phys.\ Rev.\ Lett.\  {\bf 96}, 131601 (2006)
  [arXiv:hep-th/0602059].
  %%CITATION = PRLTA,96,131601;%%

  %\cite{Balasubramanian:1999zv}
\bibitem{Balasubramanian:1999zv}
  V.~Balasubramanian and S.~F.~Ross,
  %``Holographic particle detection,''
  Phys.\ Rev.\  D {\bf 61} (2000) 044007
  [arXiv:hep-th/9906226].
  %%CITATION = PHRVA,D61,044007;%%

\bibitem{Louko:2000tp}
  J.~Louko, D.~Marolf, S.~F.~Ross,
  %``On geodesic propagators and black hole holography,''
  Phys.\ Rev.\  {\bf D62}, 044041 (2000).
  [hep-th/0002111].

%\cite{Fidkowski:2003nf}
\bibitem{Fidkowski:2003nf}
  L.~Fidkowski, V.~Hubeny, M.~Kleban and S.~Shenker,
  %``The black hole singularity in AdS/CFT,''
  JHEP {\bf 0402}, 014 (2004)
  [arXiv:hep-th/0306170].
  %%CITATION = JHEPA,0402,014;%%

\bibitem{Festuccia:2005pi}
  G.~Festuccia, H.~Liu,
  %``Excursions beyond the horizon: Black hole singularities in Yang-Mills theories. I.,''
  JHEP {\bf 0604}, 044 (2006).
  [hep-th/0506202].

  %\cite{Maldacena:1998im}
\bibitem{Maldacena:1998im}
  J.~M.~Maldacena,
  ``Wilson loops in large N field theories,''
  Phys.\ Rev.\ Lett.\  {\bf 80} (1998) 4859
  [arXiv:hep-th/9803002];
  %%CITATION = PRLTA,80,4859;%%
%\cite{Rey:1998ik}
%\bibitem{Rey:1998ik}
  S.~J.~Rey, J.~T.~Yee,
  %``Macroscopic strings as heavy quarks in large N gauge theory and anti-de Sitter supergravity,''
  Eur.\ Phys.\ J.\  {\bf C22 } (2001)  379-394.
  [hep-th/9803001].

\bibitem{Ryu:2006bv}
  S.~Ryu, T.~Takayanagi,
  %``Holographic derivation of entanglement entropy from AdS/CFT,''
  Phys.\ Rev.\ Lett.\  {\bf 96}, 181602 (2006).
  [hep-th/0603001].

%\cite{Nishioka:2009un}
\bibitem{Nishioka:2009un}
  T.~Nishioka, S.~Ryu, T.~Takayanagi,
  %``Holographic Entanglement Entropy: An Overview,''
  J.\ Phys.\ A {\bf A42}, 504008 (2009).
  [arXiv:0905.0932 [hep-th]].

%\cite{Balasubramanian:2010ce}
\bibitem{Balasubramanian:2010ce}
  V.~Balasubramanian, A.~Bernamonti, J.~de Boer, N.~Copland, B.~Craps,
	E.~Keski-Vakkuri, B.~M\"uller, A.~Sch\"afer, M.~Shigemori,
	W.~Staessens,
  %``Thermalization of Strongly Coupled Field Theories,''
  arXiv:1012.4753 [hep-th].
  %%CITATION = ARXIV:1012.4753;%%

%\cite{Hatta:2008tx}
\bibitem{Hatta:2008tx}
  Y.~Hatta, E.~Iancu and A.~H.~Mueller,
  %``Jet evolution in the N=4 SYM plasma at strong coupling,''
  JHEP {\bf 0805}, 037 (2008)
  [arXiv:0803.2481 [hep-th]].
  %%CITATION = JHEPA,0805,037;%%

%\cite{Iancu:2008sp}
\bibitem{Iancu:2008sp}
  E.~Iancu,
  %``Partons and jets in a strongly-coupled plasma from AdS/CFT,''
  Acta Phys.\ Polon.\  B {\bf 39}, 3213 (2008)
  [arXiv:0812.0500 [hep-ph]].
  %%CITATION = APPOA,B39,3213;%%

\bibitem{Son:2002sd}
  D.~T.~Son and A.~O.~Starinets,
  %``Minkowski-space correlators in AdS/CFT correspondence: Recipe and applications,''
  JHEP {\bf 0209}, 042 (2002)
  [arXiv:hep-th/0205051].
  %%CITATION = JHEPA,0209,042;%%

\bibitem{Balasubramanian:1998sn}
  V.~Balasubramanian, P.~Kraus, A.~E.~Lawrence,
  %``Bulk versus boundary dynamics in anti-de Sitter space-time,''
  Phys.\ Rev.\  {\bf D59}, 046003 (1999).
  [hep-th/9805171].

%\cite{McGreevy:2009xe}
\bibitem{McGreevy:2009xe}
  J.~McGreevy,
  %``Holographic duality with a view toward many-body physics,''
  arXiv:0909.0518 [hep-th].
  %%CITATION = ARXIV:0909.0518;%%

%\cite{Berenstein:1998ij}
\bibitem{Berenstein:1998ij}
  D.~E.~Berenstein, R.~Corrado, W.~Fischler and J.~M.~Maldacena,
  ``The operator product expansion for Wilson loops and surfaces in the  large
  N limit,''
  Phys.\ Rev.\  D {\bf 59} (1999) 105023
  [arXiv:hep-th/9809188].
  %%CITATION = PHRVA,D59,105023;%%

\bibitem{Figueras:2009iu}
  P.~Figueras, V.~E.~Hubeny, M.~Rangamani and S.~F.~Ross,
  %``Dynamical black holes and expanding plasmas,''
  JHEP {\bf 0904}, 137 (2009)
  [arXiv:0902.4696 [hep-th]].

%\cite{Hubeny:2011yk}
\bibitem{Hubeny:2011yk}
  V.~E.~Hubeny,
  %``Holographic insights and puzzles,''
  [arXiv:1103.1999 [hep-th]].
  
%\cite{Calabrese:2004eu}
\bibitem{Calabrese:2004eu}
  P.~Calabrese, J.~L.~Cardy,
  %``Entanglement entropy and quantum field theory,''
  J.\ Stat.\ Mech.\  {\bf 0406}, P002 (2004).
  [hep-th/0405152].

\bibitem{Calabrese:2009qy}
  P.~Calabrese, J.~Cardy,
  %``Entanglement entropy and conformal field theory,''
  J.\ Phys.\ A {\bf A42}, 504005 (2009).
  [arXiv:0905.4013 [cond-mat.stat-mech]].

\bibitem{Srednicki:1993im}
  M.~Srednicki,
  %``Entropy and area,''
  Phys.\ Rev.\ Lett.\  {\bf 71}, 666 (1993)
  [arXiv:hep-th/9303048].
  %%CITATION = PRLTA,71,666;%%

%\cite{Bombelli:1986rw}
\bibitem{Bombelli:1986rw}
  L.~Bombelli, R.~K.~Koul, J.~Lee and R.~D.~Sorkin,
  %``A Quantum Source of Entropy for Black Holes,''
  Phys.\ Rev.\  D {\bf 34}, 373 (1986).
  %%CITATION = PHRVA,D34,373;%%

\bibitem{Takayanagi:2010wp}
  T.~Takayanagi and T.~Ugajin,
  %``Measuring Black Hole Formations by Entanglement Entropy via Coarse-Graining,''
  arXiv:1008.3439 [hep-th].
  %%CITATION = ARXIV:1008.3439;%%

\bibitem{Sinai}
  A.~N.~Kolmogorov,
  Dokl.\ Akad.\ Nauk. SSSR {\bf 119}, 861 (1958),
  {\em ibid.} {\bf 124}, 754 (1959);
  Ya.~G.~Sina\"i
  %``On the concept of entropy for a dynamic system,''
  Dokl.\ Akad.\ Nauk SSSR {\bf 124}, 768 (1959).

\bibitem{Bolte:1999th}
  J.~Bolte, B.~M\"uller, and A.~Sch\"afer,
  %``Ergodic properties of classical SU(2) lattice gauge theory,''
  Phys.\ Rev.\ D {\bf 61}, 054506 (2000);
%\bibitem{Kunihiro:2010tg}
  T.~Kunihiro {\em et al.},
  %``Chaotic behavior in classical Yang-Mills dynamics,''
  Phys.\ Rev.\ D {\bf 82}, 114015 (2010).

\bibitem{Latora:1999xx}
  V.~Latora and M.~Baranger,
  %``Kolmogorov-Sinai entropy rate versus physical entropy,''
  Phys.\ Rev.\ Lett.\ {\bf 82}, 520 (1999).

%\cite{Calabrese:2006xx}
\bibitem{Calabrese:2006xx}
  P.~Calabrese and J.~Cardy,
  %``Quantum Quenches in Extended Systems,''
  Phys.\ Rev. Lett.\  {\bf 96}, 136801 (2006)
  [cond-mat/0601225].

%\cite{Calabrese:2007rg}
\bibitem{Calabrese:2007rg}
  P.~Calabrese and J.~Cardy,
  %``Quantum Quenches in Extended Systems,''
  J.\ Stat.\ Mech.\  {\bf 0706}, P06008 (2007)
  [arXiv:0704.1880 [cond-mat.stat-mech]].
  %%CITATION = JSTAT,0706,P06008;%%

%\cite{Ebrahim:2010ra}
\bibitem{Ebrahim:2010ra}
  H.~Ebrahim and M.~Headrick,
  %``Instantaneous Thermalization in Holographic Plasmas,''
  arXiv:1010.5443 [hep-th].
  %%CITATION = ARXIV:1010.5443;%%

\bibitem{Sekino:2008he}
  Y.~Sekino and L.~Susskind,
  %``Fast Scramblers,''
  JHEP {\bf 0810}, 065 (2008)
  [arXiv:0808.2096 [hep-th]].

\end{thebibliography}
\end{document}